# Facebook's Architecture Undermines Vaccine Misinformation Removal Efforts

**Authors:** David A. Broniatowski[1,2]*, Jiayan Gu[3], Amelia M. Jamison[4], Joseph R. Simons[5]**,

Lorien C. Abroms[2,3]

**Affiliations:**

[1]Department of Engineering Management and Systems Engineering, The George Washington University; Washington, DC, 20052, USA.

[2]Institute for Data, Democracy, and Politics, The George Washington University; Washington, DC, 20052, USA.

[3]Department of Prevention and Community Health, The George Washington University; Washington, DC, 20052, USA.

[4]Department of Health, Behavior, and Society, Bloomberg School of Public Health, Johns Hopkins University; Baltimore, MD, 21205, USA.

[5]Office of the Assistant Secretary for Financial Resources, United States Department of Health & Human Services; Washington, DC, 20543, USA.

*Corresponding author. Email: broniatowski@gwu.edu

**The views expressed are those of the author and do not reflect the official position of the U.S. Department of Health and Human Services, or the United States

**Misinformation promotes distrust in science,[1] undermines public health,[2] and may drive civil unrest.[3–5] Vaccine misinformation, in particular, has stalled efforts to overcome the COVID-19 pandemic,[6,7] prompting social media platforms' attempts to reduce it. Some have questioned whether "soft" content moderation remedies[8,9] –e.g., flagging and downranking misinformation – were successful, suggesting that the addition of "hard" content remedies – e.g., deplatforming and content bans – is necessary.[10–14] We therefore examined whether Facebook's vaccine misinformation content removal policies were effective. Here, we show that Facebook's policies reduced the number of anti-vaccine posts but also caused several perverse effects: pro-vaccine content was also removed, engagement with remaining anti-vaccine content repeatedly recovered to pre-policy levels, and this content became more misinformative, more politically polarised, and more likely to be seen in users' newsfeeds. We explain these results as an unintended consequence of Facebook's design goal: promoting community formation. Members of communities dedicated to vaccine refusal appear to seek out misinformation from multiple sources. Community administrators make use of several channels afforded by the Facebook platform to disseminate misinformation. Our findings suggest the need to address how social media platform architecture enables community formation and mobilisation around misinformative topics when managing the spread of online content.**

**Main**

Online misinformation undermines trust in scientific evidence[1] and medical recommendations.[2] It has been linked to harmful offline behaviours including stalled public health efforts,[6] civil unrest,[5] and mass violence.[4] The COVID-19 pandemic has exacerbated misinformation online, resulting in widespread concern about low vaccine uptake rates,[11,15] even as government officials continue to urge vaccination in response to new disease variants.[16] Social media has enabled misinformation to be quickly and widely disseminated. Therefore, policymakers and public officials have put significant pressure on social media platforms to curtail misinformation spread.[17–19]

Several major platforms have taken steps intended to reduce misinformation spread. Years of "soft" content remedies – such as warning labels and downranking objectionable content in search – have demonstrated some success;[20,21] however, misinformation continues to spread widely online, leading many to question the impact and efficacy of these interventions.[22] Some have suggested that combining these "soft" remedies with "hard" content remedies[7,8] – content removal and deplatforming objectionable accounts[10,23,24] – could significantly curtail misinformation spread.[14] In practice, social media platforms already use a combination of "hard" and "soft" content remedies; however, evidence for the short-term efficacy of "hard" remedies is mixed,[25–28] and the long-term efficacy of such strategies has not been systematically examined. "Hard" remedies have also spurred accusations of censorship and concrete legal action.[29,30] There is therefore a critical need to understand whether this combination of remedies is effective, and if not, why not.

Here, we conduct a comprehensive evaluation of the world's largest social media platform's – Facebook's – attempts to remove anti-vaccine misinformation as COVID-19 vaccines were rolled out in late 2020 and early 2021. We estimate the causal impacts of Facebook's remedies on anti- and pro-vaccine content before and after Facebook implemented "hard" remedies targeting anti-vaccine misinformation. Our dataset consists of roughly 1 million posts from 488 English-language public Facebook pages and groups identified on November 15, 2020, and covers a time period beginning immediately prior to the outbreak of the COVID-19 pandemic through the end of the first Omicron variant wave in the United States: November 15, 2019 through February 28, 2022. (A second dataset of about 1 million posts from 297 venues identified on July 21, 2021, was also collected for replication purposes.) As expected, we found evidence that such policies, when added to existing "soft" content remedies reduced the number of anti-vaccine posts. However, we also observed several unexpected findings that, taken together, call into question the long-term efficacy of Facebook's policies, and suggest that they may have even been counterproductive. We examine the mechanisms that may underlie these observations, explaining them as a consequence of Facebook's core design goals and system architecture. Facebook is designed to build online communities. (Facebook's mission statement is to "give people the power to build community, and to bring the world closer together"[31].) In implementing this goal, Facebook has created a system that affords anti-vaccine content providers several ways to circumvent the intent of content moderation efforts. These content providers use Facebook to construct communities around vaccine scepticism, encouraging and enabling users to seek out anti-vaccine content online, but also encouraging these communities to take action offline action, such as participating in political activism. Taken together, our findings suggest the need to address a new dimension – how social media platform design features enable community formation and mobilisation around misinformative topics – to the problem of managing the spread of misinformation online.

**Anti- and Pro-Vaccine Posts Were Removed**

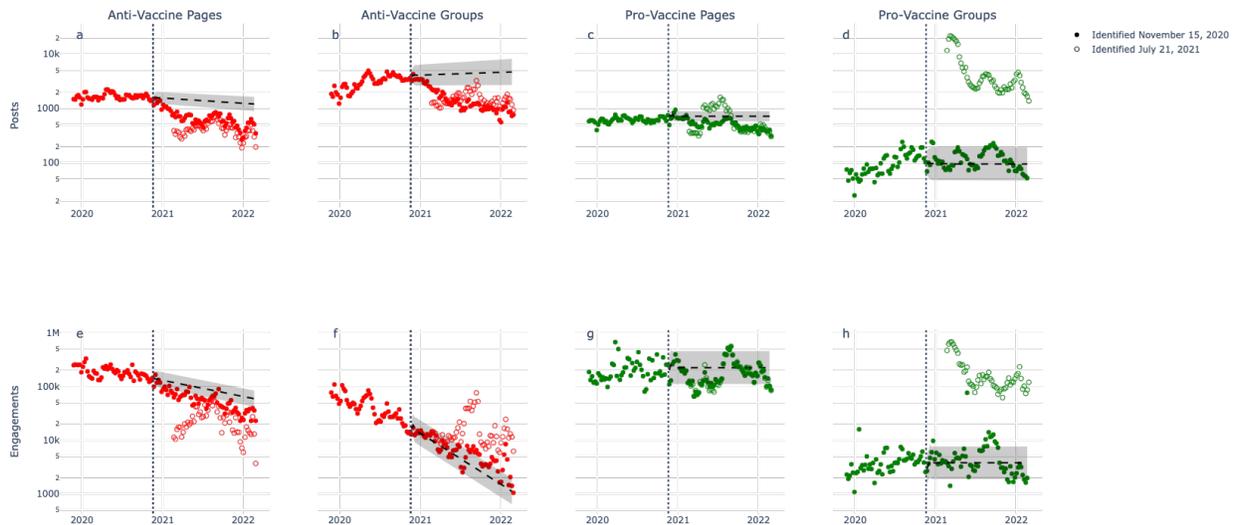

**Fig 1. Facebook's policies did not lead to a sustained decrease in engagement with anti-vaccine content despite a reduction in the total number of posts.**

ARIMA model results show that, compared to projections (black dashed line), Facebook's policies led to a decline in posts in **a.** anti-vaccine pages (51% average decrease, $\chi^2(66)= 1425.15, p<0.001$) **b.** anti-vaccine groups (63% average decrease, $\chi^2(66)= 1030.71, p<0.001$), and **c.** pro-vaccine pages (26% average decline, $\chi^2(66)= 612.17, p<0.001$). We did not detect a significant difference in **d.** pro-vaccine groups (25% average increase, $\chi^2(66)= 61.15, p=0.65$). Additionally, engagements declined with content in **e.** anti-vaccine pages (37% average decrease, $\chi^2(66)= 468.37, p<0.001$). Despite this average decline, engagements fell within the prediction's 90% confidence intervals for the months of November 2020-January 2021, April-May 2021, August – September, 2021, and November 2021. Engagements increased significantly beyond pre-policy predictions in **f.** anti-vaccine groups (52% average increase, $\chi^2(66)= 164.27, p<0.001$). Engagements declined with content in **g.** pro-vaccine pages (13% average decrease,

$x^2(66)= 111.43, p<0.001$) but **h.** increased for the months of August – October 2021 in pro-vaccine groups (52% average increase in the first sample, $x^2(66)= 164.64, p<0.001$). Error bars reflect 90% confidence intervals. Data from the second sample is shown for comparison and qualitatively replicates findings from the first sample. In anti-vaccine groups, maximum engagements exceeded 75,000 in September, 2021, and in pro-vaccine groups several new "vaccine hunter" groups had formed to raise awareness about COVID vaccine test sites.

We found that Facebook reduced the number of posts in anti-vaccine pages by 51%, $x^2(66)= 1425.15, p<0.001$, and in groups by 63%, $x^2(66)= 1030.71, p<0.001$, on average (Fig. 1a-b). However, Facebook's policy also appears to have had a notable "unintended consequence", having significantly reduced the number of posts in pro-vaccine pages, $x^2(66)= 612.17, p<0.001$, by 26% on average (Fig. 1c).

**Anti-Vaccine Engagement Recovered**

We expected to see that a reduction in the number of posts ("supply") led to a sustained reduction in the number of engagements ("demand").[32] We found that overall user engagement (Fig. 1e) and especially reactions (e.g., "likes" indicating demand; Extended Data Fig. 1i) to content in anti-vaccine pages repeatedly returned to pre-policy trends by May, 2021, and again by September, 2021. Compared to pre-policy trends, engagements with content in groups increased by 52% on average, $x^2(66)= 164.64, p<0.001$ (Fig. 1f). Engagement with content in the second sample of groups, collected in July 2021, exceeded the pre-policy trend by an average of 628%, reaching levels comparable to the pre-policy maximum in May, 2021. This means that

Facebook's content removals may not have reduced overall exposure to anti-vaccine content compared to pre-policy trends. We therefore posited that users actively "demand" – i.e., seek out – new misinformation despite Facebook attempting to curtail "supply". In order to more fully examine this supposition, we constructed a simulation model that we calibrated to pre-policy data (see Supplementary Information, S1). We best reproduced post-policy data when we assumed that prior content removals lead to an increase in demand for anti-vaccine content in remaining venues. Additionally, in our model, Facebook's current "hard" content remedies were most effective if coupled with a decrease in demand for misinformation.

**Facebook's Architecture Enables Demand**

An analysis of Facebook's system architecture explains why their content removal policies did not lead to a sustained decrease in engagement. Users interact with anti-vaccine content in three ways: 1) as followers of pages – venues run by anti-vaccine opinion leaders where only the administrator can post; 2) as members of groups – venues for public discussion where any group member (including page administrators) can post, comment, or engage; and 3) by interacting with posts on users' newsfeeds, which may originate from pages, groups, or other sources. The resulting layered hierarchy (Fig. 2) is resilient to disruption[33] – e.g., content or venue removal – while still allowing page administrators some degree of control over information flow.[34] Each layer in this hierarchy contributes to the system's resilience in different, cumulative, ways.

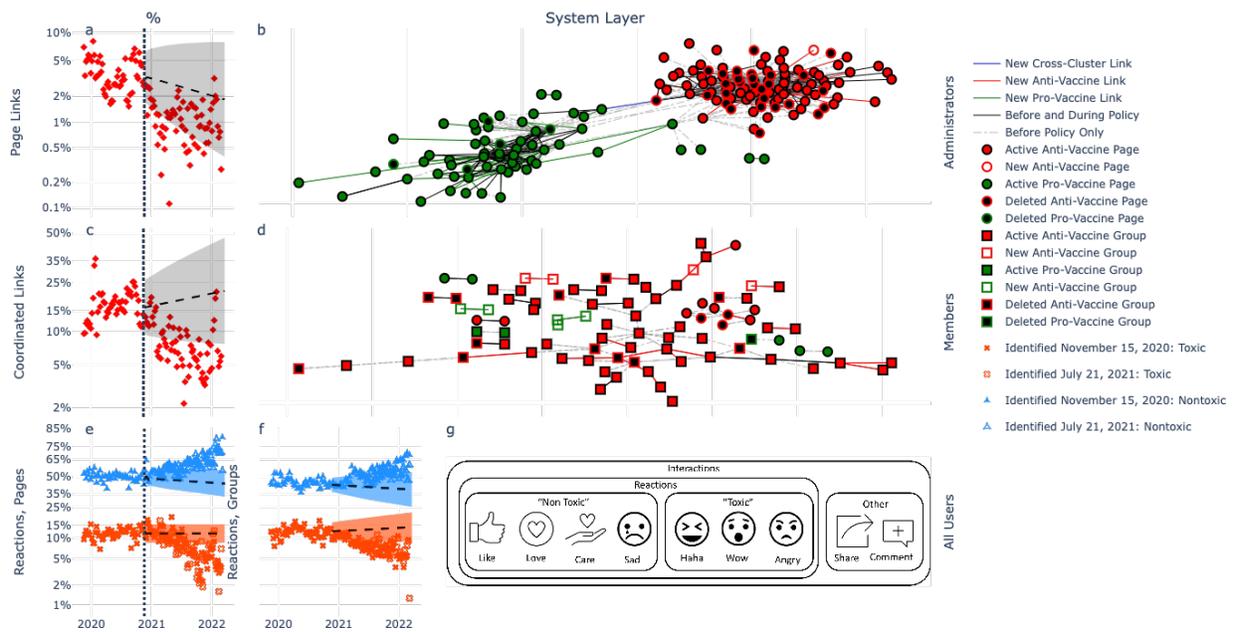

**Fig 2. Pages, groups, and newsfeeds form a layered hierarchy that facilitates access to, and demand for, anti-vaccine content despite "hard" remedies.**

In the "top" layer, administrators link to one another's pages, enabling users to circumvent Facebook's search algorithm. **a.** Combining data across samples to allow us to detect links between both, we found that page proportions decreased by 54% on average, $\chi^2(66)= 192.43$ $p<0.001$, but returned to the 90% confidence intervals by April 2021, and reached pre-policy projection (black dashed line) by June, 2021. **b.** Anti-vaccine (red) and pro-vaccine (green) administrators use these links to enable audiences to find new pages. In the "middle" layer, pages administrators, and other users, can create and administer groups, facilitating information exchange. Although intended to be organic communities, repeated simultaneous link sharing suggests coordinated behaviour between nominally separate venues. **c.** Combining data across samples, Facebook's policies did not lead to a sustained reduction in coordinated URL proportions in anti-vaccine venues compared to projections (black dashed line). Despite an average weekly decrease of 54%, $\chi^2(66)= 267.00$, $p<0.001$, coordinated link proportions returned to the 90% confidence intervals throughout 2021, reaching the pre-policy projection by January, 2022. **d.** Several anti-vaccine (red) venues engaged in this coordinated behaviour, forming a connected component including new groups. In contrast, coordination between pro-vaccine venues (green) was rare. In the "bottom" layer, users access content via their newsfeeds, including posts from pages and groups, but also other sources. **e.** In pages, Facebook's policies led to a 30% average increase in "non-toxic" reactions, $\chi^2(66)= 463.39$, $p<0.001$ and a 26% average decrease in "toxic", $\chi^2(66)= 816.74$, $p<0.001$. **f.** In groups – Facebook's policies led to a 33% average increase in "non-toxic" reactions, $\chi^2(66)= 280.21$, $p<0.001$ and a 40% average decrease in "toxic" reactions, $\chi^2(66)= 423.24$, $p<0.001$. **g.** Facebook's newsfeed algorithm reportedly[35] promotes content garnering these "non-toxic" reactions ("likes", "love", "care", and "sad") over "toxic" reactions ("angry", "haha", and "wow"). Proportions are shown on a logit scale.

First, page administrators can explicitly collaborate to share followers and content in case some pages are removed (e.g., by linking to or "liking"[36] one another's' pages). Consistent with this explanation, we observe that both anti-vaccine and pro-vaccine pages frequently posted links to other aligned pages, forming densely connected clusters (Fig 2b). Users accessing these links likely follow multiple pages; thus, removing some pages is unlikely to prevent access to anti-vaccine content. Specifically, for Facebook's policy to be successful, it would need to have led to a sustained reduction in links between anti-vaccine pages, eroding the system's resilience. Instead, we found that the proportion of links between anti-vaccine pages repeatedly returned to the pre-policy baseline by June, 2021, despite having decreased by an average of 54% following the policy, $\chi^2(66)= 192.43$ $p<0.001$ (Fig. 2a; in contrast, pro-vaccine pages were unaffected; Extended Data Fig. 2b). This means that the anti-vaccine page network was "self-healing", with page administrators enabling users to satisfy their demand for misinformation by directing them to new pages as older pages were removed.

A second source of resilience stems from covert coordination between venues that are nominally independent. Page administrators and other active users can reach larger audiences – and potentially recruit these audiences to follow pages – by sharing the same content in multiple different venues, and especially groups, simultaneously. Before the policy, we found evidence that several anti-vaccine groups routinely posted the same content at the same time – a sign of potential "coordinated inauthentic behaviour", which is explicitly prohibited by Facebook[37] (Extended Data Figure 3b). This behaviour was comparatively rare in pro-vaccine venues. We expected to see a sustained reduction in such behaviour in anti-vaccine venues. Although we observed an average decrease of 54% in the proportion of "near-simultaneous" link pairs, $\chi^2(66)= 267.00$, $p<0.001$ (Fig. 2c), levels of coordination returned to pre-policy levels by early 2022. Furthermore, several new groups created after November, 2020, appear to have coordinated with existing groups (Fig. 2d). Qualitatively, we observed that these coordinated links often promoted anti-vaccine Facebook pages, anti-vaccine content on other social media platforms (e.g., YouTube), known anti-vaccine websites (e.g., ageofautism.com), and websites promoting political calls to action, such as petitions opposing mandatory vaccines. In contrast, coordinated pro-vaccine content pointed to websites facilitating COVID vaccination, such as vaccinefinder.org.

A third source of resilience arises from Facebook's newsfeed algorithm, which is designed to promote content that has generated "meaningful social interaction".[35] Although Facebook states that vaccine misinformation is not eligible for recommendation using this algorithm,[38] engagement levels with anti-vaccine content might recover from post removals if this remaining content is not marked as misinformation, and if the newsfeed algorithm determines this content is more likely to generate "meaningful social interaction". Facebook reportedly[35] uses the number of reactions (e.g., likes", "love", "angry", etc.) content has already spurred to make this determination (Fig. 2g). Specifically, at the time Facebook's policies were first implemented, the newsfeed algorithm reportedly downranked content spurring angry, wow, and haha reactions because they believed these to be associated with "toxic" content and misinformation. In contrast, safer "non-toxic" content reportedly spurred likes, love, and sad reactions, leading Facebook to increase their relative weights.[35] We found that "non-toxic" likes, love, care, and sad reactions to anti-vaccine content have increased in proportion since Facebook's policy was implemented by a weekly average of 30% in pages, $\chi^2(66)= 463.39, p<0.001$, (Fig. 2e) and 33% in groups, $\chi^2(66)= 280.21, p<0.001$ (Fig. 2f), whereas "toxic" angry, wow, and haha reactions have decreased, by a weekly average of 26% in pages ($\chi^2(66)= 816.74, p<0.001$) and 40% in groups ($\chi^2(66)= 423.24, p<0.001$), with results replicating across samples. We did not observe significant changes in these reactions for pro-vaccine content, except for "toxic" reactions in pro-vaccine pages, which increased by 271% ($\chi^2(66)= 398.87, p<0.001$; Extended Data Fig. 4). Thus, it appears that Facebook's policies led anti-vaccine content producers, but not pro-vaccine content producers, to increasingly frame posts in a manner that was weighted as more meaningful by Facebook's content recommendation algorithms.

## Misinformation & Polarisation Increased

One might think that Facebook's policies were successful in removing explicit misinformation, while retaining topics promoting discussion among vaccine hesitant individuals. We therefore examined how the topics that were discussed changed after Facebook's policies, using a probabilistic topic model to summarise the posts in our datasets (Fig. 3; Supplementary Materials, Table S2-1).

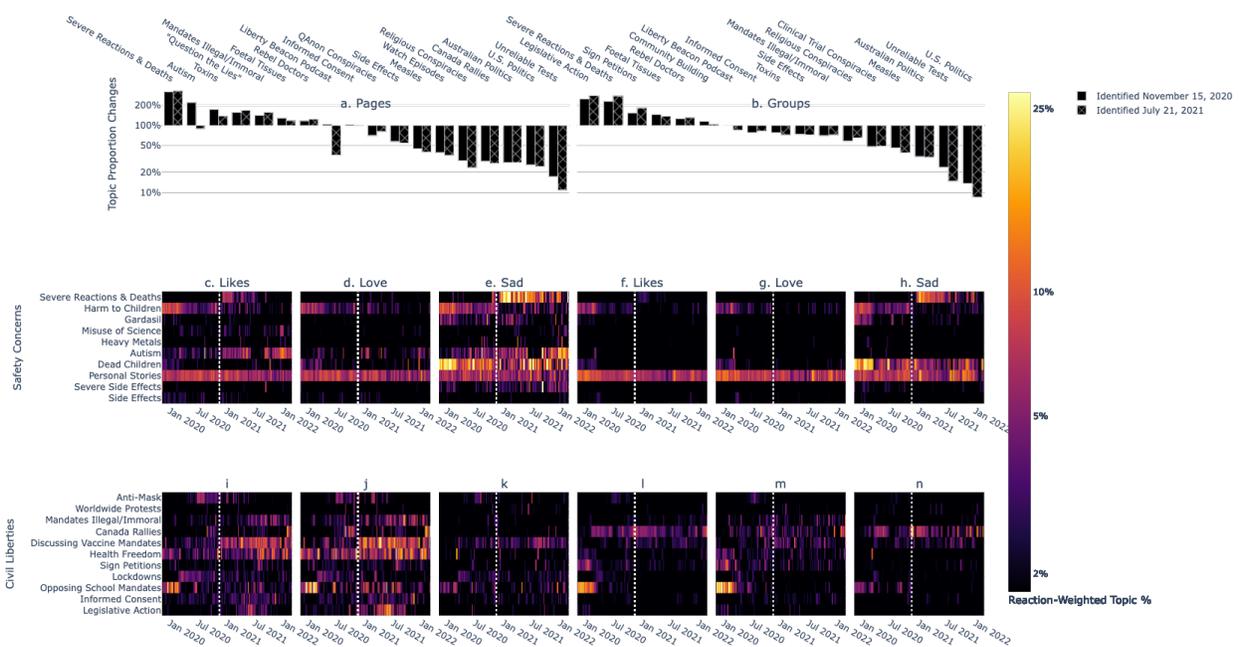

**Fig 3. Topics emphasising severe adverse reactions and calls to political action increased in pages and groups, garnering "non-toxic" reactions.**

**a.** In pages, 18 topics differed significantly from pre-policy projections. Topics pertaining to severe adverse reactions (including the canard that vaccines cause autism) and allegedly harmful or unethical vaccine ingredients increased in proportion, as did topics condemning vaccine mandates and those from "rebel doctors" providing pseudoscientific rationales for vaccine refusal. In contrast, topics pertaining to national politics, COVID testing, measles, and specific conspiracy theories decreased in proportion. **b.** In groups, 17 topics differed significantly from pre-policy projections. Topics calling for political action, including support for legislation and requests to sign petitions, and building community by welcoming new group members, increased. Topics alleging severe adverse reactions , unethical vaccine ingredients, and promoting "rebel doctors" also increased. In contrast, several topics that decreased in pages also decreased in groups. Several topics containing misinformation alleging vaccine harms garnered "non-toxic" reactions, likely promoting them in users' newsfeeds, including: **c.** In pages, topics alleging severe adverse events and deaths from COVID-vaccine, and claiming that vaccines cause autism garnered "likes" and **d.** to a lesser extent, also garnered "love" reactions. **e.** These topics especially garnered "sad" reactions as did topics alleging that vaccine kill children. Although these topics did not garner many **e.** "likes" or **f.** "love" reactions in groups, **g.** they did garner "sad" reactions in groups. Beyond this misinformation,  topics pertaining to political action also garnered "non-toxic" reactions, including: **c.** In pages, topics opposing vaccine mandates and calling for or celebrating political action garnered "likes" **d.** and "love" reactions. Additionally, **e.** topics reporting on vaccine mandates garnered "sad" reactions. In groups, topics opposing vaccine mandates and discussing Canadian politics – ultimately culminating in the "freedom convoy" rallies of early 2022 – also garnered **e.** "likes", **f.** "love" reactions, and **g.** "sad" reactions.

Although posts expressing some misinformative topics about vaccines (e.g., conspiracy theories; see Supplementary Materials, Table S2-2) decreased in proportion, we also observed increases in several misinformative topics that explicitly violated Facebook's community standards. Specifically, a topic containing reports of allegedly severe COVID-19 side effects and death increased significantly compared to pre-policy projections in both anti-vaccine pages, 314%, $\chi^2(66)= 518.69, \ p<0.001$, and groups, 270%, $\chi^2(66)= 841.56, \ p<0.001$. This increase occurred during a crucial time period for vaccine uptake: the initial COVID vaccine rollout. In anti-vaccine pages, this increase goes beyond what can be attributed to the news cycle, having increased 16% more on average than the corresponding pro-vaccine topic over the span of our study, $\chi^2(120) = 620.23, p<0.001$ and 88% more on average between November 18, 2021, and July 4, 2021 (when the Biden administration set a goal for at least 70% of the US population to have received at least one COVID vaccine shot; Extended Data Fig. 5).

Several other topics also increased in proportion relative to the pre-policy trend, including those spreading conspiracy theories from "rebel doctors" who claim to have been silenced by the medical establishment (Supplementary Materials, Table S2-2), alleging that vaccines expose people to toxins and/or "unhealthy radiation" (e.g., 5G wireless), claiming that vaccines contain foetal tissues, and claiming that vaccines cause autism. These misinformative topics are all prohibited under Facebook's community standards,[39] calling into question Facebook's ability to carry out its content removal policies in a comprehensive and consistent manner with their existing technology and approach.

Beyond these misinformative topics, we also observed proportional increases in several topics calling for political or legislative action (which, when associated with anti-vaccine content, were also prohibited by Facebook's community standards).[39] These topics especially increased in Facebook groups, which are designed to promote user interaction and community formation. In pages, these topics spurred "non-toxic" reactions, promoting them in newsfeeds, and included posts framing vaccine refusal as "health freedom", discussing vaccine mandates, arguing that these mandates are illegal or immoral, and celebrating successful legislation opposing these mandates.

Facebook is embedded in a larger information ecosystem comprising several different social media platforms and external websites. Furthermore, Facebook's policies called for increasing the quality of information shared.[39] We therefore examined whether the quality of information from external sources changed as consequence of content removal policies. If successful, Facebook's policies should have led to a decrease in the proportion of links to sources that are known to be "low credibility" because they routinely share misinformation, and an increase in "high credibility" sources that adhere to strict standards of fact-checking.[40,41]. Instead, weekly links to websites that are known to spread misinformation increased by an average of 8% in pages, $\chi^2(66)= 934.55, p<0.001$, (Fig. 4a) and by 32% between November 18, 2020 and July 4, 2021. Overall, users engaged 308% more with posts containing these misinformative links $\chi^2(66)= 247.64, p<0.001$ (Fig. 4b). Furthermore, links to high-quality academic and government sources decreased by a weekly average of 44% in groups compared to pre-policy projections, $\chi^2(64)= 161.72, p<0.001$ (Extended Data Fig. 6b). Finally, on a factual accuracy scale ranging from 0 = "Very Low" to 5 = "Very High" we found that the average factual accuracy of user engagements decreased by 0.13 points in pages, $\chi^2(65)= 183.36, p<0.001$, and 0.80 points in groups, $\chi^2(66)= 130.82, p<0.001$ even though the overall accuracy of rated posts did not change significantly (Extended Data Figure 7).

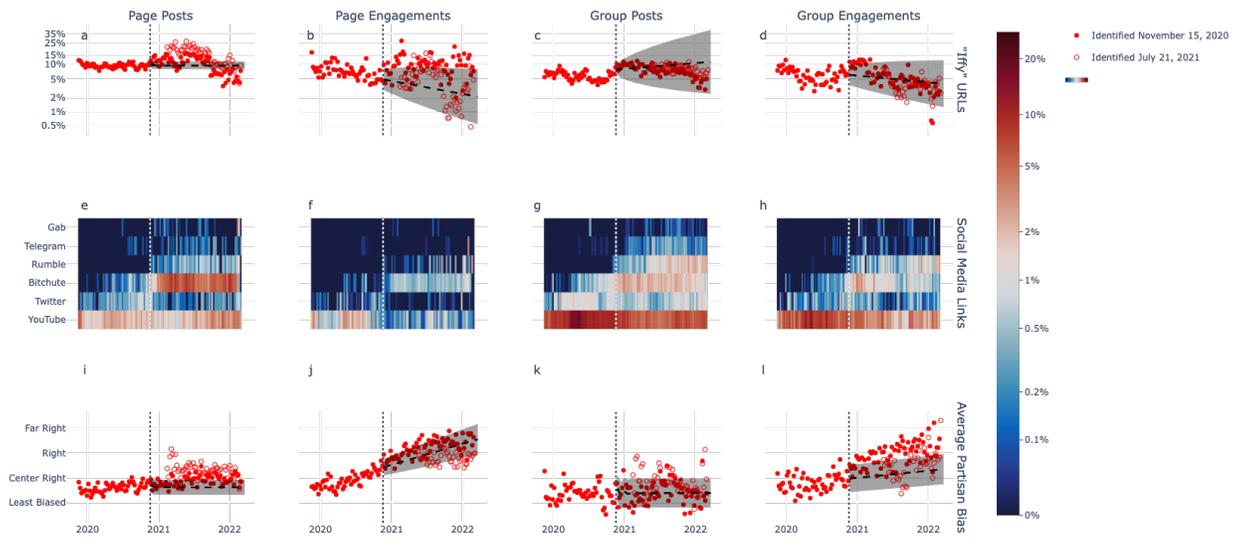

**Fig 4. Facebook's policy led to an increase in off-platform links promoting misinformation and polarisation.**

**a.** In pages, the proportion of links to "iffy" sources (https://iffy.news/index/#methodology) increased by an average of 8% in pages, $\chi^2(66)= 934.55$, $p<0.001$, and by 32% between November 18, 2020 and July 4, 2021, compared to pre-policy projections. **b.** These links spurred 308% more engagements per week, on average, $\chi^2(66)= 247.64, p<0.001$ than pre-policy projections. **c.** In contrast, we did not detect significant differences in links to misinformative sources in groups, $\chi^2(66)= 24.43, p=1.00$ **d.** or in engagements with these links, $\chi^2(66)= 70.96, p=0.32$. Among misinformative links were pointers to "alternative" social media platforms, such as Bitchute, Rumble, Telegram, and Gab, all of which are known to host politically polarized content, especially from the far right. **e.** The weekly proportion of URLs pointing to these platforms, and especially to Bitchute, increased by a factor of 10 in pages after November 18, 2020. **f.** These URLs garnered a larger share of engagements, whereas engagements with mainstream platforms Twitter and YouTube decreased. **g.** We observed a similar increase in links to "alternative" platforms in groups, **h.** and these links garnered a larger share of weekly engagements. In general, we observed that the average partisan bias of rated URLs **i.** increased in Facebook pages, by an average of 0.27 points, $\chi^2(66)= 222.24, p<0.001$, **j.** leading to a slight (0.02 points), yet significant increase in engagements, $\chi^2(66)= 114.22, p<0.001$. **k.** This effect was even more pronounced in groups, where an average increase of 0.24 points, $\chi^2(66)= 106.48, p=0.001$ **l.** led to a 0.78 point increase in right-wing polarisation of the average engagement, $\chi^2(66)= 402.10, p<0.001$. All proportions are shown on a logit scale.

One reason for this change may be that Facebook's policies led anti-vaccine content producers to link to external sources of misinformative content rather than posting it explicitly on Facebook in an attempt to circumvent Facebook's content moderation efforts. Consistent with this explanation, we observed that anti-vaccine content producers increasingly linked to alternative social media platforms, such as Bitchute, Rumble, Gab, and Telegram in lieu of more mainstream platforms that had implemented similar content removal policies (e.g., YouTube and Twitter; Fig 4e-h).

In practice, alternative platforms often host politically-extreme right wing content.[42] This means that Facebook's content removal policies may have the unintended consequence of radicalising their audiences. On the other hand, some have alleged that Facebook disproportionately targets right wing content for removal, and downranks content that has not been removed.[43] If so, then one would expect to see a decrease in right-wing bias among links shared on the platform, and a decrease in engagement with right-wing content. Instead, on a partisan bias scale ranging from -4 = "Extreme Left" to +4 = "Extreme Right", we observed that rated links shared in anti-vaccine venues became more polarised towards the political right wing by an average of 0.27 points in pages (Fig. 4i), $\chi^2(66)= 222.24, p<0.001$ and 0.24 points groups, $\chi^2(66)= 106.48, p=0.001$ (Fig. 4k). The average user engagement also became more polarised in the same direction by an average of 0.02 points in pages, $\chi^2(66)= 114.21, p<0.001$ (Fig 4j), and 0.78 points in groups, $\chi^2(66)= 402.10, p<0.001$ (Fig 4l). In contrast, content in pro-vaccine pages became 0.13 points more polarised towards the political left wing, $\chi^2(66)= 95.86, p=0.01$, and we did not detect a significant change in the polarisation of engagements (Extended Data Fig. 8).

**Conclusion**

Taken together, our findings demonstrate that Facebook's policy reduced the number of posts in anti-vaccine venues, but was not successful in inducing a sustained reduction in engagement with anti-vaccine content, including misinformation. This underscores a need to account for, and address, the forces driving users' engagement with – i.e., "demand" for – misinformative content. Social media platforms are engineered systems, whose explicit design goal is to enable the formation of online communities. By definition, Facebook groups and pages are communities made up of people sharing the same interests – in this case, rationales for vaccine refusal. Communities provide members the ability to make sense of overwhelming amounts of online content,[44] while also offering social support and a feeling of solidarity.[45] These are attractive attributes that drive up demand for this content.

Users satisfy this demand when they seek out content and community through alternate venues, including on other social media platforms.[46] Furthermore, active anti-vaccine community members appear to be using Facebook's technological affordances to expand their peers' access to new venues and content, undermining attempts to curtail supply. Specifically, page administrators and group members coordinate with one another when using the platform to circumvent content and venue removals. This enables users to find aligned anti-vaccine content and like-minded peers, even if their specific venue has been removed. Furthermore, anti-vaccine content producers appear to be spurring increased engagement, and especially reactions that are more likely to be promoted in users' newsfeeds. Thus, resistance to "hard" content remedies may stem from the fact that these remedies directly oppose social media platforms' core design function of facilitating community engagement.

Furthermore, Facebook's system architecture appears to have facilitated unintended consequences. To garner engagements, anti-vaccine content producers make use of topics and links to websites promoting "diagnostic frames": specific narratives that identify putative problems and grievances – e.g., sensationalised vaccine harms. Other content offers "prognostic frames": offering solutions to these problems, e.g., calls for concrete political action, such as encouraging people to vote on specific legislative bills, circulating petitions, or joining protests against mask-wearing, mandates, and lockdowns. Finally, a third set of content offers "motivational frames": making successful concrete action seem possible, e.g., messages celebrating legislative victories or emphasising widespread support for shared values such as "health freedom" or opposition to vaccine mandates. Taken together, these findings suggest that Facebook's architecture facilitates, and perhaps even accelerates, the creation of some of the features of a social movement[47] in a digital space. When combined with increased political polarisation, this could, in turn, facilitate coordinated offline behaviour.[48] (See also work in preparation by C. Bailard, R. Tromble, W. Zhong, F. Bianchi, P. Hosseini, and DAB.)

Limitations to this study include that only public Facebook pages and groups were studied. This leaves lacunae in the extent to which external evaluations can infer Facebook's success. Although we cannot make claims about behaviours in private spaces, the data available through tools such as CrowdTangle constitute a critical window into the largest, and most public, venues on the platform, which are most likely to achieve high numbers of exposures. Indeed, significant prior work has shown that important misinformation about vaccines is often found in public data,[49,50] in part because anti-vaccine advocates seek to recruit the vaccine hesitant in public forums. Similarly, we cannot rule out the possibility that pro-vaccine venues might have contained some sensationalist stories of vaccine harm; however, these stories do not appear to have been sufficiently prevalent to have been detected by our topic models. Finally, our data cannot distinguish between posts or engagements made by unique individuals or the same individuals repeatedly posting. However, this concern is mitigated by the fact that, in pages, only administrators may post, whereas users in any venue may only react to a given post once. Taken as a whole, our findings emphasise the critical need for platforms to continue to provide researchers with access to such public data.

In conclusion, Facebook's three-layered architecture facilitates its resilience in the face of disruption from Facebook's "hard" content removal remedies. This engineered system possesses a structure that might promote "participatory"[51] misinformation and could even lead to social mobilisation. Our results therefore suggest that attempts to address misinformation must rely on a multifaceted strategy that goes beyond "hard" and "soft" content remedies. Analogous strategies have been effective in other domains. For example, effective tobacco control interventions focus on both prevention and treatment. By analogy, our findings and model suggest that different approaches may be needed for those who actively seek out misinformation versus those who simply casually encounter it. Just as tobacco control prevention strategies combine both demand-side and supply-side interventions (e.g., educational campaigns, but also taxation),[52] both "hard" and "soft" content remedies have a role to play in moderating the online ecosystem, as do technological advances which may help improve detection of problematic content. However, beyond these interventions, tobacco control makes use of community-based resources supporting tobacco cessation for current smokers. By analogy, in the short term, Facebook or public health officials might consider using social media platforms to engage directly with users who have been repeatedly exposed to misinformation, and may attempt to recruit vaccine hesitant individuals into other venues where their needs for community can be met, while ensuring that their questions about vaccines are answered in a manner that is most consistent with public health guidelines.[25,53] In the longer term, platforms may incorporate explicit features designed to promote communities that organically resist misinformation.[54] There is nevertheless a dearth of scientific literature regarding specific interventions to reduce demand for misinformation among active seekers. Thus, future work is needed to develop interventions that can work in tandem with both "hard" and "soft" content remedies to reduce misinformation seeking behaviour.

# References

1. West, J. D. & Bergstrom, C. T. Misinformation in and about science. *Proc. Natl. Acad. Sci. U.S.A.* **118**, e1912444117 (2021).

2. Chou, W.-Y. S., Oh, A. & Klein, W. M. P. Addressing Health-Related Misinformation on Social Media. *JAMA* **320**, 2417–2418 (2018).

3. Hearing on 'Disinformation Nation: Social Media's Role in Promoting Extremism and Misinformation'. *Democrats, Energy and Commerce Committee* https://energycommerce.house.gov/committee-activity/hearings/hearing-on-disinformation-nation-social-medias-role-in-promoting (2021).

4. Gowen, A. As mob lynchings fueled by WhatsApp messages sweep India, authorities struggle to combat fake news. *Washington Post* (2018).

5. Silverman, B., Mac, R. & Lytvynenko, J. How Facebook Failed To Prevent Stop The Steal. *BuzzFeed News* (2021).

6. Loomba, S., de Figueiredo, A., Piatek, S. J., de Graaf, K. & Larson, H. J. Measuring the impact of COVID-19 vaccine misinformation on vaccination intent in the UK and USA. *Nat Hum Behav* (2021) doi:10.1038/s41562-021-01056-1.

7. Pierri, F. *et al.* Online misinformation is linked to early COVID-19 vaccination hesitancy and refusal. *Sci Rep* **12**, 5966 (2022).

8. Goldman, E. Content Moderation Remedies. *Mich. Tech. L. Rev.* **28**, 1 (2021).

9. Grimmelmann, J. The Virtues of Moderation. *Yale J.L. & Tech.* **17**, 42–109 (2015).

10. Jhaver, S., Boylston, C., Yang, D. & Bruckman, A. Evaluating the Effectiveness of Deplatforming as a Moderation Strategy on Twitter. *Proc. ACM Hum.-Comput. Interact.* **5**, 381:1-381:30 (2021).

11. Ball, P. & Maxmen, A. The epic battle against coronavirus misinformation and conspiracy theories. *Nature* **581**, 371–374 (2020).

12. Facebook's Latest Attempt To Address Vaccine Misinformation—And Why It's Not Enough | Health Affairs Blog. https://www.healthaffairs.org/do/10.1377/hblog20201029.23107/full/.

13. Rodriguez, S. 12 attorneys general call on Facebook and Twitter to remove anti-vaxxers from their services. *CNBC* https://www.cnbc.com/2021/03/24/attorneys-general-call-on-facebook-and-twitter-to-remove-anti-vaxxers-off-their-services.html (2021).

14. Bak-Coleman, J. B. *et al.* Combining interventions to reduce the spread of viral misinformation. *Nat Hum Behav* 1–9 (2022) doi:10.1038/s41562-022-01388-6.

15. Tyson, A., Johnson, C. & Funk, C. U.S. Public Now Divided Over Whether To Get COVID-19 Vaccine. *Pew Research Center Science & Society* https://www.pewresearch.org/science/2020/09/17/u-s-public-now-divided-over-whether-to-get-covid-19-vaccine/ (2020).

16. Sun, L. H. Biden officials urge use of booster shots, antivirals against BA.5. *Washington Post* (2022).

17. Senators Klobuchar, Baldwin, Peters Urge Tech Industry Leaders to Combat Coronavirus Vaccine Misinformation. *U.S. Senator Amy Klobuchar* https://www.klobuchar.senate.gov/public/index.cfm/2021/1/senators-klobuchar-baldwin-peters-urge-tech-industry-leaders-to-combat-coronavirus-vaccine-misinformation.

18. Call to Action: CSIS-LSHTM High-Level Panel on Vaccine Confidence and Misinformation. https://www.csis.org/analysis/call-action-csis-lshtm-high-level-panel-vaccine-confidence-and-misinformation.

19. Donovan, J. Social-media companies must flatten the curve of misinformation. *Nature* (2020) doi:10.1038/d41586-020-01107-z.

20. Gu, J. *et al.* The impact of Facebook's vaccine misinformation policy on user endorsements of vaccine content: An interrupted time series analysis. *Vaccine* **40**, 2209–2214 (2022).

21. Pennycook, G. *et al.* Shifting attention to accuracy can reduce misinformation online. *Nature* **592**, 590–595 (2021).

22. Roozenbeek, J., Freeman, A. L. J. & van der Linden, S. How Accurate Are Accuracy-Nudge Interventions? A Preregistered Direct Replication of Pennycook et al. (2020). *Psychol Sci* 095679762110245 (2021) doi:10.1177/09567976211024535.

23. Ali, S. *et al.* Understanding the Effect of Deplatforming on Social Networks. in *13th ACM Web Science Conference 2021* 187–195 (Association for Computing Machinery, 2021). doi:10.1145/3447535.3462637.

24. Chandrasekharan, E. *et al.* You Can't Stay Here: The Efficacy of Reddit's 2015 Ban Examined Through Hate Speech. *Proc. ACM Hum.-Comput. Interact.* **1**, 31:1-31:22 (2017).

25. Broniatowski, D. A., Dredze, M. & Ayers, J. W. 'First Do No Harm': Effective Communication About COVID-19 Vaccines. *Am J Public Health* **111**, 1055–1057 (2021).

26. Gorwa, R., Binns, R. & Katzenbach, C. Algorithmic content moderation: Technical and political challenges in the automation of platform governance. *Big Data & Society* **7**, 2053951719897945 (2020).

27. Gillespie, T. Content moderation, AI, and the question of scale. *Big Data & Society* **7**, 2053951720943234 (2020).

28. Ribeiro, M. H. *et al.* Do Platform Migrations Compromise Content Moderation? Evidence from r/The_Donald and r/Incels. *Proc. ACM Hum.-Comput. Interact.* **5**, 1–24 (2021).

29. Senate Bill 7072 (2021) - The Florida Senate. https://www.flsenate.gov/Session/Bill/2021/7072/.

30. 87(2) HB 20 - Senate Committee Report version - Bill Text. https://capitol.texas.gov/tlodocs/872/billtext/html/HB00020S.htm.

31. Company Info | Meta. https://about.facebook.com/company-info/.

32. Munger, K. & Phillips, J. Right-Wing YouTube: A Supply and Demand Perspective. *The International Journal of Press/Politics* **27**, 186–219 (2022).

33. Broniatowski, D. A. Flexibility Due to Abstraction and Decomposition. *Systems Engineering* **20**, 98–117 (2017).

34. Broniatowski, D. A. & Moses, J. Measuring Flexibility, Descriptive Complexity, and Rework Potential in Generic System Architectures. *Systems Engineering* **19**, 207–221 (2016).

35. Merrill, J. B. & Oremus, W. Five points for anger, one for a 'like': How Facebook's formula fostered rage and misinformation. *Washington Post* (2021).

36. Johnson, N. F. *et al.* Hidden resilience and adaptive dynamics of the global online hate ecology. *Nature* **573**, 261–265 (2019).

37. Inauthentic Behavior | Transparency Center. https://transparency.fb.com/policies/community-standards/inauthentic-behavior/.

38. What are recommendations on Facebook? | Facebook Help Center. https://www.facebook.com/help/1257205004624246.

39. COVID-19 and Vaccine Policy Updates & Protections | Facebook Help Center. https://www.facebook.com/help/230764881494641.

40. Grinberg, N., Joseph, K., Friedland, L., Swire-Thompson, B. & Lazer, D. Fake news on Twitter during the 2016 US presidential election. *Science* **363**, 374–378 (2019).

41. Shao, C. *et al.* The spread of low-credibility content by social bots. *Nat Commun* **9**, 4787 (2018).

42. Zhou, Y., Dredze, M., Broniatowski, D. A. & Adler, W. D. Elites and foreign actors among the alt-right: The Gab social media platform. *First Monday* **24**, (2019).

43. Jordan, J., Collins, D., Gaetz, M. & Steube, W. G. *Reining in Big Tech's Censorship of Conservatives*. (2020).

44. Reyna, V. F., Broniatowski, D. A. & Edelson, S. M. Viruses, Vaccines, and COVID-19: Explaining and Improving Risky Decision-making. *Journal of Applied Research in Memory and Cognition* **10**, 491–509 (2021).

45. Snow, D. A., Soule, S. A. & Kriesi, H. *The Blackwell Companion to Social Movements*. (John Wiley & Sons, 2008).

46. Mitts, T., Pisharody, N. & Shapiro, J. Removal of Anti-Vaccine Content Impacts Social Media Discourse. in *14th ACM Web Science Conference 2022* 319–326 (Association for Computing Machinery, 2022). doi:10.1145/3501247.3531548.

47. Benford, R. D. & Snow, D. A. Framing Processes and Social Movements: An Overview and Assessment. *Annual Review of Sociology* **26**, 611–639 (2000).

48. Giugni, M., McAdam, D. & Tilly, C. *How social movements matter*. (University of Minnesota Press, 1999).

49. Broniatowski, D. A. *et al.* Facebook Pages, the "Disneyland" Measles Outbreak, and Promotion of Vaccine Refusal as a Civil Right, 2009–2019. *Am J Public Health* **110**, S312–S318 (2020).

50. Jamison, A. M. *et al.* Not just conspiracy theories: Vaccine opponents and pro-ponents add to the COVID-19 'infodemic' on Twitter. *The Harvard Kennedy School Misinformation Review* **1**, (2020).

51. Starbird, K., Arif, A. & Wilson, T. Disinformation as Collaborative Work: Surfacing the Participatory Nature of Strategic Information Operations. *Proc. ACM Hum.-Comput. Interact.* **3**, 127:1-127:26 (2019).

52. U.S. National Cancer Institute and World Health Organization. *The Economics of Tobacco and Tobacco Control.* (2016).

53. Larson, H. J. & Broniatowski, D. A. Why Debunking Misinformation Is Not Enough to Change People's Minds About Vaccines. *Am J Public Health* **111**, 1058–1060 (2021).

54. Forestal, J. *Designing for Democracy: How to Build Community in Digital Environments*. (Oxford University Press, 2021).

## Methods

*Data Source*

We downloaded data from CrowdTangle,[55] a public insights tool owned and operated by Facebook which has been called "perhaps the most effective transparency tool in the history of social media".[56,57] CrowdTangle tracks interactions with public content from Facebook pages and groups ("venues"). It does not include activity on private accounts, or posts made visible only to specific groups of followers.

*Identifying vaccine-related pages and groups*

Similar to prior work[58,59] that relies upon iterative procedures to reduce biases associated with keyword selection:

1. We first identified a large set of pages and groups that mentioned vaccines at least once within the most recent ~300,000 posts. We did not include venues with earlier timestamps since our aim was to capture activity in the most prolific Facebook groups in the leadup to Facebook's policy implementation. To do so, we searched CrowdTangle on November 15, 2020, identifying and downloading all posts containing at least one keyword from the following list: "vaccine, vaxx, vaccines, vaccination, vaccinated, vax, vaxxed, vaccinations, jab". Several of these posts contained content pertaining to guns, and to pet and other animal vaccines. Thus, we ran a second search excluding posts containing the following keywords: "gun, dog, cat, foster, adopt, shelter, vet, kennel, pet, chicken, livestock, kitten, puppy, paw, cow". We conducted this search on November 15, 2020 and retrieved the 299,981 most recent page posts and 299,904 group posts meeting search criteria before hitting CrowdTangle's download limit, with the earliest posts timestamped September 7, 2020, for pages, and July 1, 2020, for groups. This procedure yielded 73,438 pages and 57,485 groups.

To ensure that we only retained venues that routinely discussed vaccines, we retained all pages and groups whose name contained at least one of the strings "vacc", "vax", or "jab", or which posted very frequently about vaccines: i.e., in the top percentile – at least 44 times for pages and 58 times for groups. This procedure yielded 1,231 pages and 773 groups.

2. Several of the venues generated in Step 1 were news organizations that did not primarily focus on vaccination. We therefore further narrowed down our initial list as follows: we retrieved as many posts as possible from these venues yielding the 299,994 most recent page posts and 299,969 group posts before hitting CrowdTangle's download limit. As above, we did not collect earlier posts since our aim was to identify venues that actively posted about vaccination in the leadup to the policy's implementation. The earliest post from these venues was timestamped November 8, 2020, for pages, and November 13, 2020, for groups. We retained all venues for which at least 20% of posts retrieved contained at least one word containing "vacc" or "vax". We selected this 20% threshold by inspection, and our results were insensitive to changing it (relaxing the threshold yielded more groups and pages that were characterized as "other" in Step 3, which were not included in our analysis. All groups and pages were checked for relevance by two authors – AMJ and JG).

3. On November 15, 2020, we retrieved all posts from the venues identified in Step 2 for a 12-month period starting on November 15, 2019, forward. We did not exceed the download limit for these venues. We repeatedly collected content from these venues from November 30, 2020 through February 28, 2022 (see Extended Data Table 1). Qualitatively, we found that these posts focused on general vaccine content, adhering closely to existing typologies of pro- and anti-vaccine topics.[60,61] We therefore manually annotated these venues as either pro-vaccine, anti-vaccine, or "other" (see Extended Data Table 2). Two independent annotators (JG and AMJ) manually assessed each group and page following a two-tiered coding scheme,[61] achieving high reliability. Entities were first labelled as either pro-vaccine, anti-vaccine, or other (Cohen's κ = 0.88, 95% CI: 0.85-0.92). Next, a sub-category was assigned based on specific concerns, adapted from a previously published coding scheme[60,61] (Cohen's κ = 0.75, 95% CI: 0.71-0.79; Extended Data Table 2). Categorization was based on content shared in the "about" section of each venue. When this section was left blank, annotators considered the venue's title, any imagery used, and recent posts to decide. All venues were double-coded, with disagreements discursively reconciled. We retained all venues that were labelled as pro-vaccine or anti-vaccine.

*Identifying a second sample of venues*

To test sensitivity to our venue selection date, we ran a second search on July 21, 2021, to identify new venues using the same technique. These new venues contained 79 (33 anti-vaccine and 46 pro-vaccine) pages and 139 (69 anti-vaccine and 70 pro-vaccine) groups. Of these, 27 (34%; 14, 42% anti-vaccine and 13, 28% pro-vaccine) pages and 114 groups (82%; 47, 68% anti-vaccine and 67, 96% pro-vaccine) were not present in the original sample. Of the venues not present in the original sample, we did not detect posts before November 15, 2020 for 5 (19%; 4, 29% anti-vaccine and 1, 8% pro-vaccine) pages and 59 groups (52%; 19, 40% anti-vaccine and 40, 60% pro-vaccine) when we conducted a CrowdTangle search on February 22, 2022. We again repeatedly collected posts in pages and groups from the second sample at multiple points in time, ranging from August 9, 2021 through February 28, 2022 (see Extended Data Table 2). Our results largely replicated across samples.

*Interrupted Time Series Analysis*

We examined changes to the weekly number of posts, and engagements with those posts using an interrupted time series design with a non-equivalent control group – one of the strongest quasi-experimental designs available.[62] This design enables us to estimate causal effects of changes to Facebook's policies because any changes to observed data affecting anti-vaccine that are not due to Facebook's policies content – e.g., external news about vaccine trials – should affect both pro- and anti-vaccine groups and pages since they are both focused on vaccination. We compared the year prior to November 15, 2020 to the remainder of the dataset, enabling us to tease apart the effects of Facebook's "hard" content remedies on anti-vaccine content.

We analysed pages separately from groups. Pages – representing the voices of opinion leaders, who tend to be explicitly pro- or anti-vaccine rather than vaccine hesitant[25,53,63] – differ from Facebook groups – designed to be discussion forums.[64] Whereas only page administrators can post in pages, any member can post in groups.

*Measures: Posts and Engagements*

For each dataset, we calculated the weekly number of posts in anti- and pro-vaccine pages and groups. We applied a logarithmic transform to post volumes to control for data skew. We also calculated the weekly number of engagements with these posts as the sum of all Likes, Shares, Comments, and other reactions (Love, Wow, Haha, Sad, Angry, and Care) reported by CrowdTangle. Facebook's algorithms reportedly uses a weighted sum of these engagements when prioritising content within users' newsfeeds.[65] After applying a logarithm transform, we found that individual engagement types are strongly correlated with one another for both pages (Cronbach's $\alpha$ = 0.97, 95% CI: 0.97-0.98) and groups (Cronbach's $\alpha$ = 0.96, 95% CI: 0.95-0.97), meaning that our results are robust to changes in weights. (We excluded the "Care" emoji from Cronbach's α calculations because they were not introduced until March, 2020, were not widely used until May, 2020, and continue to make up less than 1% of all engagements).

*Fitting ARIMA Models*

We conducted interrupted time series analyses using Autoregressive Integrated Moving Average (ARIMA) models fit to weekly sums of posts and engagements from November 15, 2019 through November 15, 2020 – immediately prior to Facebook's November 18th, 2020, removal of "Stop Mandatory Vaccination" – one of the largest anti-vaccine fan pages.[66] On December 3, 2020, Facebook announced its intention to remove false claims about COVID-19.[67] Finally, on February 8, 2021, Facebook extended this policy to vaccine misinformation in general,[68] while promising to increase the credibility of information shared about vaccines more generally. Thus, November 18th, 2020, marks the beginning of a series of policies targeting anti-vaccine content and combining prior "soft" anti-vaccine content removal remedies with new "hard" content removal remedies. Our results are nevertheless insensitive to our selection of this specific date (see Supplementary Information S3).

To control for the formation of new pages and groups between November 15, 2019, and November 15, 2020, we divided weekly post and engagement counts by the total number of weekly venues in each dataset prior to fitting ARIMA models (results were similar when analysing raw post counts, see Supplementary Information S3).

We fit all ARIMA models to pre-policy data using the auto_arima function in the pmdarmia python package.[69] When time series were not stationary, as determined by an Augmented Dickey Fuller test, data were detrended using differencing. We selected the number of autoregressive and moving average terms using a parallel grid search to minimise the corrected Akaike Information Criterion (AICc).

*Generating Counterfactual Predictions*

We used these models to generate counterfactual projections for weekly posts or engagements that would have been present assuming no "hard" remedies. We calculated the percent difference between these counterfactual projections and observed post and engagement counts. We consider a policy to have been effective if it consistently reduced content beyond the 90% confidence bounds of these projections. The residual values of an ARIMA model are normally distributed and independently identically distributed (i.i.d.). meaning the residual sum of squares values follows a $\chi^2$ distribution with the degrees of freedom equal to the number of predicted datapoints. We conducted $\chi^2$ tests to assess the goodness of fit of the model's pre-policy projections against post-policy data.

*Using Pro-Vaccine Venues as a Non-Equivalent Control Group*

Pro-vaccine venues make an ideal non-equivalent control group because, like anti-vaccine venues, they contain users who are motivated to post about vaccines and would therefore respond to exogenous factors, such as the news cycle, in the same way; however, platforms' policies were not designed to target pro-vaccine content. A statistically significant difference in anti-vaccine, but not pro-vaccine, venues, or between anti- and pro-vaccine venue effect sizes – indicates that it is more likely to have been Facebook's policies, and not some contemporaneous event, which caused the observed change.

*Analysis of Facebook's System Architecture*

*Overt Links between Anti-Vaccine and Pro-Vaccine Pages*

We examined whether Facebook's new policies reduced links between the anti- and pro-vaccine pages in our sample. (The number of links to, and between, groups was zero in most weeks.) We identified all posts containing a link starting with www.facebook.com. We then calculated the proportion of these links pointing to the anti- and pro-vaccine pages in each of our samples. Specifically, we identified and extracted the unique numerical Facebook ID and username for each pro- and anti-vaccine page in each dataset. We considered a source page to be linked to a target page if a URL posted on the source page began with www.facebook.com/<Facebook ID>/ or www.facebook.com/<target username>/. We next calculated the weekly proportion of all such links that pointed from anti-vaccine pages to other anti-vaccine pages (excluding self-links), and from pro-vaccine pages to other pro-vaccine pages. We next conducted interrupted time series analyses on these weekly proportions. After applying a logit transform to our data to control for floor and ceiling effects, we fit ARIMA models to pre-policy data, comparing post-policy data to model projections (Extended Data Fig. 2). Results were similar across venues identified November 15, 2020, and July 21, 2021; therefore, we combined these data since doing so allowed us to also examine links between pages that were in separate datasets. We also used the links that we extracted to construct unweighted networks. In practice, these networks qualitatively resemble prior work based upon mutual "likes" between pages, but which currently require access to Facebook's commercial APIs to replicate at scale.[36,70] Like this prior work, our networks were displayed using a force-directed layout algorithm.[71]

*Coordinated Link-Sharing*

Facebook's community standards disallow "us[ing] fake accounts, artificially boost[ing] the popularity of content, or engag[ing] in behaviours designed to enable other violations under our Community Standards."[37] Prior work[72,73] suggests that this type of "coordinated inauthentic behaviour" may be detected under the assumption that "near-simultaneous link sharing" is a signal of coordination.[73–76] Building upon prior work,[73] we operationalized "near-simultaneous" link sharing in a manner that was intended to be robust to the specific query being used. We conducted three "blank search" queries between March 30 and March 31, 2020 on CrowdTangle to identify the ~300,000 most recent posts each for pages and groups available on the platform, combining across pages and groups. We calculated the time difference in seconds between each successive share of the same URLs. To distinguish between coordinated and uncoordinated behaviours, we modelled the distribution of these interarrival times as a mixture of exponential distributions with components corresponding to "near simultaneous" sharing and "non-simultaneous" sharing. We used the exp-mixture-model[77] package in Python to fit these exponential mixture models. The best fitting model (see Supplemental Materials) was made up of two components, with mean interarrival times of $\mu_{\text{near-simultaneous}}$=9.95 seconds and $\mu_{\text{non-simultaneous}}$=227.45 seconds, meaning that a URL shared by two venues in under 33 seconds is more likely to have originated from the "near-simultaneous" component than the "non-simultaneous" component. This number is comparable to thresholds defined heuristically in prior work.[73–76] We considered venues to be routinely coordinated if their empirical frequency of "near simultaneous" links significantly exceeded 13.34% – the expected likelihood that links were drawn from the non-simultaneous distribution – using binomial tests. Venues were linked if were significantly coordinated at the $p<0.05$ level after controlling for multiple comparisons using the Holm-Bonferroni procedure. We also tested several threshold values ranging from 25 - 41 seconds and found that results qualitatively replicated.

*"Toxic" and "Non-Toxic" Reactions*

For each dataset, we calculated the weekly proportion of engagements that were reportedly "toxic" (angry, haha, and wow), and "non-toxic" (likes, love, sad, care), and conducted interrupted time series analyses on these proportions. We fit ARIMA models to logit-transformed pre-policy data, comparing post-policy data to pre-policy projections. Results replicated when examining each type of engagement separately (Supplementary Information 5).

**Topic Modelling**

We extracted the text from each Facebook post by combining the "Message", "Image Text", "Link Text" and "Description" fields returned by CrowdTangle. After identifying and removing non-English posts using the langdetect Python package,[78] we used Latent Dirichlet Allocation (LDA[79]) – a popular text summarization algorithm – implemented in the MALLET software package,[80] to fit two separate topic models to the text of the 268,875 unique English anti-vaccine posts and 76,954 pro-vaccine posts collected in venues identified November 15, 2020. In each case, we selected a model with 50 topics, using Bayesian hyperparameter optimization[81] to ensure coherent topics. We next fit the data from venues collected on July 21, 2021, to these topics using post-hoc topic model inference. After reading the 50 most representative posts from each topic, two annotators (AJ, JG) assigned descriptive labels to each topic with a third author (DAB) summarising them for brevity. Based upon these labels, topics were assigned into the same typology categories that were used to categorise pro- and anti-vaccine venues (see

Supplementary Materials). Next, for each dataset, we calculated the average weekly proportion of each topic. Finally, we fit ARIMA models to logit-transformed weekly topic proportions, comparing observations to pre-policy projections (see Supplementary Materials).

*Reaction-Weighted Topic Analysis*

For each of the "non-toxic" reaction, we calculated the weekly proportion, ρ, of those reactions associated with a given topic, as follows:

$$\rho_{r,t|w} = \frac{\sum_{p \in w} \pi_{t,p} \times n_{r,p}}{\sum_{p \in w} n_{r,p}}$$

where $\pi_{t,p}$ is the proportion of post p assigned to topic t, $n_{r,p}$ is the number of reactions of type r to post p, and $p \in w$ is the set of all posts, p, in week w. Although we combined across the first and second samples for this analysis, results replicate when analysing each sample independently (Supplementary Materials).

*URL Quality*

For each dataset, we extracted all URLs in the "Link" field returned by CrowdTangle, or the "Final Link" field if it was non-empty (meaning that the "Link" field used a URL shortener). Next, using the TLDExtract Python module,[82] we extracted the top-level domain (TLD) and suffix for each URL (for example, the top-level domain of www.example.com/this-is-an-example.html is example.com).

*URL Source Credibility and Partisanship*

Significant prior work[41,83–88] shows that posts with links to misinformative sources may serve as a proxy for content quality.[88–90] After removing links to facebook.com, we calculated the weekly proportion in pro- and anti-vaccine venues of all TLDs listed on iffy.news – a list of publishers identified by MediaBiasFactCheck.com as having "low" or "very low" factual reporting scores – on April 30, 2022. MediaBiasFactCheck are known to be strongly correlated with several other URL credibility ratings.[84,87] We also calculated the weekly proportion of engagements with low credibility URLs by weighting each post containing an off-platform URL by the total number of engagements with it.

We separately examined posts from "high quality health sources"[84,87] which we operationalized as academic and government sources. All TLDs ending in .gov, .gc.ca, .mil, .nhs.uk, starting with gov., mygov., government., containing .govt. or .gov., or matching who.int, paho.org, un.org, canada.ca, ontario.ca, toronto.ca, or alberta.ca were coded as "government". Similarly, all TLDs ending in .edu, containing .edu., .ac., thelancet.com, sciencedirect.com., medrxiv.org, pnas.org, apa.org, nature.com, sciencemag.org, nejm.org, bmj.com, mayoclinic.org, aaas.org, healthdata.org, researchgate.net, or rand.org were coded as "academic". We calculated the weekly proportion of all such TLDs.

We also examined links to news sources. Specifically, we collated a list of 2,340 publishers that had been given a "Fact Rating" score by MediaBiasFactCheck.com from "Very Low" (0) to "Very High" (5) as of February 3, 2022. Links to these publishers made up 49% and 23% of links to pages and groups among venues identified November 15, 2020, garnering 57% and 25% of all engagements, respectively. Similarly, these links made up 47% and 20% of links to pages and groups among venues identified July 21, 2021, garnering 55% and 14% of all engagements, respectively.

MediaBiasFactCheck.com also scores websites by their partisan bias ranging from far right (-4) to far left (4). We collated a list of 1,314 publishers that had been scored by MediaBiasFactCheck.com as of February 3, 2022. Links to these publishers made up 34% and 17% of off-platforms links to pages and groups, among venues identified November 15, 2020, garnering 41% and 19% of engagements, respectively. Similarly, links to these publishers made up 34% and 15% of off-platforms links to pages and groups, among venues identified July 21, 2021, garnering 55% and 14% of engagements.

We calculated the weekly average "Fact Rating" and "Bias Rating" for all rated links, and examined how this average changed over time in both pro- and anti-vaccine venues. We conducted interrupted time series ARIMA analyses and chi-square tests on these weekly averages.

*Links to Alternative Social Media Platforms*

We examined the prevalence of links to both mainstream social media platforms YouTube (youtube.com; youtu.be) and Twitter (twitter.com; t.co) but also alternative platforms such as Bitchute (bitchute.com), Rumble (rumble.com), Gab (gab.com), and Telegram (t.me). We conducted interrupted time series ARIMA analyses and chi-square tests on logit-transformed weekly proportions of both mainstream and alternative platforms.

**Data and code availability**

All processed data required to produce the reports in this study are available as data files included in supplementary information. Code required to generate these HTML files are also included in supplementary materials. Raw data in this study were obtained from CrowdTangle for Academics and Researchers, a third-party data provider owned and operated by Facebook. CrowdTangle list IDs are provided in the references. Anyone with a CrowdTangle account may access these lists and the corresponding raw data. Researchers may request CrowdTangle access at https://help.crowdtangle.com/en/articles/4302208-crowdtangle-for-academics-and-researchers. CrowdTangle's terms of service prohibit providing raw data to anyone outside of a CrowdTangle user's account. The user can share the findings, but not the data. If a journal asks for data to verify findings, the CrowdTangle user may send a .csv, but it cannot be posted publicly, and the journal must delete it after verification. CSV files containing the raw data used in this study are therefore available upon request subject to these terms.

**Methods References**

55. CrowdTangle Team. *CrowdTangle*. (2021). List IDs: 1475046, 1475046, 1584315, 1584316

56. Platform Transparency: Understanding the Impact of Social Media. (2022).

57. Smith, B. A Former Facebook Executive Pushes to Open Social Media's 'Black Boxes'. *The New York Times* (2022).

58. King, G., Lam, P. & Roberts, M. E. Computer-Assisted Keyword and Document Set Discovery from Unstructured Text. *American Journal of Political Science* **61**, 971–988 (2017).

59. Dredze, M., Broniatowski, D. A., Smith, M. C. & Hilyard, K. M. Understanding Vaccine Refusal: Why We Need Social Media Now. *American Journal of Preventive Medicine* (2016) doi:10.1016/j.amepre.2015.10.002.

60. Kata, A. A postmodern Pandora's box: anti-vaccination misinformation on the Internet. *Vaccine* **28**, 1709–1716 (2010).

61. Jamison, A. *et al.* Adapting and Extending a Typology to Identify Vaccine Misinformation on Twitter. *Am J Public Health* **110**, S331–S339 (2020).

62. Shadish, W. R., Cook, T. D. & Campbell, D. T. *Experimental and quasi-experimental designs for generalized causal inference*. (Wadsworth Cengage learning, 2002).

63. Larson, H. J. & Broniatowski, D. A. Volatility of vaccine confidence. *Science* **371**, 1289–1289 (2021).

64. Facebook. Facebook Tips: What's the Difference between a Facebook Page and Group? https://www.facebook.com/notes/facebook/facebook-tips-whats-the-difference-between-a-facebook-page-and-group/324706977130/ (2010).

65. Oremus, W., Alcantara, C., Merrill, J. B. & Galocha, A. How Facebook shapes your feed. *Washington Post* https://www.washingtonpost.com/technology/interactive/2021/how-facebook-algorithm-works/ (2021).

66. Sulleyman, A. Facebook bans one of the anti-vaccine movement's biggest pages for violating QAnon rules. *Newsweek* https://www.newsweek.com/facebook-bans-anti-vaccine-group-violating-qanon-rules-1548408 (2020).

67. Keeping People Safe and Informed About the Coronavirus. *About Facebook* https://about.fb.com/news/2020/12/coronavirus/ (2020).

68. Reaching Billions of People With COVID-19 Vaccine Information. *About Facebook* https://about.fb.com/news/2021/02/reaching-billions-of-people-with-covid-19-vaccine-information/ (2021).

69. Smith, T. G. & others. pmdarima: ARIMA estimators for Python. (2017).

70. Johnson, N. F. *et al.* The online competition between pro- and anti-vaccination views. *Nature* **582**, 230–233 (2020).

71. Fruchterman, T. M. J. & Reingold, E. M. Graph drawing by force-directed placement. *Software: Practice and Experience* **21**, 1129–1164 (1991).

72. Ayers, J. W. *et al.* Spread of Misinformation About Face Masks and COVID-19 by Automated Software on Facebook. *JAMA Intern Med* (2021) doi:10.1001/jamainternmed.2021.2498.

73. Giglietto, F., Righetti, N., Rossi, L. & Marino, G. It takes a village to manipulate the media: coordinated link sharing behavior during 2018 and 2019 Italian elections. *Information, Communication & Society* **23**, 867–891 (2020).

74. Pacheco, D. *et al.* Uncovering Coordinated Networks on Social Media: Methods and Case Studies. *ICWSM* **21**, 455–466 (2021).

75. Nizzoli, L., Tardelli, S., Avvenuti, M., Cresci, S. & Tesconi, M. Coordinated Behavior on Social Media in 2019 UK General Election. in *ICWSM* 443–454 (2021).

76. Weber, D. & Neumann, F. Amplifying influence through coordinated behaviour in social networks. *Social Network Analysis and Mining* **11**, 1–42 (2021).

77. Okada, M., Yamanishi, K. & Masuda, N. Long-tailed distributions of inter-event times as mixtures of exponential distributions. *Royal Society Open Science* **7**, 191643.

78. Danilak, M. M. langdetect: Language detection library ported from Google's language-detection.

79. Blei, D. M., Ng, A. Y. & Jordan, M. I. Latent dirichlet allocation. *the Journal of machine Learning research* **3**, 993–1022 (2003).

80. McCallum, A. K. Mallet: A machine learning for language toolkit. (2002).

81. Wallach, H. M., Mimno, D. M. & McCallum, A. Rethinking LDA: Why priors matter. in *Advances in neural information processing systems* 1973–1981 (2009).

82. Kurkowski, J. john-kurkowski/tldextract. (2020).

83. Cinelli, M. *et al.* The COVID-19 social media infodemic. *Sci Rep* **10**, 16598 (2020).

84. Singh, L. *et al.* Understanding high-and low-quality URL Sharing on COVID-19 Twitter streams. *Journal of Computational Social Science* 1–24 (2020).

85. DeVerna, M. R. *et al.* CoVaxxy: A Collection of English-Language Twitter Posts About COVID-19 Vaccines. in *ICWSM* 992–999 (2021).

86. Yang, K.-C. *et al.* The COVID-19 Infodemic: Twitter versus Facebook. *Big Data & Society* **8**, 20539517211013860 (2021).

87. Broniatowski, D. A. *et al.* Twitter and Facebook posts about COVID-19 are less likely to spread misinformation compared to other health topics. *PLoS ONE* **17**, e0261768 (2022).

88. Grinberg, N., Joseph, K., Friedland, L., Swire-Thompson, B. & Lazer, D. Fake news on Twitter during the 2016 U.S. presidential election. *Science* **363**, 374–378 (2019).

89. Lazer, D. M. *et al.* The science of fake news. *Science* **359**, 1094–1096 (2018).

90. Pennycook, G. & Rand, D. G. Fighting misinformation on social media using crowdsourced judgments of news source quality. *Proc Natl Acad Sci USA* **116**, 2521–2526 (2019).

**Acknowledgments**

This work was supported in part by the John S. and James L. Knight Foundation through the GW Institute for Data, Democracy, and Politics

**Author Contributions**

    Conceptualization: DAB, JG, JRS, LCA

    Methodology: DAB, JG, AMJ, JRS, LCA,

    Investigation: DAB, JG, JRS AMJ

    Visualization: DAB, JRS

    Funding acquisition: DAB, LCA

    Project administration: DAB

    Supervision: DAB

    Writing – original draft: DAB

    Writing – review & editing: DAB, JG, AMJ, JRS, LCA

**Competing Interests Declaration**

DAB has received consulting fees from Merck & Co. for participating in the 2021 Merck Global Vaccine Confidence Expert Input Forum, and has received a speaking honorarium from the United Nations Shot@Life Foundation. JG, AMJ, JRS, and LCA declare no competing interests.

**Supplementary Information** is available for this paper.

**Extended Data Fig. 1**

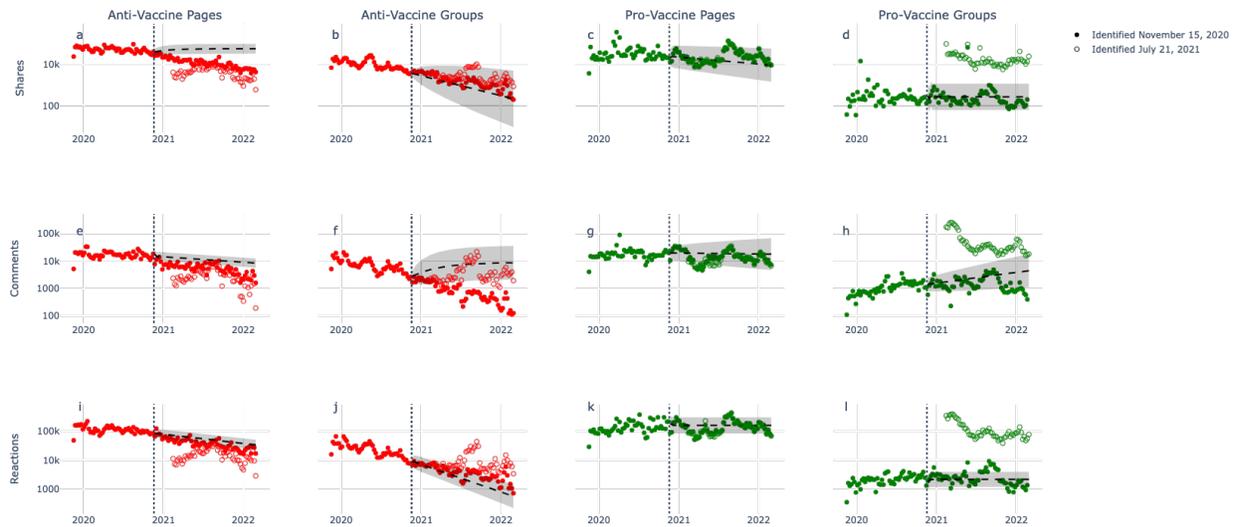

**Extended Data Fig. 1. Reactions (e.g., "likes") were especially insensitive to Facebook's content removal policies.**

ARIMA model results show that, compared to projections (black dashed line), Facebook's policies led to a decline in shares in **a.** anti-vaccine pages (74% average decrease, $\chi^2(66)= 1666.28, p<0.001$) but not **b.** anti-vaccine groups (84% average increase, $\chi^2(66)= 17.72, p=1.00$), **c.** pro-vaccine pages (202% average increase, $\chi^2(66)= 50.53, p=0.92$) or **d.** pro-vaccine groups (465% average increase, $\chi^2(66)= 76.17, p=0.18$). Furthermore, comments declined in **e.** anti-vaccine pages (49% average decrease, $\chi^2(66)= 838.72, p<0.001$) but returned to within the prediction's 90% confidence intervals for the months of March, May, June, and July-September, 2021. Comments also declined in **f.** anti-vaccine groups (82% average decrease, $\chi^2(66)= 570.58, p<0.001$), although comments in groups in the second sample returned to pre-policy levels, possibly reflecting a transition to newer groups. In contrast, we did not detect a change in comments **g.** pro-vaccine pages (16% average decrease, $\chi^2(66)= 70.82, p=0.32$) although they did decrease in **h.** pro-vaccine groups (29% average increase, $\chi^2(66)= 120.35$, $p<0.001$). Finally, reactions declined in **i.** anti-vaccine pages (28% average decrease, $\chi^2(66)= 215.75, p<0.001$) but also repeatedly returned to within the prediction's 90% confidence intervals. In contrast, reactions increased in **j.** anti-vaccine groups (83% average increase, $\chi^2(66)= 114.09, p<0.001$). Reactions also decreased slightly in **k.** pro-vaccine pages (5% average decrease, $\chi^2(66)= 107.38, p<0.001$) and briefly increased in **h.** pro-vaccine groups (30% average increase, $\chi^2(66)= 174.55, p<0.001$). Error bars reflect 90% confidence intervals. Data from the second sample is shown for comparison and qualitatively replicates findings from the first sample.

**Extended Data Fig. 2**

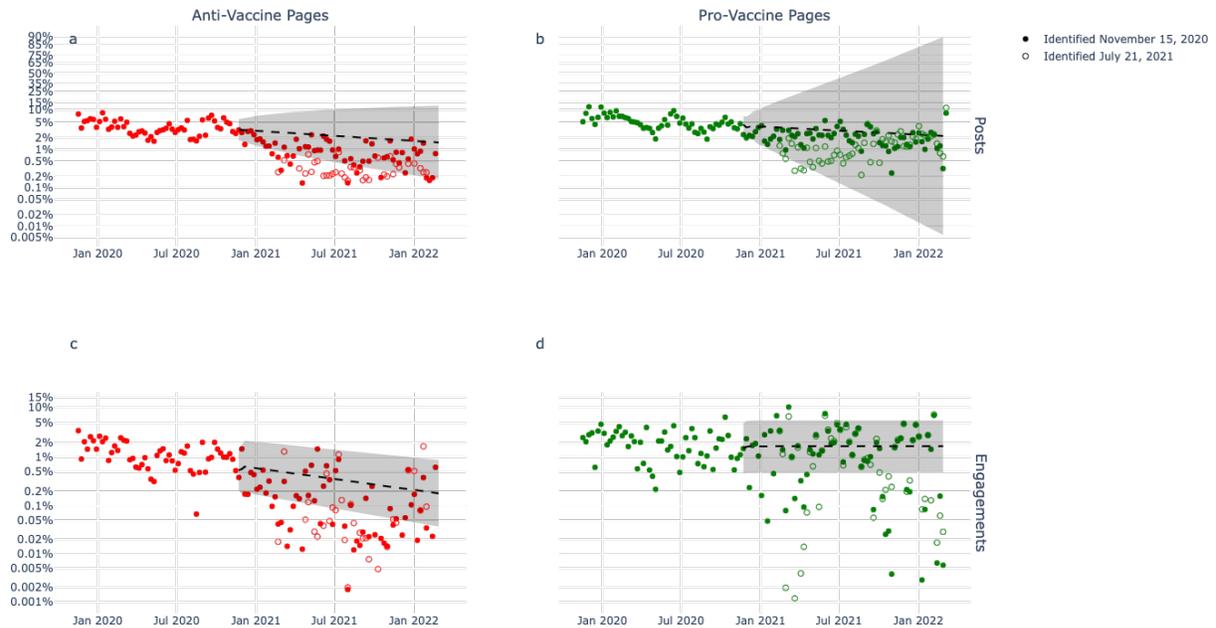

**Extended Data Fig. 2. Proportions of links between Facebook pages declined in anti-vaccine pages, but repeatedly returned to baseline levels.**

ARIMA model results show that, compared to projections (black dashed line), Facebook's policies led to a decline in in **a.** the proportion of links to in-sample pages anti-vaccine pages in the first sample, consisting of venues identified on Nov. 15, 2020, (57% average decrease, $\chi^2(62)= 114.05, p<0.001$) and **b.** the proportion of engagements with those links (42% average decrease, $\chi^2(60)= 279.38\ p<0.001$). In contrast, **c.** we did not detect a change in the proportion of links to pro-vaccine pages from other pro-vaccine pages (27% average decrease, $\chi^2(66)= 17.24, p=1.00$) and **d.** engagements with these links increased significantly (14% average increase, $\chi^2(65)= 459.44, p<0.001$). Error bars reflect 90% confidence intervals. Data from the second sample consisting of venues identified on Jul. 21, 2021, is shown for comparison and qualitatively replicates findings from the first sample. Weeks in which no links to other pages were detected are treated as missing data.

**Extended Data Fig. 3**

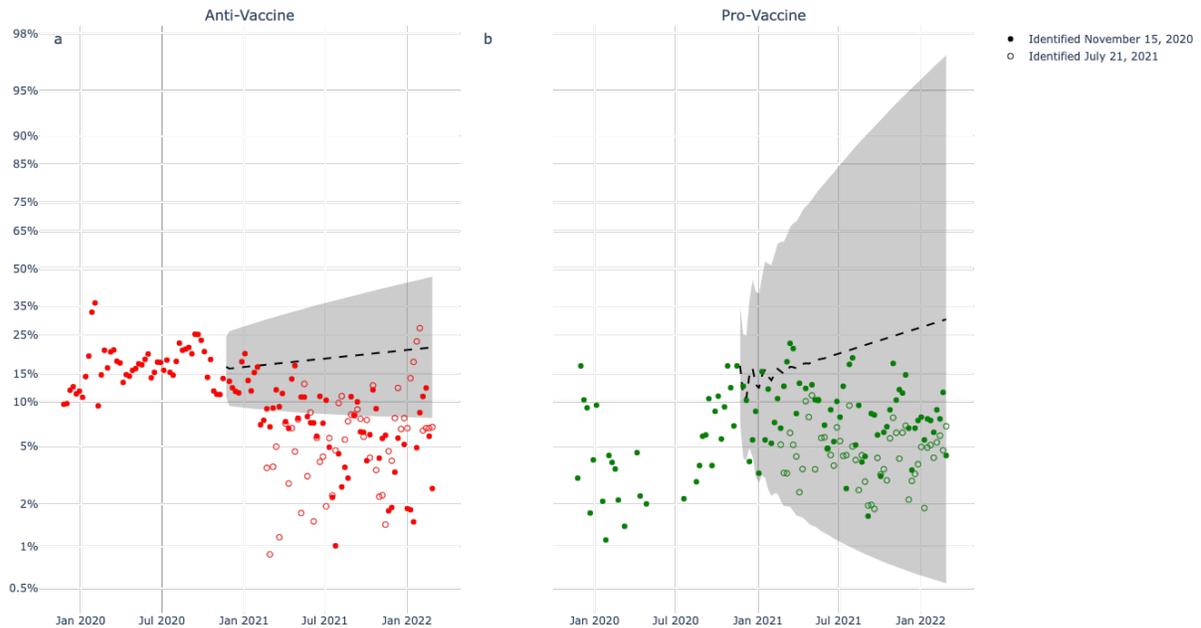

**Extended Data Fig. 3. Weekly proportions of successive URL pairs that were in under 33 seconds, suggesting coordinated behaviour.**

ARIMA model results show that, compared to projections (black dashed line), Facebook's policies led to a decline in **a.** the proportion of coordinated URLs among anti-vaccine venues in the first sample, (64% average decrease, $\chi^2(64)= 321.96, p<0.001$), despite a repeated return to within 90% confidence levels from the pre-policy projection. In the second sample, proportions exceeded the pre-policy projection. In contrast, **b.** the proportion of coordinated URLs among pro-vaccine venues in the first sample did not differ significantly from pre-policy projections, (63% average decrease, $\chi^2(62)= 30.01, p=1.00$). Qualitatively, we observe an increase in pro-vaccine coordinated behaviour, consistent with the emergence of "vaccine hunter" groups that promoted awareness of COVID-19 vaccine availability during the first half of 2021. Error bars reflect 90% confidence intervals. Data from the second sample consisting of venues identified on Jul. 21, 2021, is shown for comparison and qualitatively replicates findings from the first sample. Weeks in which no links to other pages were detected are treated as missing data.

**Extended Data Fig. 4**

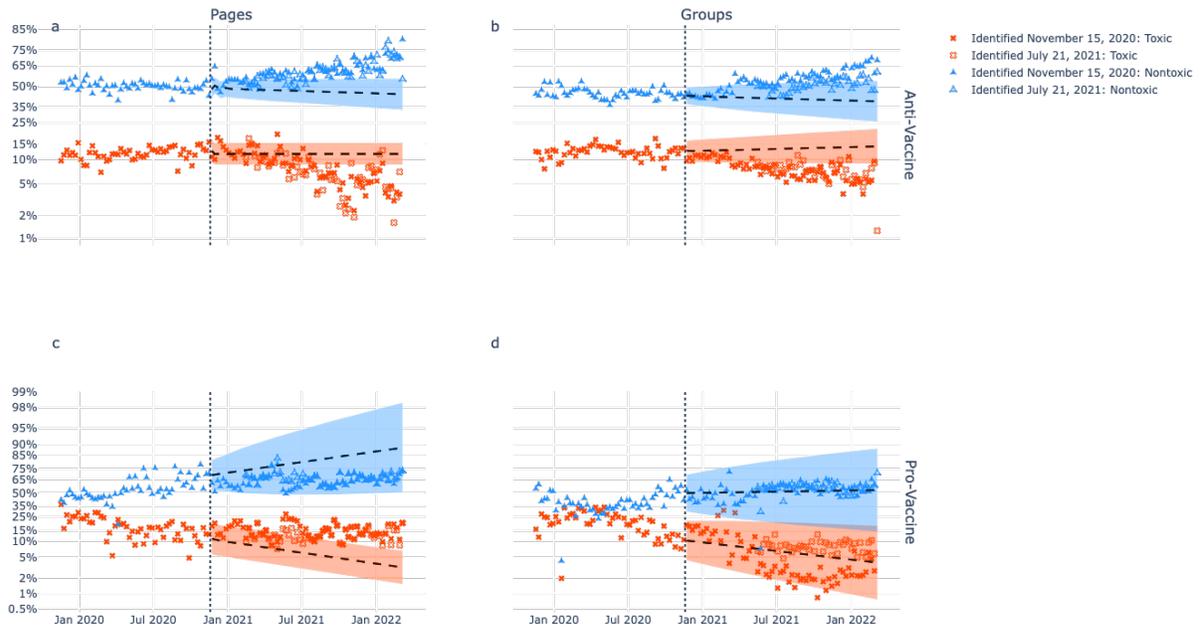

**Extended Data Fig. 4. Weekly proportions of "toxic" and "nontoxic" reactions for pro- and anti-vaccine pages and groups.**

All proportions are calculated as the percent of all interactions. ARIMA model results show that, compared to projections (black dashed line), Facebook's policies led to a decline in the proportion of "toxic" reactions and an increase in the proportion of "non-toxic" reactions in **a.** anti-vaccine pages: 26% average decrease, $\chi^2(66)= 816.74, p<0.001$ and 30% average increase, $\chi^2(66)= 463.39, p<0.001$, respectively, and **b.** anti-vaccine groups: 40% average decrease, $\chi^2(66)= 423.24, p<0.001$ and 33% average increase, $\chi^2(66)= 280.21, p<0.001$, respectively. In contrast, we observed **c.** a significant increase – 271% in average – in the proportion of "toxic" reactions in pro-vaccine pages, $\chi^2(66)= 398.87, p<0.001,$ but no significant changes in the proportion of "non-toxic" reactions in pro-vaccine pages – 22% decrease on average – $\chi^2(66)=$

73.53, $p=0.25$, **d.** "toxic" reactions in pro-vaccine groups, – 12% decrease on average – $\chi^2(66)=$ 86.34, $p=0.05$ and "non-toxic" reactions in pro-vaccine pages, – 0% decrease on average – $\chi^2(66)= 21.07$, $p=1.00$.

**Extended Data Fig. 5**

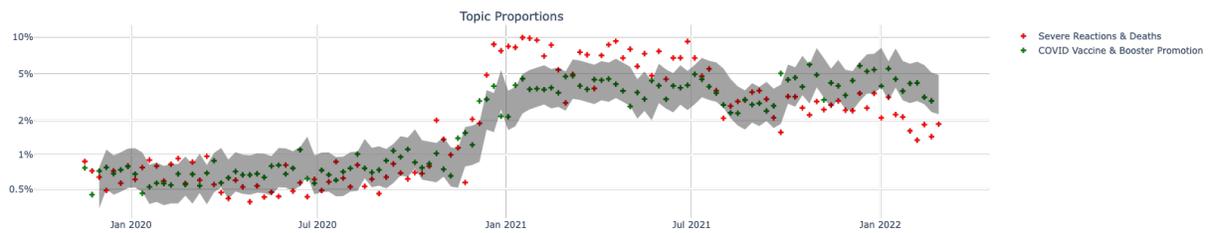

**Extended Data Fig. 5. Following Facebook's policies, a topic promoting sensationalised reports of severe adverse events and deaths from COVID-vaccines grew faster than the corresponding pro-vaccine topic.**

Facebook's policy coincided with the worldwide rollout of COVID-vaccines, spurring increased discussion in both anti-vaccine and pro-vaccine venues. We found that content in anti-vaccine pages was, on average, 16% more likely to report misinformative stories alleging that COVID vaccines caused severe adverse reactions and deaths, compared to content in pro-vaccine promoting vaccine uptake, $\chi^2(120)= 620.23, p<0.001$. This difference was most pronounced between November 18, 2020 through July 4, 2021, during the initial vaccine rollout, when content in anti-vaccine pages was, on average, 88% more likely than corresponding pro-vaccine content. All example messages shown are from this time period.

**Extended Data Fig. 6**

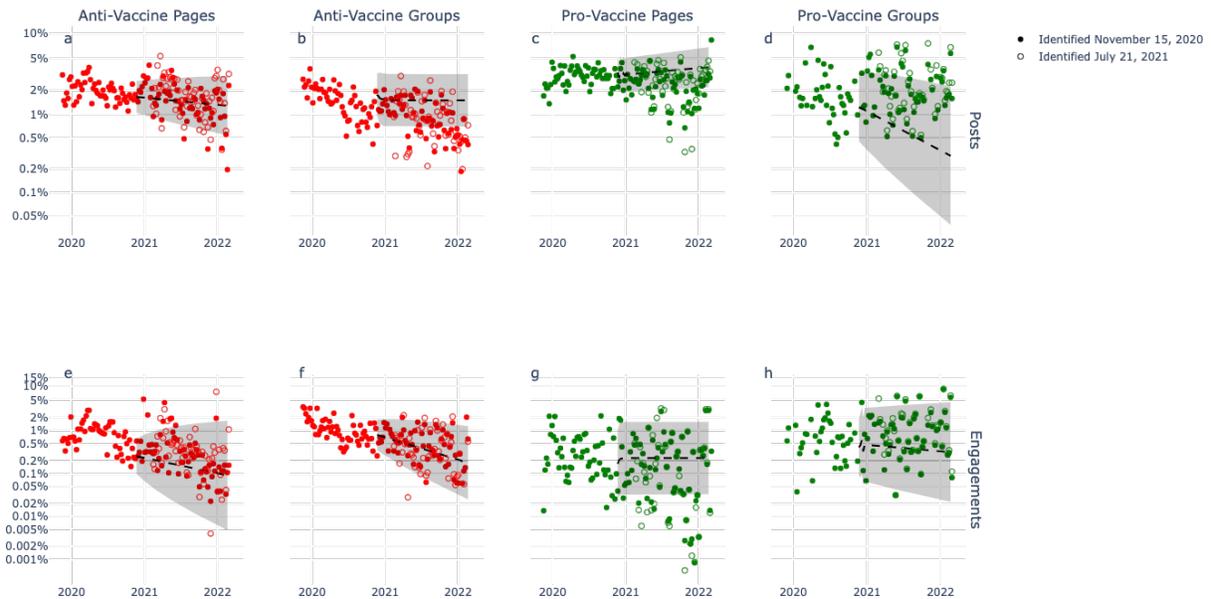

**Extended Data Fig. 6. Weekly proportions of URLs pointing to high-quality academic and government sources for pro- and anti-vaccine pages and groups.**

We examined whether Facebook's policies led to a significant change in the proportion of URLs pointing to high-quality academic and government sources. We found **a.** a slight, yet significant, increase in anti-vaccine pages: 10% average increase, $\chi^2(66)= 94.44, p=0.01$, but **b.** a large decrease in anti-vaccine groups: 44% average decrease, $\chi^2(64)= 161.72, p<0.001$ (one datapoint was dropped due to having zero URLs). **c.** We also detected a significant decrease of 29% in pro-vaccine pages, $\chi^2(66)= 246.88, p<0.001$, **d.** Although posts in pro-vaccine groups were significantly more likely – 362% on average compared to pre-policy projections – to contain these links, $\chi^2(55)= 100.06, p<0.001$. Although the proportion of high-quality posts in anti-vaccine pages increased slightly, **e.** we did not detect a significant change in engagement with these posts $\chi^2(66)= 72.88, p=0.26$. **f.** Similarly, we did not detect a significant change in engagements in anti-vaccine groups $\chi^2(63)= 50.32, p=0.88$. **g.** Followers of pro-vaccine pages were 182% more likely to engage with high-quality sources, $x^2(66)= 187.26, p<0.001$. **h.** In contrast, engagement rates did not increase in pro-vaccine groups $x^2(55)= 54.75, p=0.48$.

**Extended Data Fig. 7**

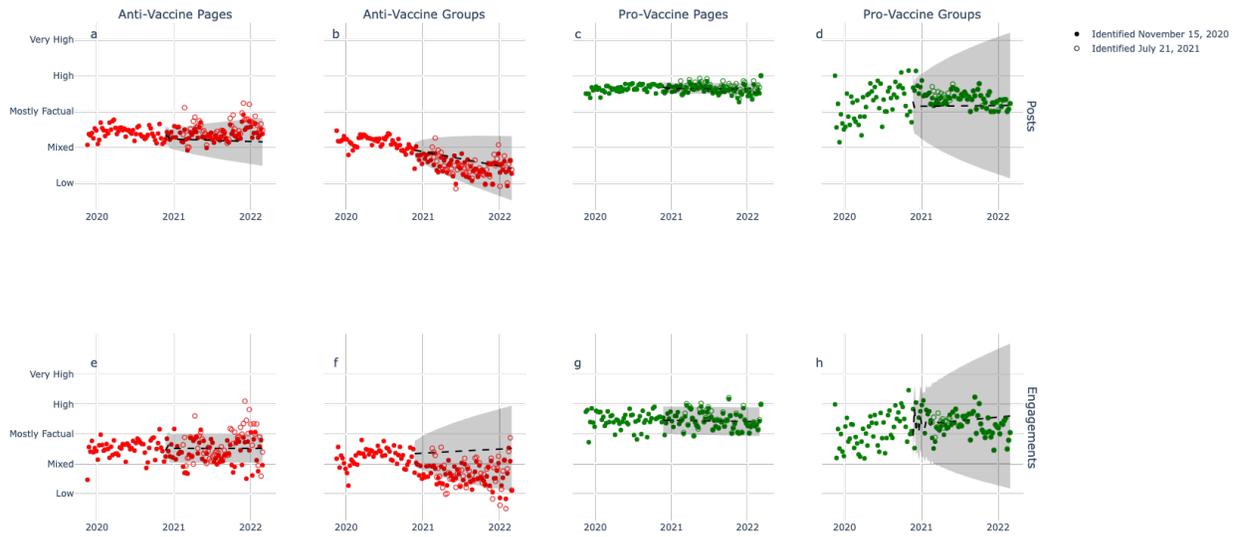

**Extended Data Fig. 7. Weekly average factual rating score of rated URLs for pro- and anti-vaccine pages and groups.**

We examined whether Facebook's policies led to a significant change in the average fact rating scores for rated URLs. Compared to pre-policy projections, **a.** we did not detect a significant change in the average ratings for posts in anti-vaccine pages, $\chi^2(66)= 48.52, p=0.94,$ or **b.** anti-vaccine groups, $\chi^2(66)= 43.76, p=0.98$, although the average factual rating score for anti-vaccine groups was already decreasing prior to November, 18, 2020. **c.** In contrast, we did detect a slight decrease (0.04 points on average) in the average factual accuracy of URLs in pro-vaccine pages, $\chi^2(66)= 110.06, p<0.001$ **d.** but not in pro-vaccine groups, $\chi^2(66)= 15.16, p=1.00$. Despite the lack of significant changes in post ratings **e.** we found that the average fact rating of user engagements decreased by 0.13 points in anti-vaccine pages, $\chi^2(66)= 183.36, p<0.001$. **f.** and by 0.80 points in anti-vaccine groups $\chi^2(66)= 130.82, p<0.001$. **g.** In contrast, we did not detect significant differences in average fact ratings for engagements in pro-vaccine pages, $\chi^2(66)= 56.73, p=0.78$. **h.** or pro-vaccine groups $\chi^2(66)= 20.43, p=1.00$.

**Extended Data Fig. 8**

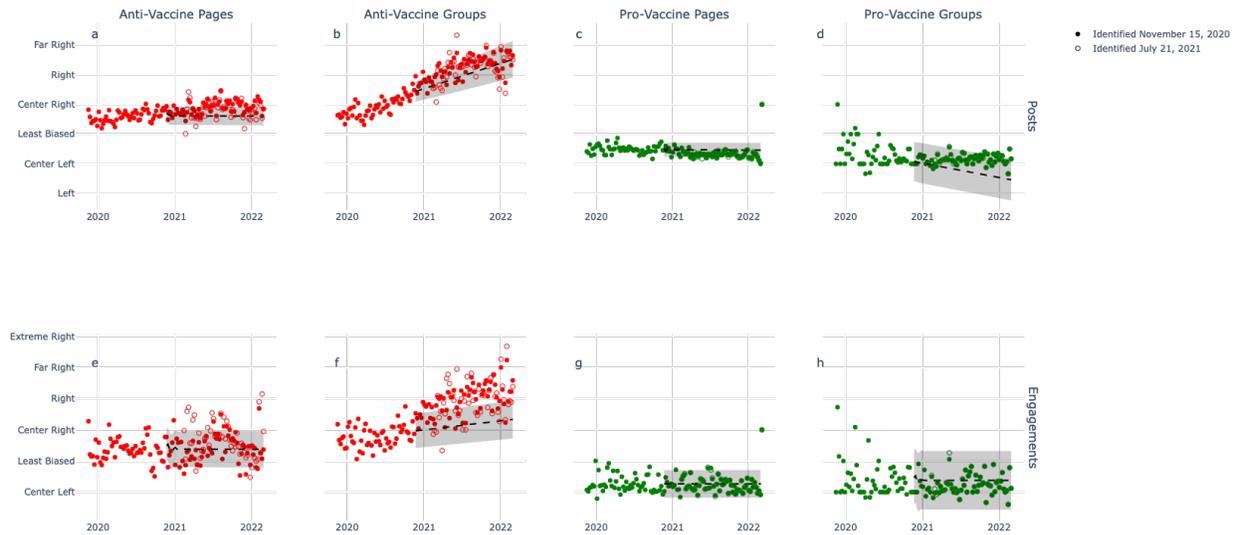

**Extended Data Fig. 8. Weekly average partisan bias score of rated URLs for pro- and anti-vaccine pages and groups.**

We examined whether Facebook's policies led to a significant change in the average partisan bias scores for rated URLs. Compared to pre-policy projections, **a.** rated posts in anti-vaccine pages became a weekly average of 0.27 points more biased towards the political right wing, $\chi^2(66)= 222.24, p<0.001,$ and **b.** rated posts in anti-vaccine groups became a weekly average of 0.24 points more biased towards the political right wing, $\chi^2(66)= 106.48, p<0.001$. **c.** In contrast, rated posts in pro-vaccine pages became a weekly average 0.13 points more biased towards the political left wing, $\chi^2(66)= 95.86, p=0.01$ **d.** We did not detect a change in pro-vaccine groups, $\chi^2(66)= 75.52, p=0.17$. Concurrently **e.** we found that the user engagements with posts containing rated links became a weekly average of 0.02 points more biased towards the political right wing, $\chi^2(66)= 114.22, p<0.001$ in anti-vaccine pages. **f.** of 0.78 points more biased towards the political right wing, $\chi^2(66)= 402.10, p<0.001$ in anti-vaccine groups. **g.** In contrast, we did not detect significant differences in average partisan for engagements in pro-vaccine pages, $\chi^2(66)= 42.21, p=0.99$. **h.** or pro-vaccine groups $\chi^2(66)= 24.02, p=1.00$.

# Extended Data Table 1

Extended Data Table 1. Dates and Sample Sizes for Each Dataset Collection from CrowdTangle

| Collection Date | First Post | Last Posts | N (Pages) | N (Groups) |
|---|---|---|---|---|
| Identified November 15, 2020 | | | | |
| 11/15/20 | 11/15/19 | 11/15/20 | 199,183 | 264,568 |
| 11/30/20 | 11/18/20 | 11/30/20 | 7,619 | 14,646 |
| 3/7/21 | 11/18/20 | 3/7/21 | 58,811 | 76,443 |
| 3/16/21 | 11/15/19 | 3/16/21 | 227,864 | 243,863 |
| 4/8/21 | 2/8/21 | 4/8/21 | 27,739 | 33,960 |
| 5/8/21 | 2/8/21 | 5/8/21 | 40,290 | 53,390 |
| 7/11/21 | 2/8/21 | 7/8/21 | 63,156 | 80,263 |
| 8/9/21 | 2/8/21 | 8/8/21 | 72,059 | 96,500 |
| 9/9/21 | 8/9/21 | 9/8/21 | 12,004 | 16,284 |
| 9/20/21 | 8/9/21 | 9/19/21 | 15,974 | 23,000 |
| 1/31/22 | 2/8/21 | 1/31/22 | 130,439 | 188,548 |
| 2/8/22 | 2/1/21 | 2/8/22 | 136,535 | 195,755 |
| 2/28/22 | 11/15/19 | 2/28/22 | 326,930 | 367,046 |
| Identified July 21, 2021 | | | | |
| 8/9/21 | 2/8/21 | 8/8/21 | 92,624 | 333,748 |
| 9/9/21 | 8/9/21 | 9/8/21 | 16,117 | 36,798 |
| 9/20/21 | 8/9/21 | 9/19/21 | 20,945 | 50,129 |
| 1/31/22 | 2/9/21 | 1/31/22 | 156,820 | 477,832 |
| 2/8/22 | 2/1/21 | 2/8/22 | 162,513 | 491,752 |
| 2/28/22 | 11/15/19 | 2/28/22 | 303,165 | 750,286 |

# Extended Data Table 2

Extended Data Table 2. **Venue counts for each sample.**

| Category | Venues Identified November 15, 2020 | Venues Identified July 21, 2021 |
|---|---|---|
| Alternative medicine | 6 | 0 |
| Civil liberties | 43 | 14 |
| Conspiracy | 24 | 4 |
| Conspiracy & Civil liberties | 0 | 0 |
| Limited info | 0 | 0 |
| Morality issues | 2 | 1 |
| Other | 7 | 0 |
| Safety concerns | 32 | 14 |
| Pro-Vaccine | | |
| Anti anti-vaccine arguments | 24 | 6 |
| Other | 4 | 4 |
| Pro science | 4 | 3 |
| Pro-policy | 3 | 1 |
| Promotion | 56 | 25 |
| Safe & effective | 11 | 7 |
| Other | | |
| Foreign language | 3 | 3 |
| General public health | 10 | 7 |
| News sharing | 12 | 12 |
| Non-relevant | 1 | 0 |
| Other | 23 | 4 |
| Pharmacy | 1 | 0 |
| Research | 17 | 11 |
| Unrelated | 29 | 14 |
| Groups | | |
| Anti-Vaccine | | |
| Alternative medicine | 1 | 0 |
| Civil liberties | 23 | 24 |
| Conspiracy | 40 | 18 |
| Conspiracy & Civil liberties | 1 | 0 |
| Limited info | 1 | 0 |
| Morality issues | 2 | 2 |
| Other | 7 | 4 |
| Safety concerns | 17 | 22 |
| Pro-Vaccine | | |
| Anti anti-vaccine arguments | 5 | 4 |
| Other | 1 | 20 |
| Pro science | 0 | 1 |
| Pro-policy | 0 | 1 |
| Promotion | 1 | 41 |
| Safe & effective | 1 | 3 |
| Unclear | | |
| Side effects | 0 | 7 |
| Other | | |
| Foreign language | 17 | 9 |
| General public health | 2 | 0 |
| News sharing | 4 | 1 |
| Non-relevant | 0 | 0 |
| Other | 5 | 0 |
| Pharmacy | 0 | 0 |
| Research | 1 | 0 |
| Unrelated | 11 | 5 |

# Supplementary Information

## 1. Simulation model

We constructed a computational simulation to clearly articulate why Facebook's policies have not permanently reduced engagement with anti-vaccine content beyond pre-policy projections, despite significant reductions in content volume.

*Agent Types*

Our model posits three types of agents:

1) Venues: To initialise our model, we generate a set of $v \in V = \{v_1, v_2, \ldots, v_n\}$ venues. In each timestep, each venue generates a random number of posts:

$$P_{v,t} \sim NegativeBinomial(a_v, b_v)$$

   where $P_v$ has mean $\mu = e^{a_v}$ and dispersion parameter $b_v$. This formulation is intended to capture the fact that venues differ greatly in their level of activity, with most venues relatively inactive and a small number of venues generating a large number of posts.

2) Posts: Each post generates a random number of user engagements:

$$E_{p\_v} \sim Poisson(\lambda_v)$$

   where $\lambda_v$, the rate parameter, captures the average number of engagements per post in that venue. As above, posts differ greatly from one another in their popularity, with a small number of posts generating a large number of engagements and vice versa. In each timestep, $E_{p\_v}$ decays at a constant rate, $d \in (0, \infty)$, capturing the empirical observation that the total number of engagements per week tends to decline over time, perhaps due to the cumulative effects of prior "soft" content remedies.[1]

3) Content Moderators: We model Facebook's "hard" content removal remedies by randomly assigning an initial number of Facebook staff who serve as content moderators,

$m_0 \in \mathbf{N}$, to each venue in proportion to the total number of posts generated in that timestep. Each timestep, each content moderator removes one post in the moderator's assigned venue unless all posts have already been removed. Additionally, each content moderator that has removed a post "reproduces" – i.e., another content moderator is created – with a fixed probability, $r \in [0,1]$. When the number of content moderators exceeds the number of posts in a venue, all content moderators that did not remove a post are randomly reassigned to new venues in proportion to the number of posts that those venues generated.

*Potential Demand*

Our model explicitly captures how removal of a venue or post creates unmet demand for anti-vaccine content. Specifically, we posit that when content or venues are removed, users are prevented from accessing content that they seek. As a consequence, these users seek out new anti-vaccine venues to follow. Our model contains a variable representing this potential demand at time t, $h_t \in \mathbf{R}^+$, where $h_0 = 0$. Each time a venue, $v$, is removed at time t, the demand for that venue's content is converted into potential demand. In addition, whenever a post is removed by a moderator, the engagements from that post are also converted into potential demand as follows:

$$h_t = h_{t-1} + \sum_{v \in w_{r,t}} \lambda_{v,t} + \frac{|P_r|}{\sum_{p \in P_r} E_p}$$

where $w_{r,t}$ is the set of all venues that have been removed in timestep t, $P_r$ is the set of all posts that have been removed in timestep t, and $E_p$ is the total number of engagements with post $p$. Finally, at the end of each timestep, a fixed percentage of this potential demand is converted to actual engagements, as follows:

$$\Lambda_{v,t} = \lambda_{v,t-1} + h_t \times p_{v,t} \times g$$

where $g$ is a constant between 0 and 1, $\lambda_{v,t}$ is the average number of engagements per post in venue $v$ at time $t$, $p_{v,t}$ is the normalised total number of posts in venue $v$ at time $t$ such that $\sum_v p_{v,t} = 1$, and $\Lambda_{v,t}$ captures the average number of engagements per post that are no longer susceptible to "soft" content remedies because they come from users who actively seek anti-vaccine content, as has been observed for some content remedies that seek to overtly change peoples' preferences.[2–5]

*Model Initialisation*

The model is initialised with the following variables:

- $m_0$: The initial number of content moderators
- $r$: The probability with which content moderators reproduce
- $g$: The rate at which potential demand is converted into actual engagements per post
- $T$: a wave of venues removal occurs when the number of content moderators at time t, $m_t$, exceeds $T$
- $w_{min}$ and $w_{max}$: when $m_t > T$, $w_t$ venues are removed, where

$$w_t \sim Uniform[w_{min}, w_{max}]$$

In addition, the model is calibrated with several parameters:

- *d:* The decay rate of engagements per post, fit to pre-policy data
- $t_{policy}$: The timestep when "hard" content removal remedies begin – content moderators begin to remove posts.
- $v_{initial}$: Several venues were created between November 15, 2019, and November 15, 2020, and were therefore not actively generating posts or engagements throughout the entire time period. At the beginning of the simulation, only $v_{initial}$ venues are actively generating posts and engagements.

- $t_{max\_venues}$: The timestep when all venues in the simulation are active.
- $v_{max}$: Between the beginning of the simulation and $t_{max\_venues}$, new venues are added uniformly at random until all $v_{max}$ venues are active.
- $t_{max}$: The total number of timesteps in a model run

*Model Execution*

In each timestep, the model carries out the following steps:

1) The average number of engagements per post in each venue, $E_{p,v}$, decays by a fixed percentage, $d$, i.e., $E_{p, v, t} = d \times E_{p, v, t-1}$
2) Each venue, $v$, randomly generates $P_{v,t} \sim NegativeBinomial(a_v, b_v)$ posts
3) For each post, $P_{v,t}$, each venue randomly generates $E_{p\_v} \sim Poisson(\lambda_v)$ engagements
4) If $t < t_{policy}$: Unless all venues are already active, one inactive venue is selected at random to become active.
5) If $t >= t_{policy}$:
   a. Each content moderators removes at most one post. Content moderators that removed one post "reproduce" with probability $r$.
   b. Potential demand is incremented by the total number of engagements per post removed:

$$h_{t*} = h_{t-1} + \frac{|P_r|}{\sum_{p \in P_r} E_p}$$

   c. Content moderators that did not remove a post are randomly assigned to new venues with probability $p_{v,t}$ where $\sum_v p_{v,t} = 1$.

d. If the total number of moderators is greater than the threshold: $m_t > T$

   i. $w_t$ venues are selected for removal, where $w_t \sim Uniform[w_{min}, w_{max}]$. For each venue that is removed, all of its posts, engagements, and content moderators are also removed.

   ii. For each venue removed, potential demand is incremented by $\lambda_{v,t}$:

$$h_t = h_{t*} + \sum_{v \in w_{r,t}} \lambda_{v,t}$$

e. Potential demand is converted to engagements per post at a fixed rate, $g$:

$$\Lambda_{v,t} = \lambda_{v,t-1} + h_t \times p_{v,t} \times g$$

The model terminates after $t_{max}$ timesteps.

*Baseline Simulation Results*

**Pages.** Based on pre-policy data, we calibrated our model's parameters for anti-vaccine pages as follows:

$$d = 1.8\%$$

$$v_{initial} = 81$$

$$t_{max\_venues} = 51$$

$$v_{max} = 114$$

$$a_v \sim Normal(\mu = 2.03, \sigma = 1.20)$$

$$b_v \sim LogNormal(\mu = -0.79, \sigma = 1.19)$$

$$\lambda_v \sim (e^{Uniform[0, 6.43]} - 1)$$

We found that our simulation reproduced post-policy data when $m_0 = 500$, $r = 10\%$, $g = 7.5\%$, $T = 5000$, $w_{min} = 5$, and $w_{max} = 15$ (Fig. S1-1)

**Groups.** We also calibrated our model's parameters to fit pre-policy group data as follows:

$$d=3.0\%$$

$$v_{initial}=81$$

$$t_{max\_venues}=50$$

$$v_{max}=92$$

$$a_v \sim Normal(\mu=3.06, \sigma=1.36)$$

$$b_v \sim LogNormal(\mu=-1.10, \sigma=0.96)$$

$$\lambda_v \sim LogNormal(\mu=1.32, \sigma=1.11)$$

We found that our simulation reproduced post-policy data when $m_0$=1500, $r$=5%, $g$=5%, $T$=5000, $w_{min}$=1, and $w_{max}$=3 (Fig. S1)

We ran our model for $t_{max}$=119 timesteps – i.e., the total number of complete weeks from November 15, 2019 through February 28, 2022. We assume Facebook's "hard" content removal policies occurred during week 53.

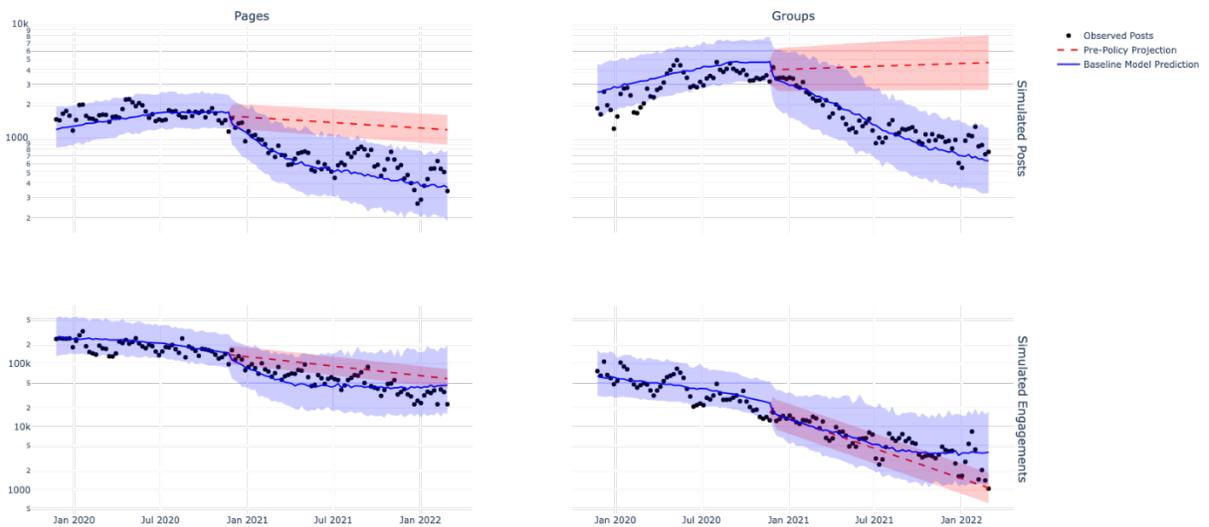

**Fig S1-1. Calibrated simulation model results for pages and groups.** Each model was run 500 times. The median number of pages or engagements in each timestep is represented by the solid blue line, and error bars reflect the 5$^{th}$ and 95$^{th}$ percentiles. In addition, projections based on the ARIMA models fit to pre-policy data are shown by the red dashed line, with error bars reflecting 90% confidence intervals.

*Examining Sensitivity of Baseline Results to Changes in Potential Demand*

The primary source of novelty in our model is our explicit representation of potential demand for misinformation, and the way in which it is converted into engagements. Prior work suggests that a combination of "hard" and "soft" content remedies can successfully reduce misinformation sharing online and, indeed, Facebook already implements both types of in its attempts to reduce anti-vaccine content. Given our findings that engagements repeatedly returned to pre-policy baseline levels, our aim was to examine the consequences of reducing the conversion rate of potential demand into actual engagements, *g,* for both "hard" and "soft" remedies. We first explored the consequences of so by a factor of 5 (Fig. S1-2). We found that this policy led to a large reduction in the number of engagements: 54% in pages and 68% in groups. (Table S1-1).

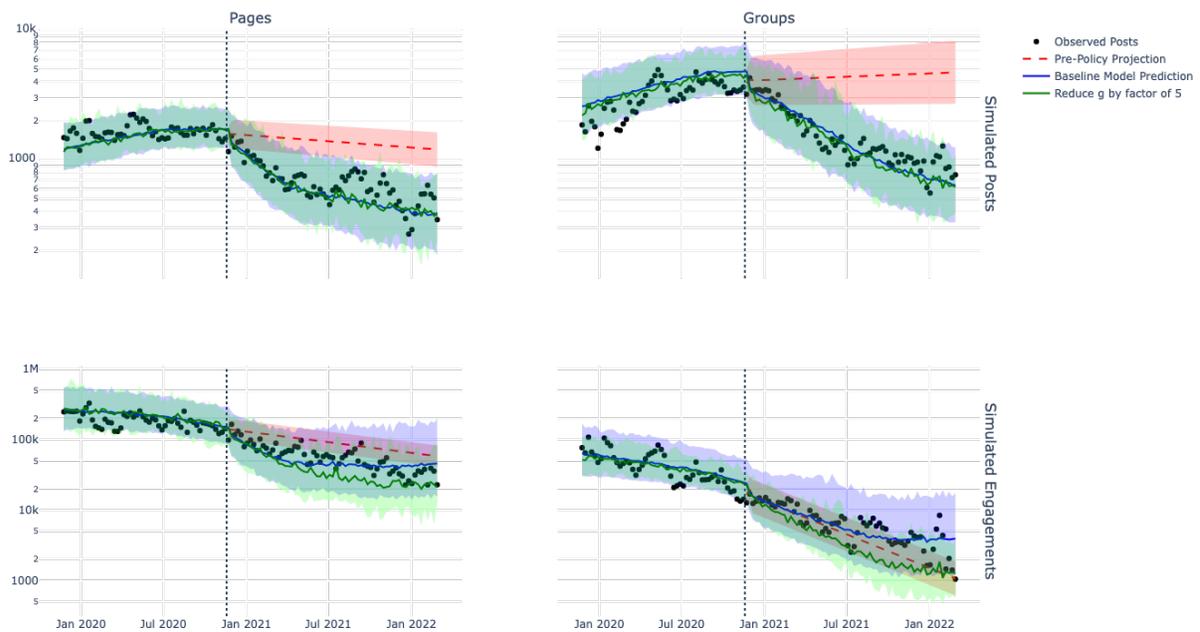

**Fig S1-2. Results of simulation model when decreasing *g*, the conversion rate of potential demand to actual engagements, by a factor of 5.** Each model was run 500 times. The median number of pages or engagements in each timestep is represented by the solid green line, and error bars reflect the 5th and 95th percentiles. The solid blue line and corresponding error bars reflect the baseline model results. Finally projections based on the ARIMA models fit to pre-policy data are shown by the red dashed line, with error bars reflecting 90% confidence intervals.

**Table S1-1.** Percent reduction in post and engagement counts at timesteps 60, 90, and 119 (the end of the simulation) compared to model baseline.

| Variable | Pages | | | | | | Groups | | | | | |
|---|---|---|---|---|---|---|---|---|---|---|---|---|
| Venue | Posts | | | Engagements | | | Posts | | | Engagements | | |
| Timestep | 60 | 90 | 119 | 60 | 90 | 119 | 60 | 90 | 119 | 60 | 90 | 119 |
| | Reducing Potential Demand, *g* | | | | | | | | | | | |
| By factor of 2 | 2% | 8% | 2% | -5% | -13% | -25% | -1% | -2% | 0% | -3% | -28% | -38% |
| By factor of 5 | -1% | 4% | 6% | 0% | -30% | -54% | -4% | -2% | -3% | -12% | -50% | -68% |
| *g* = 0 | 0% | 18% | 6% | -8% | -44% | -79% | -8% | -6% | -2% | -6% | -64% | -89% |
| | "Hard" Content Remedies That Change Moderator Behaviour | | | | | | | | | | | |
| 5 x *r* | -45% | -69% | -71% | -49% | -61% | -59% | -56% | -77% | -65% | -58% | -72% | -45% |
| *T* / 5 | 2% | -57% | -63% | 1% | -47% | -44% | -40% | -70% | -73% | -34% | -46% | -44% |

| | | | | | | | | | | | |
|---|---|---|---|---|---|---|---|---|---|---|---|
| 5 x $m_0$ | -46% | -25% | -15% | -52% | -24% | 10% | -72% | -53% | -44% | -69% | -44% | -4% |
| 5 x $w_{max}$ | -2% | -1% | -28% | -2% | 3% | -25% | -9% | -14% | -4% | -3% | -13% | 15% |
| No moderator limit | -4% | 14% | 25% | 8% | 15% | 42% | -4% | 3% | 73% | -2% | 11% | 113% |
| | | | | | | "Soft" Remedies | | | | | |
| 10% "nudge" | 3% | 5% | 1% | -8% | -3% | -15% | -3% | -1% | 5% | -10% | -10% | -7% |
| 20% "nudge" | 4% | 5% | 0% | -23% | -14% | -21% | -2% | 0% | 7% | -20% | -25% | -14% |
| 40% "nudge" | 8% | 1% | 16% | -35% | -38% | -30% | -4% | -4% | -3% | -45% | -43% | -45% |
| VCB, 100 engagements | -2% | 11% | 15% | -48% | -30% | -36% | 1% | -7% | -6% | -6% | -9% | -22% |
| VCB, 500 engagements | 1% | 4% | 4% | -6% | 3% | 4% | -6% | -10% | 9% | 2% | -7% | 2% |

*Note.* * =p<0.05. **=p<0.01. ***=p<0.001, *g:* Conversion rate of potential demand to actual engagements per post, *r* = moderator growth rate, $m_0$ = initial number of content moderators, *T* = threshold above which a venue removal wave is triggered, $w_{max}$ = maximum number of venues removed per wave, VCB = "Virality Circuit Breaker".

Reducing *g* to zero had an even stronger effect (Table S1-3; 79% in pages and 89% in groups). These findings suggest that existing content remedies employed by Facebook could be significantly strengthened if they were combined with efforts to reduce the conversion of potential demand into actual engagements – e.g., if early gains due to content removal were not reversed due to users explicitly seeking out anti-vaccine content.

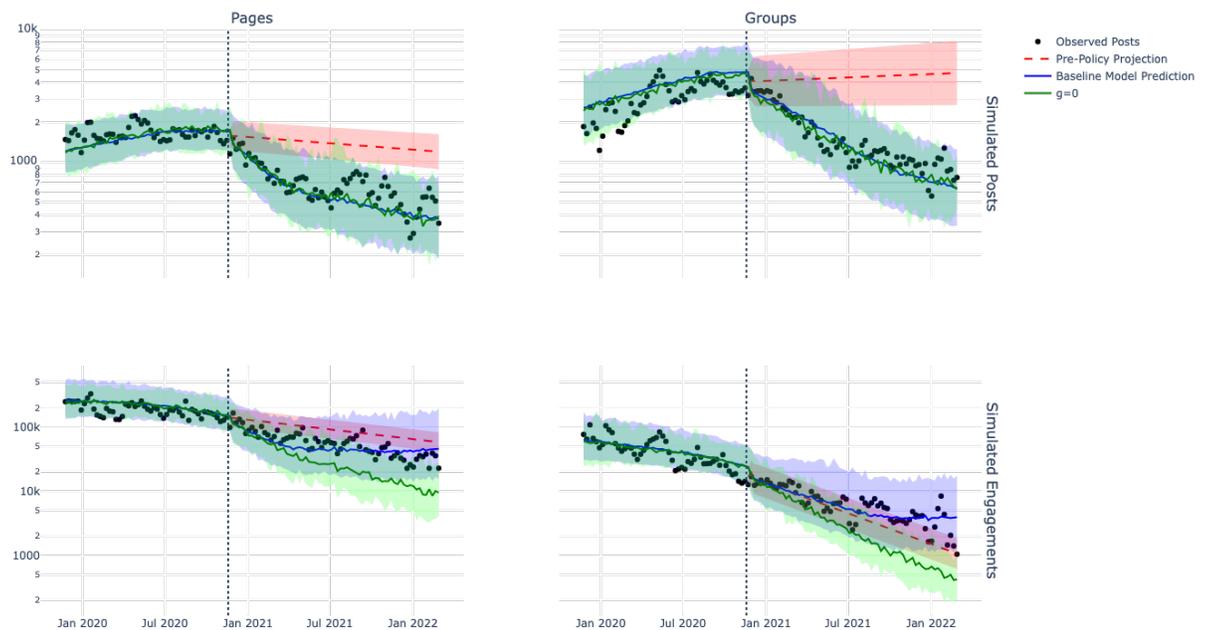

**Fig S1-3. Results of simulation model when decreasing the conversion rate of potential demand to actual engagements to 0.** Each model was run 500 times. The median number of pages or engagements in each timestep is represented by the solid green line, and error bars reflect the 5th and 95th percentiles. The solid blue line and corresponding error bars reflect the baseline model results. Finally projections based on the ARIMA models fit to pre-policy data are shown by the red dashed line, with error bars reflecting 90% confidence intervals.

*Examining Sensitivity of Baseline Results to Changes in "Hard" Content Remedies*

We next examined whether existing content remedies could be amplified to achieve similar results, and what the costs of doing so might be.

**Employing more content moderators.** We first examined whether an initial "pulse" of additional moderators would be effective (Fig. S1-4). We implemented this in our model by increasing the initial number of moderators, $m_0$, by a factor of 5 (see Table S1-1). We found that this approach was effective in the short term, but that this efficacy decreased over time since the total number of moderators recovered more slowly than demand did, despite the initial removal of several venues.

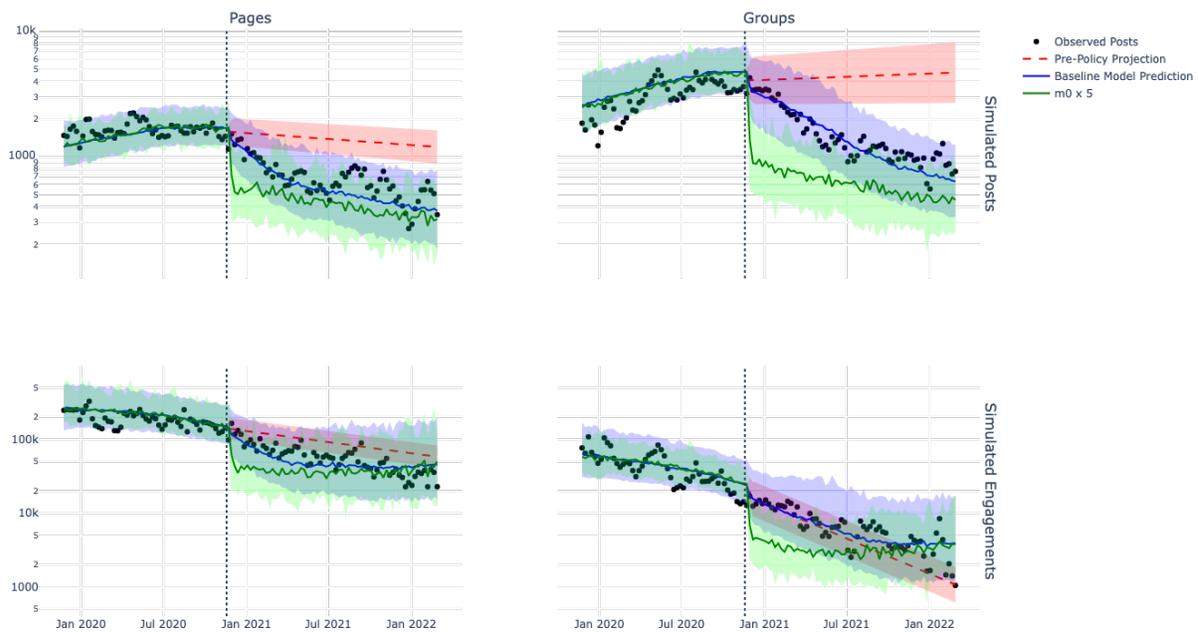

**Fig S1-4. Results of simulation model when increasing initial number of content moderators, $m_0$, by a factor of 5.** Each model was run 500 times. The median number of pages or engagements in each timestep is represented by the solid green line, and error bars reflect the 5$^{th}$ and 95$^{th}$ percentiles. The solid blue line and corresponding error bars reflect the baseline model results. Finally projections based on the ARIMA models fit to pre-policy data are shown by the red dashed line, with error bars reflecting 90% confidence intervals.

We therefore next examined the consequences of increasing the moderator growth rate, $r$, by a factor of 5 (Fig. S1-5). We found that this was an effective approach to reduce the total number of engagements with anti-vaccine content, (by 59% in pages and 45% in groups). However, this approach essentially requires maintaining a "standing army" of moderators that can be quickly deployed to remove content as it appears. In practice, implementing this intervention would be expensive and likely not scalable.

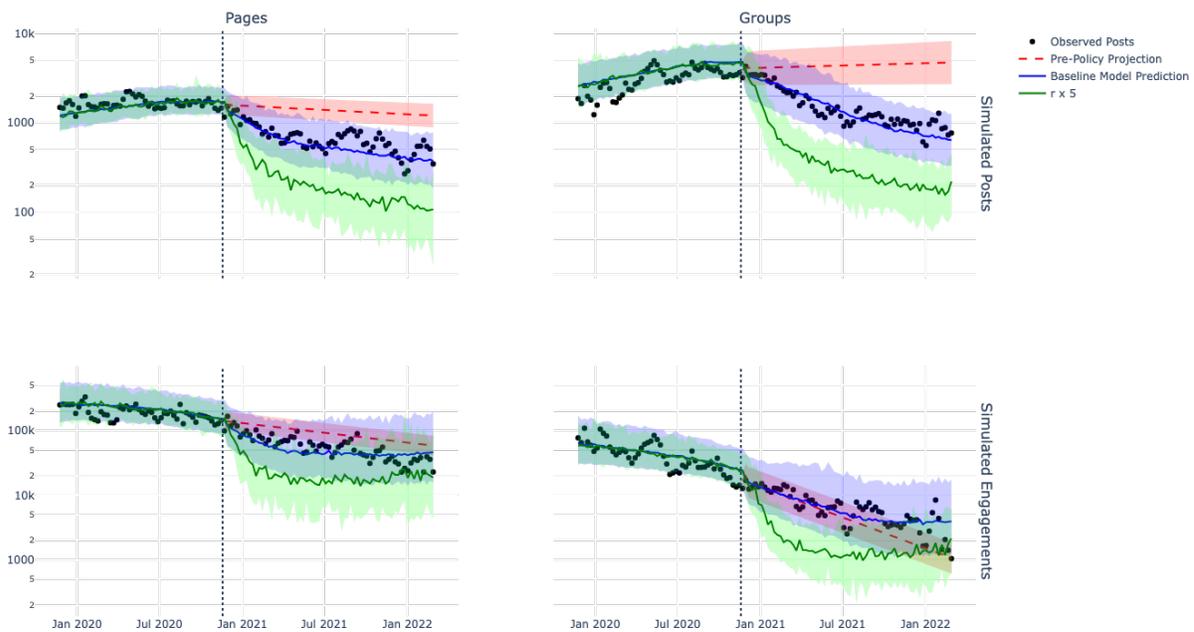

**Fig S1-5. Results of simulation model when increasing the growth rate of content moderators, *r*, by a factor of 5.** Each model was run 500 times. The median number of pages or engagements in each timestep is represented by the solid green line, and error bars reflect the 5[th] and 95[th] percentiles. The solid blue line and corresponding error bars reflect the baseline model results. Finally projections based on the ARIMA models fit to pre-policy data are shown by the red dashed line, with error bars reflecting 90% confidence intervals.

**Removing more venues.** Rather than maintaining a large number of moderators, we next examined the consequences of removing more venues. Specifically, we explored the consequences of removing more venues in each removal wave by increasing $w_{max}$ by a factor of 5 (Fig. S1-6). We found that this approach reduced engagements in the long-term, but not in the short-term.

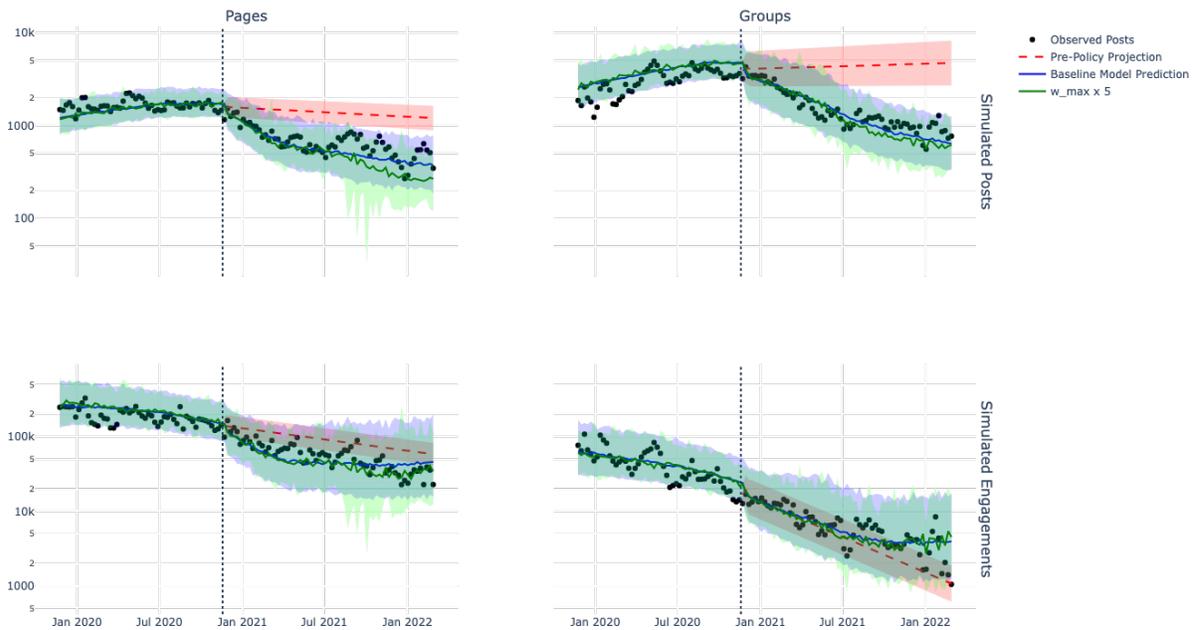

**Fig S1-6. Results of simulation model when the maximum number venues removed in each wave, $w_{max}$, is increased by a factor of 5.** Each model was run 500 times. The median number of pages or engagements in each timestep is represented by the solid green line, and error bars reflect the 5$^{th}$ and 95$^{th}$ percentiles. The solid blue line and corresponding error bars reflect the baseline model results. Finally projections based on the ARIMA models fit to pre-policy data are shown by the red dashed line, with error bars reflecting 90% confidence intervals. We therefore next explored the consequences of removing venues more frequently. Since removals occur when the total number of moderators exceeds $T$, we reduced $T$ by a factor of 5 (Fig. S1-7).

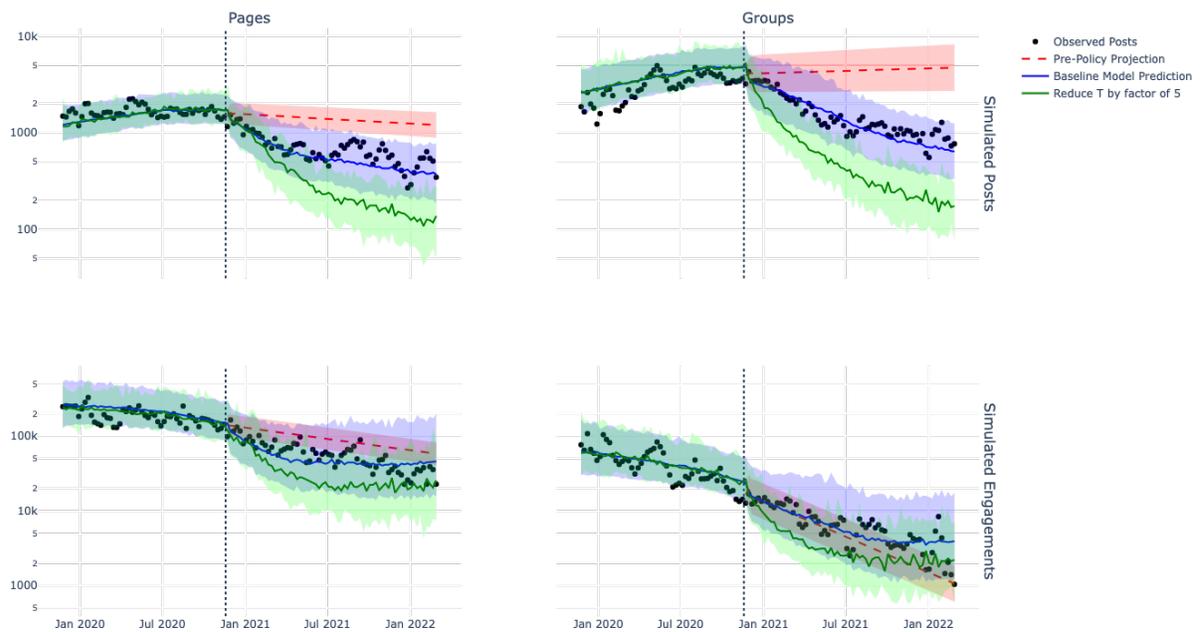

**Fig S1-7. Results of simulation model when reducing content moderators threshold, *T*, by a factor of 5.** Each model was run 500 times. The median number of pages or engagements in each timestep is represented by the solid green line, and error bars reflect the 5th and 95th percentiles. The solid blue line and corresponding error bars reflect the baseline model results. Finally projections based on the ARIMA models fit to pre-policy data are shown by the red dashed line, with error bars reflecting 90% confidence intervals.

We found that this approach also led to a sustained reduction in engagement with anti-vaccine content compared to the baseline; however, this approach is also likely to be expensive since it requires frequently detecting and removing several venues. Furthermore, compared to post removal, venue removal is a blunt instrument that may remove significant amounts of inoffensive content. (We also explored the consequences of increasing *T*, but found it to be ineffective; Fig. S1-8.)

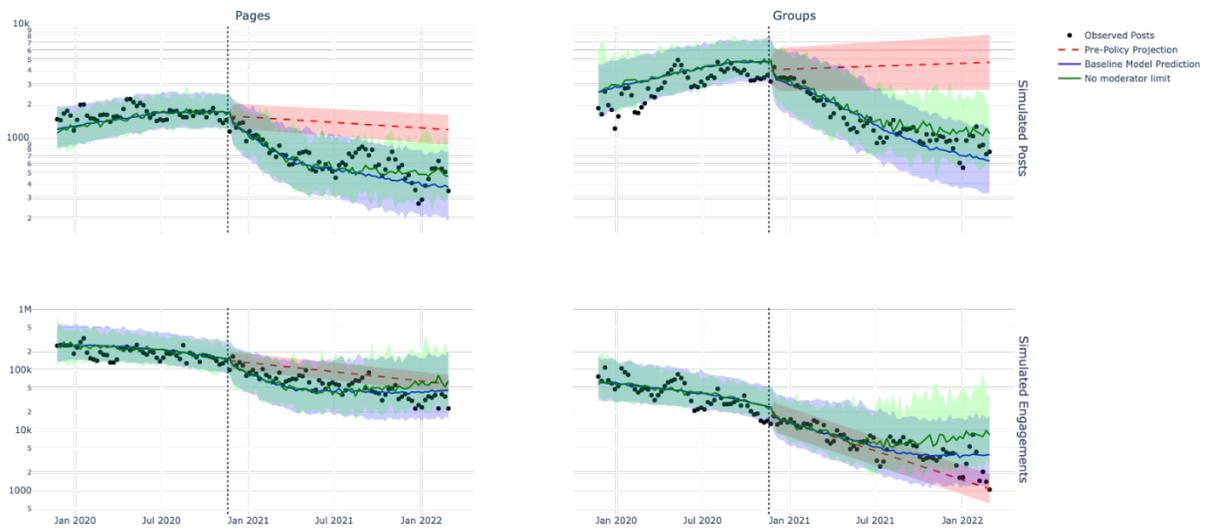

**Fig S1-8. Results of simulation model when allowing content moderators to grow without removing venues.** Each model was run 500 times. The median number of pages or engagements in each timestep is represented by the solid green line, and error bars reflect the 5[th] and 95[th] percentiles. The solid blue line and corresponding error bars reflect the baseline model results. Finally projections based on the ARIMA models fit to pre-policy data are shown by the red dashed line, with error bars reflecting 90% confidence intervals.

*"Soft" Remedies From Prior Literature*

For consistency with prior work,[6] we also examined the effects of using additional "soft" content remedies to reduce engagement with misinformative posts (e.g., by placing a banner on vaccine-related content[1] or by employing nudges,[7] both of which have been shown to reduce engagements) leading the weekly number of engagements to decay at a higher rate. We implemented these in our model by multiplying $\lambda_v$ by a fixed constant, *n*, between 0 and 1. Facebook already employs banners for anti-vaccine content; however, they do not currently employ nudges. Thus, we examined the consequences of setting the value of *n* to 0.1, 0.2, and 0.4 based upon values used in prior work.[6]  0.1 is an estimate derived from experimental data, and 0.4 is an optimistic upper bound based upon modelling studies.[6] As in prior work, we found that such nudges do indeed reduce engagements (Table S1); however, these results should be interpreted with caution since there is some correlational evidence suggesting that the efficacy of such nudges may decrease over time[8] – a factor that is not captured in our model.

Finally, we examined the effects of employing a "virality circuit breaker"[9] which would prevent viral content from obtaining more than a fixed number of engagements without content moderator approval.  We implemented this in our model by changing the functional form of the probability distribution generating engagements per post. Specifically:

$$E_{p\_v} \sim TruncatedPoisson(\lambda_v, \kappa)$$

where $\kappa$ is a maximum value above which posts are not allowed to garner more engagements. We found that setting $\kappa=100$ led to a sustained reduction in engagement with anti-vaccine content; however, $\kappa=1000$ and $\kappa=500$ did not (Table S1). In practice, requiring content moderators to review all posts that have garnered 100 engagements would be burdensome; thus we conclude that this may not be a viable strategy to implement at scale.

*General Discussion*

In conclusion, our model suggests that Facebook's "hard" content remedies appear to have been undermined in the long-term by a slow increase in the number of engagements per post in each venue. Our model attributes this increase to the conversion of potential demand to actual engagements – e.g., when users actively seek out anti-vaccine content after it has been denied to them by prior content and venue removals. Our model's results show that a sustained decline in engagements is possible under the existing paradigm, but that implementing this reduction would require Facebook to make changes. The first possible change would be costly, requiring Facebook to develop a "standing army" of content moderators that can swiftly be deployed to remove venues on relatively short notice. Alternatively, social media platforms such as Facebook could invest in techniques to reduce potential demand or develop techniques that restrict conversion into actual engagements.

## 2. Topic Models

*Table S2-1.* Top 5 terms associated with each topic automatically extracted using Latent Dirichlet Allocation topic model

| Position | Category | Topic Label | Top 5 Terms | | | | |
|---|---|---|---|---|---|---|---|
| Anti-Vaccine | Alternative Medicines | Natural Health | vitamin | health | cancer | body | natural |
| | Civil Liberties | Anti-Mask | mask | masks | wearing | wear | face |
| | | Canada Rallies | canada | protest | freedom | covid | lockdown |
| | | Discussing Vaccine Mandates | vaccine | covid | vaccination | vaccinated | get |
| | | Health Freedom | freedom | vaccine | health | medical | choice |
| | | Informed Consent | health | medical | public | science | consent |
| | | Legislative Action | gov | state | governor | bill | senator |
| | | Lockdowns | coronavirus | covid | health | pandemic | cdc |
| | | Mandates Illegal/Immoral | law | court | rights | act | government |
| | | Opposing School Mandates | school | bill | new | children | state |
| | | Sign Petitions | please | help | information | email | share |
| | | Worldwide Protests | world | south | countries | africa | germany |
| | Conspiracy Theories | "Question the Lies" | anti | vaccine | vaccines | science | truth |
| | | Big Pharma | vaccine | vaccines | pharma | companies | pharmaceutical |
| | | Bill Gates | gates | bill | world | foundation | global |
| | | Clinical Trial Conspiracies | vaccine | covid | vaccines | trials | trial |
| | | COVID Not Deadly | covid | deaths | death | cases | people |
| | | Fauci Conspiracies | fauci | china | coronavirus | virus | wuhan |
| | | General Conspiracies | one | would | many | even | also |
| | | Liberty Beacon Podcast | twitter | follow | article | tlb | liberty |
| | | Measles | measles | vaccine | polio | gates | vaccines |
| | | Medical Harm | covid | patients | doctors | hospital | care |
| | | Please Share Truth! | share | post | video | please | watch |
| | | QAnon Conspiracies | earth | world | children | child | black |
| | | Rebel Doctors | medical | university | professor | medicine | research |
| | | RSB Podcast | robert | show | kennedy | health | del |
| | | Social Media Censorship | media | news | facebook | social | fact |
| | | Totalitarian Government | people | world | government | fear | control |
| | | Toxins | water | food | radiation | technology | health |
| | | Unreliable Tests | test | covid | positive | tests | testing |
| | Morality Issues | Foetal Tissues | vaccines | vaccine | dna | human | cells |
| | | Religious Conspiracies | god | jesus | earth | shall | lord |
| | Safety Concerns | Autism | autism | vaccines | children | study | vaccine |
| | | Dead Children | old | year | baby | son | family |
| | | Gardasil | vaccine | hpv | adverse | gardasil | women |
| | | Harm to Children | vaccine | vaccines | children | flu | child |
| | | Heavy Metals | aluminum | mercury | vaccines | ingredients | mcg |
| | | Misuse of Science | study | covid | sars | cov | risk |
| | | Personal Stories | people | get | know | like | going |
| | | Severe Reactions & Deaths | vaccine | covid | pfizer | died | people |
| | | Severe Side Effects | brain | disease | symptoms | syndrome | disorders |
| | | Side Effects | virus | immune | immunity | disease | viruses |
| | Other/Unknown | Australian Politics | australia | government | australian | lockdown | news |
| | | Business Impacts | home | food | business | new | businesses |
| | | Community Building | james | paul | david | john | andrew |
| | | Los Angeles | police | county | city | arrested | california |
| | | New Age | love | life | truth | world | time |
| | | Tagalog | ang | fai | philippines | mga | news |
| | | U.S. Politics | trump | president | biden | election | joe |
| | | Watch Episodes | live | watch | truth | com | www |
| Pro-Vaccine | Anti Anti-Vaccine Arguments | Anti-Vax Myths | anti | science | people | vaccines | vaxxers |
| | | Debunking | misinformation | covid | anti | vaccine | media |
| | | Measles Outbreaks | measles | samoa | outbreak | children | years |
| | | Mocking Anti-vax | like | post | reply | photos | antivax |
| | | Mocking Natural Health | food | water | organic | vitamin | natural |
| | | Refuting Anti-Vax | anti | vaccine | vaxxers | vax | memes |
| | | Viruses Kill | year | old | died | family | children |
| | Pro Science & Biomedicine | Immunity | covid | vaccinated | people | get | immunity |
| | | MRNA Vaccine | vaccines | immune | mrna | system | vaccine |
| | | Science Promotion | would | people | many | one | even |
| | Pro Vaccine Policy | Vaccine Mandates | vaccine | law | vaccines | mandates | covid |
| | | Vaccine Regulation | vaccine | covid | fda | cdc | use |
| | Promotion | Canadian Promotion | timeline | photos | vaccine | canada | immunization |
| | | Childhood Vaccines | children | vaccines | child | kids | parents |
| | | COVID Vaccine & Booster Promotion | vaccine | covid | dose | first | doses |
| | | Flu Shot Promotion | flu | shot | get | influenza | vaccine |
| | | HPV Vaccine | hpv | cancer | vaccine | cervical | cancers |
| | | International Childhood Vaccination | pakistan | polio | immunization | health | children |
| | | NaijaVax | side | effects | naijavax | vaccine | shot |
| | | Phillipines Vaccine Promotion | vaxcen | vaccination | hotline | com | today |
| | | Pregnancy & Vaccines | preg | women | baby | covid | vaccine |

| | | | | | | |
|---|---|---|---|---|---|---|
| | Pro-Vaccine Organizations | vaccine | health | institute | join | director |
| | Vaccine Clinics | covid | vaccine | clinic | nevada | flu |
| | Vaccine Equity | health | world | covid | vaccines | vaccine |
| Vaccine Campaign | Healthcare Workers | care | patients | covid | health | workers |
| | Personal Stories | get | people | know | like | one |
| | Stories from Nurses | one | nurse | time | day | vaxxed |
| Vaccine Safe and Effective | Adverse Event Rates | vaccine | covid | vaccines | adverse | safety |
| | Clinical Trials | vaccine | covid | vaccines | coronavirus | trials |
| | Research Results | study | vaccines | autism | evidence | research |
| | School Mandates | bill | state | school | vaccine | colorado |
| | Vaccine Efficacy | covid | variant | virus | new | sars |
| | Vaccines Work | vaccine | measles | vaccines | disease | polio |
| Other/Unknown | Australian News | australia | said | vaccination | new | australian |
| | Brazil Rotary Club | polio | event | plus | rotary | fund |
| | Canadian News | covid | news | cbc | canada | immunizeusa |
| | Click to Subscribe | vaccines | covid | vaccine | questions | information |
| | COVID Deaths | covid | cases | deaths | number | new |
| | COVID in Australia | test | covid | tested | nsw | testing |
| | COVID Symptoms | covid | symptoms | disease | heart | risk |
| | Facts & Hesitancy | vaccine | covid | vaccines | get | public |
| | Indian Polio Campaign | polio | plus | event | fund | club |
| | Job Recruitment | information | vaccination | help | online | school |
| | Masks, Handwashing, Distancing | mask | masks | wear | covid | face |
| | Pandemics | coronavirus | pandemic | covid | virus | disease |
| | Politics | trump | twitter | president | white | iphone |
| | Public Health Messaging | health | state | covid | department | county |
| | Religion & Morality | life | lives | freedom | people | love |
| | Thank you | photos | day | timeline | vaccine | happy |
| | UK NHS | nhs | dorset | photos | timeline | health |

**Table S2-2.** Percent changes in each topic.

| Position | Venue | Category | Label | % Change, Nov. 2020 Sample | $\chi^2$(dof) | p |
|---|---|---|---|---|---|---|
| Anti-Vaccine | Pages | Alternative Medicines | Natural Health | 90% | 65.93 (66) | 0.479 |
| | | Civil Liberties | Legislative Action | 130% | 51.25 (66) | 0.909 |
| | | | Informed Consent | 103% | 144.62 (66) | 0.000 |
| | | | Opposing School Mandates | 283% | 72.10 (66) | 0.283 |
| | | | Lockdowns | 58% | 25.21 (66) | 1.000 |
| | | | Sign Petitions | 110% | 62.50 (66) | 0.600 |
| | | | Health Freedom | 126% | 18.93 (66) | 1.000 |
| | | | Discussing Vaccine Mandates | 138% | 75.88 (66) | 0.190 |
| | | | Canada Rallies | 28% | 845.35 (66) | 0.000 |
| | | | Mandates Illegal/Immoral | 142% | 221.11 (66) | 0.000 |
| | | | Worldwide Protests | 95% | 77.56 (66) | 0.156 |
| | | | Anti-Mask | 202% | 3.20 (66) | 1.000 |
| | | Conspiracy Theories | Toxins | 174% | 314.84 (66) | 0.000 |
| | | | RSB Podcast | 110% | 9.67 (66) | 1.000 |
| | | | Medical Harm | 127% | 10.97 (66) | 1.000 |
| | | | Fauci Conspiracies | 146% | 5.48 (66) | 1.000 |
| | | | Question the Lies | 158% | 150.83 (66) | 0.000 |
| | | | General Conspiracies | 115% | 90.77 (66) | 0.023 |
| | | | Social Media Censorship | 111% | 33.39 (66) | 1.000 |
| | | | Big Pharma | 125% | 26.14 (66) | 1.000 |
| | | | COVID Not Deadly | 203% | 8.62 (66) | 1.000 |
| | | | Clinical Trial Conspiracies | 111% | 34.92 (66) | 0.999 |
| | | | Bill Gates | 45% | 57.36 (66) | 0.767 |
| | | | Unreliable Tests | 17% | 116.72 (66) | 0.000 |
| | | | Please Share Truth! | 108% | 88.74 (66) | 0.032 |
| | | | Measles | 44% | 116.51 (66) | 0.000 |
| | | | Rebel Doctors | 117% | 221.03 (66) | 0.000 |
| | | | QAnon Conspiracies | 70% | 212.94 (66) | 0.000 |
| | | | Liberty Beacon Podcast | 102% | 123.02 (66) | 0.000 |
| | | | Totalitarian Government | 63% | 30.24 (66) | 1.000 |
| | | Morality Issues | Religious Conspiracies | 30% | 395.61 (66) | 0.000 |
| | | | Foetal Tissues | 131% | 133.49 (66) | 0.000 |
| | | Other/Unknown | Community Building | 150% | 44.20 (66) | 0.982 |

| Groups | | | | | |
|---|---|---|---|---|---|
| | | Tagalog | 143% | 98.51 (66) | 0.006 |
| | | U.S. Politics | 25% | 532.31 (66) | 0.000 |
| | | Australian Politics | 27% | 220.29 (66) | 0.000 |
| | | Watch Episodes | 40% | 860.36 (66) | 0.000 |
| | | Los Angeles | 114% | 68.73 (66) | 0.385 |
| | | Business Impacts | 94% | 36.38 (66) | 0.999 |
| | | New Age | 135% | 74.18 (66) | 0.229 |
| | Safety Concerns | Side Effects | 58% | 445.47 (66) | 0.000 |
| | | Severe Side Effects | 194% | 43.38 (66) | 0.986 |
| | | Personal Stories | 99% | 21.90 (66) | 1.000 |
| | | Dead Children | 162% | 11.39 (66) | 1.000 |
| | | Autism | 219% | 540.90 (66) | 0.000 |
| | | Heavy Metals | 75% | 8.09 (66) | 1.000 |
| | | Misuse of Science | 94% | 20.99 (66) | 1.000 |
| | | Gardasil | 5270% | 35.47 (66) | 0.999 |
| | | Harm to Children | 157% | 22.60 (66) | 1.000 |
| | | Severe Reactions & Deaths | 315% | 479.58 (66) | 0.000 |
| | Alternative Medicines | Natural Health | 155% | 31.53 (66) | 1.000 |
| | Civil Liberties | Legislative Action | 260% | 131.98 (66) | 0.000 |
| | | Informed Consent | 78% | 138.05 (66) | 0.000 |
| | | Opposing School Mandates | 351% | 25.40 (66) | 1.000 |
| | | Lockdowns | 43% | 80.54 (66) | 0.108 |
| | | Sign Petitions | 155% | 129.79 (66) | 0.000 |
| | | Health Freedom | 135% | 10.00 (66) | 1.000 |
| | | Discussing Vaccine Mandates | 92% | 69.66 (66) | 0.355 |
| | | Canada Rallies | 77% | 40.31 (66) | 0.995 |
| | | Mandates Illegal/Immoral | 73% | 360.56 (66) | 0.000 |
| | | Worldwide Protests | 90% | 18.23 (66) | 1.000 |
| | | Anti-Mask | 341% | 9.50 (66) | 1.000 |
| | Conspiracy Theories | Toxins | 77% | 139.74 (66) | 0.000 |
| | | RSB Podcast | 66% | 28.46 (66) | 1.000 |
| | | Medical Harm | 77% | 25.48 (66) | 1.000 |
| | | Fauci Conspiracies | 92% | 3.47 (66) | 1.000 |
| | | Question the Lies | 121% | 58.65 (66) | 0.728 |
| | | General Conspiracies | 101% | 80.64 (66) | 0.106 |
| | | Social Media Censorship | 86% | 35.05 (66) | 0.999 |
| | | Big Pharma | 181% | 36.72 (66) | 0.999 |
| | | COVID Not Deadly | 72% | 52.58 (66) | 0.885 |
| | | Clinical Trial Conspiracies | 48% | 617.99 (66) | 0.000 |

| | | | | | | |
|---|---|---|---|---|---|---|
| | | | Bill Gates | 44% | 56.19 (66) | 0.800 |
| | | | Unreliable Tests | 23% | 340.10 (66) | 0.000 |
| | | | Please Share Truth! | 86% | 47.89 (66) | 0.954 |
| | | | Measles | 45% | 124.19 (66) | 0.000 |
| | | | Rebel Doctors | 127% | 121.90 (66) | 0.000 |
| | | | QAnon Conspiracies | 78% | 28.42 (66) | 1.000 |
| | | | Liberty Beacon Podcast | 103% | 197.69 (66) | 0.000 |
| | | | Totalitarian Government | 54% | 62.15 (66) | 0.612 |
| | | Morality Issues | Religious Conspiracies | 57% | 657.64 (66) | 0.000 |
| | | | Foetal Tissues | 149% | 124.55 (66) | 0.000 |
| | | Other/Unknown | Community Building | 116% | 111.73 (66) | 0.000 |
| | | | Tagalog | 84% | 6.55 (66) | 1.000 |
| | | | U.S. Politics | 13% | 609.94 (66) | 0.000 |
| | | | Australian Politics | 33% | 380.64 (66) | 0.000 |
| | | | Watch Episodes | 110% | 35.55 (66) | 0.999 |
| | | | Los Angeles | 104% | 4.76 (66) | 1.000 |
| | | | Business Impacts | 98% | 45.15 (66) | 0.977 |
| | | | New Age | 94% | 17.30 (66) | 1.000 |
| | | Safety Concerns | Side Effects | 74% | 157.62 (66) | 0.000 |
| | | | Severe Side Effects | 191% | 14.13 (66) | 1.000 |
| | | | Personal Stories | 119% | 60.50 (66) | 0.668 |
| | | | Dead Children | 1394% | 19.14 (66) | 1.000 |
| | | | Autism | 190% | 76.84 (66) | 0.170 |
| | | | Heavy Metals | 324% | 10.34 (66) | 1.000 |
| | | | Misuse of Science | 88% | 52.55 (66) | 0.885 |
| | | | Gardasil | 568% | 84.99 (66) | 0.058 |
| | | | Harm to Children | 999% | 16.38 (66) | 1.000 |
| | | | Severe Reactions & Deaths | 226% | 470.96 (66) | 0.000 |
| Pro-Vaccine | Pages | Anti Anti-Vaccine | Measles Outbreaks | 4028% | 107.02 (66) | 0.001 |
| | | | Debunking | 89% | 112.20 (66) | 0.000 |
| | | | Anti-Vax Myths | 115% | 117.79 (66) | 0.000 |
| | | | Refuting Anti-Vax | 308% | 92.96 (66) | 0.016 |
| | | | Mocking Anti-vax | 106% | 77.77 (66) | 0.152 |
| | | | Mocking Natural Health | 119% | 44.40 (66) | 0.981 |
| | | | Viruses Kill | 170% | 66.26 (66) | 0.468 |
| | | Other/Unknown | Thank you | 117% | 64.37 (66) | 0.534 |
| | | | Indian Polio Campaign | 78% | 153.79 (66) | 0.000 |
| | | | Facts & Hesitancy | 93% | 38.80 (66) | 0.997 |
| | | | Click to Subscribe | 200% | 75.32 (66) | 0.202 |
| | | | Pandemics | 53% | 30.89 (66) | 1.000 |

| Category | Subcategory | % | Stat (df) | p |
|---|---|---|---|---|
| | Politics | 22% | 1280.22 (66) | 0.000 |
| | COVID in Australia | 152% | 176.01 (66) | 0.000 |
| | COVID Deaths | 95% | 44.20 (66) | 0.982 |
| | UK NHS | 88% | 13.83 (66) | 1.000 |
| | Canadian News | 81% | 198.58 (66) | 0.000 |
| | Religion & Morality | 90% | 27.24 (66) | 1.000 |
| | Public Health Messaging | 85% | 9.49 (66) | 1.000 |
| | Masks, Handwashing, Distancing | 53% | 6.54 (66) | 1.000 |
| | Australian News | 143% | 79.04 (66) | 0.130 |
| | Job Recruitment | 108% | 55.34 (66) | 0.822 |
| | Brazil Rotary Club | 102% | 73.69 (66) | 0.241 |
| | COVID Symptoms | 52% | 684.01 (66) | 0.000 |
| Pro Science & Biomedicine | Immunity | 227% | 870.74 (66) | 0.000 |
| | MRNA Vaccine | 129% | 171.31 (66) | 0.000 |
| | Science Promotion | 87% | 48.17 (66) | 0.952 |
| Pro Vaccine Policy | Vaccine Mandates | 242% | 93.92 (66) | 0.014 |
| | Vaccine Regulation | 83% | 17.44 (66) | 1.000 |
| Promotion | Childhood Vaccines | 133% | 11.16 (66) | 1.000 |
| | Vaccine Clinics | 469% | 487.24 (66) | 0.000 |
| | Pregnancy & Vaccines | 177% | 92.51 (66) | 0.017 |
| | COVID Vaccine & Booster Promotion | 307% | 2450.16 (66) | 0.000 |
| | Pro-Vaccine Organizations | 133% | 209.52 (66) | 0.000 |
| | Vaccine Equity | 50% | 189.61 (66) | 0.000 |
| | International Childhood Vaccination | 682% | 29.15 (66) | 1.000 |
| | Flu Shot Promotion | 416% | 4.05 (66) | 1.000 |
| | Phillipines Vaccine Promotion | 20% | 122.85 (66) | 0.000 |
| | Canadian Promotion | 125% | 46.36 (66) | 0.968 |
| | NaijaVax | 50% | 168.45 (66) | 0.000 |
| | HPV Vaccine | 184% | 9.12 (66) | 1.000 |
| Vaccine Campaign | Personal Stories | 88% | 147.02 (66) | 0.000 |
| | Healthcare Workers | 19% | 13.25 (66) | 1.000 |
| | Stories from Nurses | 94% | 75.63 (66) | 0.195 |
| Vaccine Safe and Effective | Adverse Event Rates | 251% | 156.14 (66) | 0.000 |
| | Vaccines Work | 83% | 69.32 (66) | 0.366 |
| | Vaccine Efficacy | 88% | 40.61 (66) | 0.994 |
| | School Mandates | 165% | 6.16 (66) | 1.000 |
| | Clinical Trials | 27% | 266.36 (66) | 0.000 |
| | Research Results | 169% | 46.39 (66) | 0.968 |

| Groups | | | | | |
|---|---|---|---|---|---|
| | Anti Anti-Vaccine | Measles Outbreaks | 78% | 27.81 (65) | 1.000 |
| | | Debunking | 210% | 59.43 (65) | 0.672 |
| | | Anti-Vax Myths | 82% | 48.75 (65) | 0.934 |
| | | Refuting Anti-Vax | 417% | 211.60 (65) | 0.000 |
| | | Mocking Anti-vax | 117% | 11.76 (65) | 1.000 |
| | | Mocking Natural Health | 107% | 99.23 (65) | 0.004 |
| | | Viruses Kill | 229% | 31.66 (65) | 1.000 |
| | Other/Unknown | Thank you | 89% | 45.68 (65) | 0.967 |
| | | Indian Polio Campaign | 36% | 74.70 (65) | 0.192 |
| | | Facts & Hesitancy | 163% | 41.91 (65) | 0.988 |
| | | Click to Subscribe | 184% | 88.19 (65) | 0.029 |
| | | Pandemics | 153% | 28.94 (65) | 1.000 |
| | | Politics | 101% | 34.53 (65) | 0.999 |
| | | COVID in Australia | 171% | 62.21 (65) | 0.575 |
| | | COVID Deaths | 901% | 1017.62 (65) | 0.000 |
| | | UK NHS | 253% | 39.31 (65) | 0.995 |
| | | Canadian News | 119% | 62.38 (65) | 0.569 |
| | | Religion & Morality | 263% | 114.03 (65) | 0.000 |
| | | Public Health Messaging | 115% | 71.17 (65) | 0.280 |
| | | Masks, Handwashing, Distancing | 60% | 25.40 (65) | 1.000 |
| | | Australian News | 208% | 131.80 (65) | 0.000 |
| | | Job Recruitment | 88% | 81.17 (65) | 0.085 |
| | | Brazil Rotary Club | 12% | 68.23 (65) | 0.368 |
| | | COVID Symptoms | 245% | 152.13 (65) | 0.000 |
| | Pro Science & Biomedicine | Immunity | 548% | 439.51 (65) | 0.000 |
| | | MRNA Vaccine | 90% | 85.87 (65) | 0.043 |
| | | Science Promotion | 209% | 91.98 (65) | 0.015 |
| | Pro Vaccine Policy | Vaccine Mandates | 149% | 77.62 (65) | 0.136 |
| | | Vaccine Regulation | 427% | 336.62 (65) | 0.000 |
| | Promotion | Childhood Vaccines | 236% | 72.02 (65) | 0.257 |
| | | Vaccine Clinics | 140% | 85.04 (65) | 0.048 |
| | | Pregnancy & Vaccines | 649% | 409.52 (65) | 0.000 |
| | | COVID Vaccine & Booster Promotion | 760% | 1527.78 (65) | 0.000 |
| | | Pro-Vaccine Organizations | 178% | 12.40 (65) | 1.000 |
| | | Vaccine Equity | 249% | 38.55 (65) | 0.996 |
| | | International Childhood Vaccination | 653% | 136.00 (65) | 0.000 |
| | | Flu Shot Promotion | 70% | 135.80 (65) | 0.000 |
| | | Phillipines Vaccine Promotion | 67% | 39.90 (65) | 0.994 |

|  |  |  |  |  |
|---|---|---|---|---|
|  | Canadian Promotion | 191% | 45.51 (65) | 0.968 |
|  | NaijaVax | 883% | 355.57 (65) | 0.000 |
|  | HPV Vaccine | 95% | 76.35 (65) | 0.159 |
| Vaccine Campaign | Personal Stories | 237% | 129.40 (65) | 0.000 |
|  | Healthcare Workers | 278% | 276.01 (65) | 0.000 |
|  | Stories from Nurses | 199% | 117.00 (65) | 0.000 |
| Vaccine Safe and Effective | Adverse Event Rates | 207% | 89.10 (65) | 0.025 |
|  | Vaccines Work | 57% | 110.42 (65) | 0.000 |
|  | Vaccine Efficacy | 334% | 145.17 (65) | 0.000 |
|  | School Mandates | 1861% | 177.66 (65) | 0.000 |
|  | Clinical Trials | 48% | 60.44 (65) | 0.637 |
|  | Research Results | 207% | 10.36 (65) | 1.000 |

## 3. Sensitivity Analysis

We examined whether our findings were sensitive to our selection of policy date by conducting several additional interrupted time series analyses. The removal of Stop Mandatory Vaccination was notable because it was the first removal of a major anti-vaccine page, but it was also an outgrowth of Facebook's policy on misinformation associated with the QAnon conspiracy theory.

*October 6, 2020*

We therefore examined whether our results replicated when the policy date was set to October 6, 2020 – 6 weeks before November 15, 2020 – when Facebook banned content associated with the QAnon conspiracy theory. We found that our results largely replicated for anti-vaccine content, and for pro-vaccine pages.

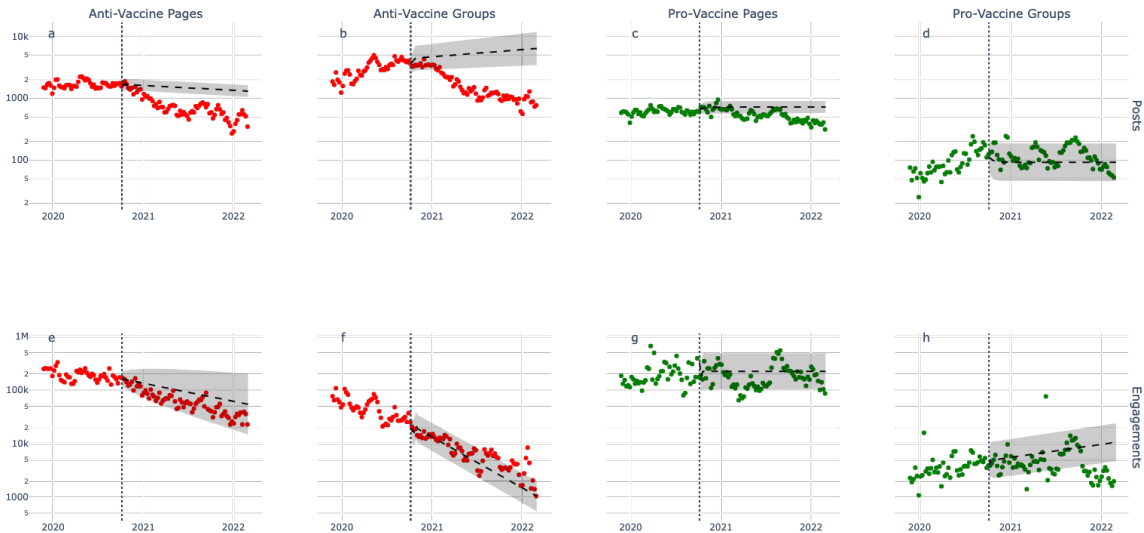

**Fig S3-1. Sensitivity analysis examining effects of policy date at October 6, 2020**

ARIMA model results show that, compared to projections (black dashed line), Facebook's policies led to a decline in posts in **a.** anti-vaccine pages (49% average decrease, $\chi^2(72)= 2760.30\ p<0.001$) **b.** anti-vaccine groups (64% average decrease, $\chi^2(72)= 1152.01, p<0.001$), and **c.** pro-vaccine pages (24% average decrease, $\chi^2(72)= 566.35, p<0.001$). **d.** We did not detect a significant difference in pro-vaccine groups ($\chi^2(71)= 72.53, p=0.43$). Unlike when the policy date was set to November 15, 2020, we did not detect a change in engagements in **e.** anti-vaccine pages ($\chi^2(72)= 62.09, p=0.79$). Engagements increased significantly beyond pre-policy predictions in **f.** anti-vaccine groups (44% average increase, $\chi^2(72)= 123.21, p<0.001$). Engagements declined with content in **g.** pro-vaccine pages (12% average decrease, $\chi^2(72)= 96.08, p=0.03$) and **h.** in pro-vaccine groups (17% average decrease, $\chi^2(71)= 230.62, p<0.001$). Error bars reflect 90% confidence intervals.

*August 19, 2020*

We next examined whether our results replicated when the policy date was set to August 19, 2020 – 13 weeks before November 15, 2020, and the day Facebook stated that they would remove explicit calls to violence associated with QAnon. We again found that our results largely replicated for anti-vaccine content (although changes to page post counts were not significant due to large error bars), and for pro-vaccine pages.

**Fig S3-2. Sensitivity analysis examining effects of policy date at August 19, 2020**

ARIMA model results show that, compared to projections (black dashed line), we did not detect a significant change in posts in **a.** anti-vaccine pages ($\chi^2(79)= 10.54, p=1.00$), **b.** although we did detect a significant decrease in anti-vaccine group posts (62% average decrease, $\chi^2(79)= 2715.71, p<0.001$), and **c.** pro-vaccine pages (27% average decrease, $\chi^2(79)= 841.03, p<0.001$). **d.** We did not detect a significant difference in pro-vaccine groups ($\chi^2(78)= 576.27, p<0.001$). Engagements in **e.** anti-vaccine pages decreased significantly, although they repeatedly returned to pre-policy levels (average decrease of 44%, $\chi^2(79)= 140.38, p<0.001$). **f.** In contrast, engagements in groups significantly exceeded pre-policy predictions (64% average increase, $\chi^2(79)= 216.53, p<0.001$). Engagements declined with content in **g.** pro-vaccine pages (14% average decrease, $\chi^2(79)= 111.01, p=0.01$), **h.** and in pro-vaccine groups (21% average decrease, $\chi^2(78)= 300.34, p<0.001$). Error bars reflect 90% confidence intervals.

*December 27, 2020*

Having established that our primary results replicated for 6 and 13 weeks before the policy, we next examined whether our results replicated when the policy date was set to the last week of 2020: December 27, 2020 – 6 weeks after November 15, 2020, and 25 days after Facebook banned COVID misinformation. We found that our results qualitatively replicated for anti-vaccine all content except anti-vaccine group engagements, where the ARIMA model predicted a recovery that did not occur.

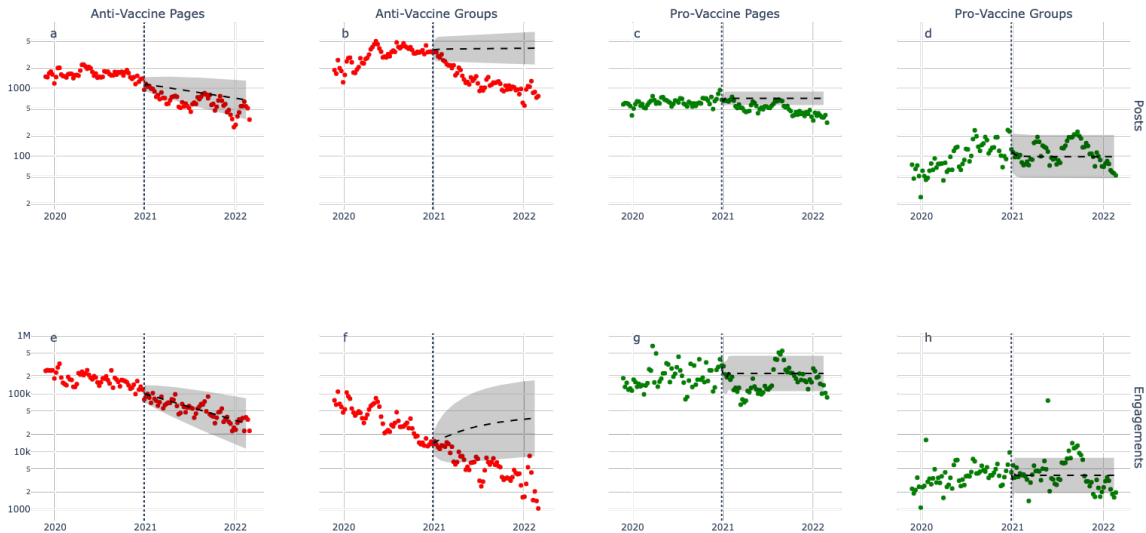

**Fig S3-3. Sensitivity analysis examining effects of policy date at December 27, 2020**

ARIMA model results show that, compared to projections (black dashed line), Facebook's policies led to a decline in posts in **a.** anti-vaccine pages (28% average decrease, $\chi^2(60)= 127.33$ $p<0.001$) **b.** anti-vaccine groups (62% average decrease, $\chi^2(60)= 861.06, p<0.001$), and **c.** pro-vaccine pages (28% average decrease, $\chi^2(60)= 509.32, p<0.001$). **d.** We did not detect a significant difference in pro-vaccine groups ($\chi^2(59)= 45.17, p=0.91$). Unlike when the policy date was set to November 15, 2020, we did not detect a change in engagements in **e.** anti-vaccine pages ($\chi^2(60)= 17.78, p=1.00$). **f.** In contrast, our model's predicted a recovery in engagements in anti-vaccine groups where, in practice, none occurred, meaning that engagements decreased significantly beyond pre-policy predictions (72% average decrease, $\chi^2(60)= 265.08, p<0.001$). Engagements declined with content in **g.** pro-vaccine pages (15% average decrease, $\chi^2(60)= 106.84, p<0.001$), **h.** but increased with content in pro-vaccine groups (51% average increase, $\chi^2(59)= 149.08, p<0.001$). Error bars reflect 90% confidence intervals.

*February 8, 2021*

Finally, we examined whether our results replicated when the policy date was set to February 8, 2021 – 12 weeks after November 15, 2020, and the day Facebook stated that they would remove explicit misinformation about all vaccine. As when we set the policy date to December 27, 2020, we found that our results qualitatively replicated for anti-vaccine all content except anti-vaccine group engagements, where the ARIMA model predicted a recovery that did not occur.

**Fig S3-4. Sensitivity analysis examining effects of policy date at February 8, 2021**

ARIMA model results show that, compared to projections (black dashed line), **a.** we did not detect a significant decrease in posts in anti-vaccine pages (16% average decrease, $\chi^2(54)= 70.28, p=0.07$) **b.** although we did observe a significant decrease in posts in anti-vaccine groups (58% average decrease, $\chi^2(54)= 481.54, p<0.001$), and **c.** pro-vaccine pages (31% average decrease, $\chi^2(54)= 545.04, p<0.001$). **d.** We did not detect a significant difference in pro-vaccine groups ($\chi^2(53)= 68.26, p=0.08$). Unlike when the policy date was set to November 15, 2020, we did not detect a change in engagements in **e.** anti-vaccine pages ($\chi^2(54)= 18.15, p=1.00$). **f.** In contrast, our model's predicted a recovery in engagements in anti-vaccine groups where, in practice, none occurred, meaning that engagements decreased significantly beyond pre-policy predictions (71% average decrease, $\chi^2(54)= 211.08, p<0.001$). Engagements declined with content in **g.** pro-vaccine pages (17% average decrease, $\chi^2(54)= 117.65, p<0.001$), **h.** but increased with content in pro-vaccine groups (53% average increase, $\chi^2(53)= 159.47, p<0.001$). Error bars reflect 90% confidence intervals.

Thus, our sensitivity analysis demonstrates that our primary results replicate qualitatively across a span of 25 weeks with a median date of roughly November 18, 2020 – when Facebook first applied its "hard" content remedies to vaccines, and bookended by changes in Facebook's "hard" remedies on either side. Although some findings – e.g., decreases in page posts and engagements – are no longer statistically significant for the selection of some policy dates, the overall qualitative conclusions remain the same: that Facebook's remedies removed posts in anti-vaccine pages and/or groups, but were unable to stop page engagements from returning to baseline predictions. In addition, pro-vaccine page post volumes declined compared to pre-policy predictions for all policy dates. In addition, engagements with posts in anti-vaccine groups appear to have exceeded pre-policy projections for all policy dates prior to and including November 18, 2020, and appear to at least have returned to within the 90% error bars by January, 2022, for all policy dates. Finally, posts in the second sample appear to have returned to pre-policy levels (Fig 1f) as predicted by models with policy dates in December 2020 and February 2021.

*Analysis Using Raw Count Data*

When conducting our ARIMA analyses, we controlled for the formation of new pages in pre-policy data by dividing weekly post and engagement counts by the total number of active venues. We examined whether our results replicated when we used raw counts without this added control. We found that our results did qualitatively replicated. Post-policy projections for posts in Facebook groups grew exponentially beyond what the data indicated because they are fit to a retrospective dataset that includes new groups appearing between November 15, 2019 and November 15, 2020. Although new groups formed after November 15, 2020, they are, by definition, not included in the dataset identified on this date (although they are included in the dataset identified on July 21, 2021).

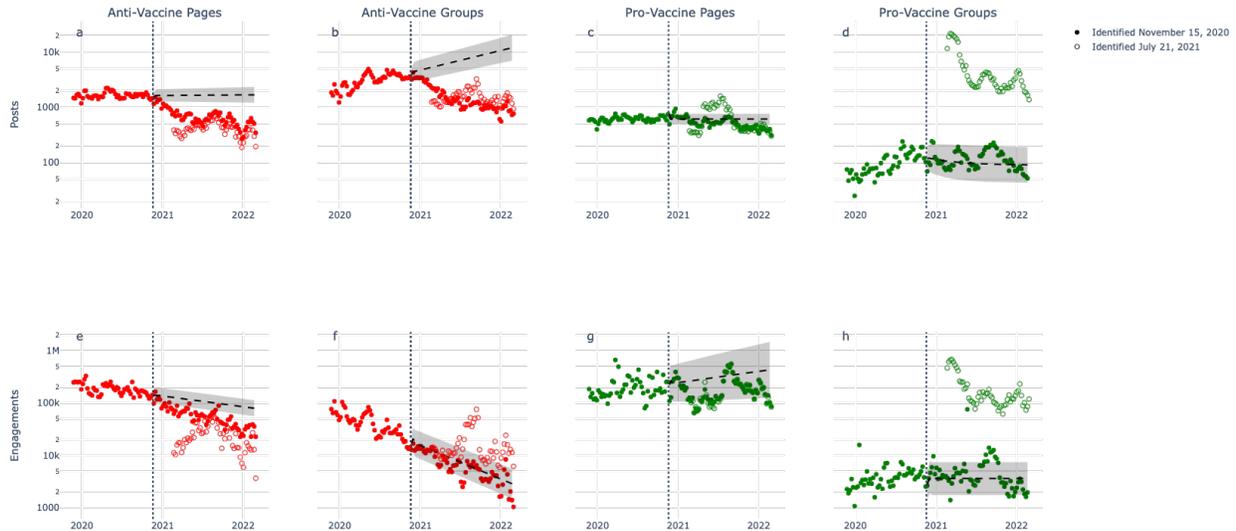

**Fig S3-5. Sensitivity analysis examining effects of fitting ARIMA models to raw count data**

ARIMA model results show that, compared to projections (black dashed line), Facebook's policies led to a decline in posts in **a.** anti-vaccine pages (58% average decrease, $\chi^2(66)= 1918.82\ p<0.001$) **b.** anti-vaccine groups (74% average decrease, $\chi^2(66)= 2301.31, p<0.001$), and **c.** pro-vaccine pages (14% average decline, $\chi^2(66)= 310.06, p<0.001$). We did not detect a significant difference in **d.** pro-vaccine groups ($\chi^2(66)= 55.94, p=0.81$). Additionally, engagements declined with content in **e.** anti-vaccine pages (46% average decrease, $\chi^2(66)= 789.70, p<0.001$). Despite this average decline, engagements fell within the prediction's 90% confidence intervals for the months of November 2020-January 2021, April-May 2021, August – September, 2021, and November 2021. **f.** In contrast, we did not detect a significant change in engagements in anti-vaccine groups ($\chi^2(66)= 73.93, p=0.86$). Engagements declined with content in **g.** pro-vaccine pages (38% average decrease, $\chi^2(66)= 108.68, p<0.001$) but **h.**

increased for the months of August – October 2021 in pro-vaccine groups (59% average increase in the first sample, $\chi^2(66)= 156.61, p<0.001$). Error bars reflect 90% confidence intervals. Data from the second sample is shown for comparison and qualitatively replicates findings from the first sample.

# 4. Detecting Coordinated Link Sharing

In order to distinguish between "near-simultaneous" and organic link sharing, we fit exponential mixture models to interarrival time data derived from three blank searches conducted on CrowdTangle between March 30-March 31, 2021. Using the exp-mixture-model[10] Python package, we found that several model goodness-of-fit measures converged on a two-component distribution for each dataset (Table S4-1).

Table S4-1. Model goodness-of-fit statistics for exponential mixture models fit to three blank searches.

| $k_i$ | $k_f$ | MLL | JLL | AIC | BIC | $AIC_{LVC}$ | $BIC_{LVC}$ | DNML |
|---|---|---|---|---|---|---|---|---|
| | | | Blank Search 1, 3/30/2021, 9:45 PM EDT (pages) and 9:50 PM EDT (groups) | | | | | |
| 1 | 1 | -436,534.30 | -436,534.30 | 436,535.30 | 436,539.95 | 436,535.30 | 436,539.95 | 436,546.98 |
| 2 | 2 | -388,698.90 | **-395,583.58*** | 388,701.90 | 388,715.84 | **395,586.58*** | **395,599.75*** | **395,608.20*** |
| 3 | 3 | -386,277.58 | -406,235.57 | 386,282.58 | 386,305.82 | 406,240.57 | 406,262.00 | 406,270.87 |
| 4 | 4 | -386,227.76 | -417,830.37 | **386,234.76*** | **386,267.30*** | 417,837.37 | 417,866.70 | 417,876.73 |
| 5 | 4 | -386,227.72 | -417,776.17 | 386,236.72 | 386,278.56 | 417,783.17 | 417,812.55 | 417,822.57 |
| 6 | 4 | -386,229.27 | -416,726.67 | 386,240.27 | 386,291.40 | 416,733.67 | 416,762.91 | 416,772.00 |
| 7 | 5 | -386,228.78 | -412,436.74 | 386,241.78 | 386,302.21 | 412,445.74 | 412,480.35 | 412,489.51 |
| 8 | 4 | -386,227.79 | -416,911.45 | 386,242.79 | 386,312.52 | 416,918.45 | 416,947.78 | 416,957.80 |
| 9 | 4 | -386,228.78 | -414,334.79 | 386,245.78 | 386,324.80 | 414,341.79 | 414,370.91 | 414,380.93 |
| 10 | 4 | **-386,227.71*** | -418,397.53 | 386,246.71 | 386,335.03 | 418,404.53 | 418,433.98 | 418,443.92 |
| | | | Blank Search 2, 3/30/2021, 11:25 PM EDT (pages) and 11:18 PM EDT (groups) | | | | | |
| 1 | 1 | -415,674.53 | -415,674.53 | 415,675.53 | 415,680.16 | 415,675.53 | 415,680.16 | 415,687.19 |
| 2 | 2 | -371,090.88 | **-377,920.43*** | 371,093.88 | 371,107.77 | **377,923.43*** | **377,936.54*** | **377,944.99*** |
| 3 | 3 | -368,675.23 | -387,924.11 | 368,680.23 | 368,703.39 | 387,929.11 | 387,950.43 | 387,959.31 |
| 4 | 3 | -368,675.23 | -387,109.26 | 368,682.23 | 368,714.65 | 387,114.26 | 387,135.55 | 387,145.17 |
| 5 | 4 | -368,641.63 | -394,604.64 | **368,650.63*** | **368,692.30*** | 394,611.64 | 394,640.21 | 394,649.30 |
| 6 | 4 | -368,641.59 | -399,063.16 | 368,652.59 | 368,703.53 | 399,070.16 | 399,099.31 | 399,108.40 |
| 7 | 3 | -368,641.21 | -386,256.08 | 368,654.21 | 368,714.41 | 386,261.08 | 386,282.31 | 386,291.93 |
| 8 | 3 | -368,641.74 | -386,353.66 | 368,656.74 | 368,726.19 | 386,358.66 | 386,379.91 | 386,389.53 |
| 9 | 4 | **-368,640.73*** | -401,787.23 | 368,657.73 | 368,736.45 | 401,794.23 | 401,823.60 | 401,833.62 |
| 10 | 4 | -368,643.01 | -401,292.54 | 368,662.01 | 368,749.99 | 401,299.54 | 401,328.91 | 401,338.93 |
| | | | Blank Search 2, 3/31/2021, 9:00 AM EDT (pages) and 8:57 AM EDT (groups) | | | | | |
| 1 | 1 | -438,915.14 | -438,915.14 | 438,916.14 | 438,920.81 | 438,916.14 | 438,920.81 | 438,927.85 |
| 2 | 2 | -396,301.16 | **-404,433.46*** | 396,304.16 | 396,318.18 | **404,436.46*** | **404,449.70*** | **404,458.15*** |
| 3 | 3 | -393,948.63 | -415,792.96 | 393,953.63 | 393,976.98 | 415,797.96 | 415,819.49 | 415,828.36 |
| 4 | 3 | -393,932.26 | -415,440.74 | 393,939.26 | **393,971.96*** | 415,445.74 | 415,467.26 | 415,476.13 |

| | | | | | | | |
|---|---|---|---|---|---|---|---|
| 5 | 4 | -393,921.77 | -428,912.47 | **393,930.77*** | 393,972.81 | 428,919.47 | 428,949.01 | 428,959.03 |
| 6 | 4 | -393,922.01 | -426,076.72 | 393,933.01 | 393,984.40 | 426,083.72 | 426,113.07 | 426,122.16 |
| 7 | 4 | **-393,921.69*** | -419,130.49 | 393,934.69 | 393,995.42 | 419,137.49 | 419,165.60 | 419,174.69 |
| 8 | 4 | -393,921.75 | -425,580.56 | 393,936.75 | 394,006.81 | 425,587.56 | 425,616.86 | 425,625.95 |
| 9 | 4 | -393,924.26 | -425,778.72 | 393,941.26 | 394,020.67 | 425,785.72 | 425,815.15 | 425,825.17 |
| 10 | 4 | -393,922.80 | -427,351.92 | 393,941.80 | 394,030.55 | 427,358.92 | 427,388.38 | 427,397.47 |

*Note.* $k_i$ = Initial number of components; $k_f$ = Final number of components; MLL = Mean log-likelihood; JLL = Joint log-likelihood; AIC = Akaike Information Criterion;[11] BIC = Bayesian Information Criterion;[12] $AIC_{LVC}$ = Akaike Information Criterion with Latent Variable Completion;[10] $BIC_{LVC}$ = Bayesian Information Criterion with Latent Variable Completion;[10] DNML = Decomposed Normalized Maximum Likelihood.[10] * = best model fit.

**Table S4-2**. Results of exponential mixture model-fitting procedure, generating two components for each of three datasets.

| | 1 | | 2 | |
|---|---|---|---|---|
| | μ | π | μ | π |
| 1 | 10.51 | 0.70 | 253.37 | 0.30 |
| 2 | 9.92 | 0.70 | 232.55 | 0.30 |
| 3 | 9.42 | 0.69 | 196.43 | 0.31 |
| Average | 9.95 | 0.70 | 227.45 | 0.30 |

*Note.* μ = mean; π = probability weight. When estimating our threshold, we did not use probability weights since these are unique to each dataset (i.e., different datasets will have different proportions of near-simultaneous and organic link-sharing, and near-simultaneous link sharing may be over-represented in blank search data due to right censoring resulting from CrowdTangle's download limit).

We next examined the goodness-of-fit of these estimated distributions by comparing them each one to its respective dataset (Figure S4-1), and using Kolmogorov-Smirnov tests. Although each test detected a significant difference ($d_1 = 0.12$, $d_2 = 0.13$, $d_3 = 0.13$, $p<0.001$ in all cases), our large dataset sizes ($n_1 = 87,000$; $n_2 = 85,346$, $n_3 = 93,075$) means that we are powered to detect very small differences. Furthermore, Figure S1 shows that our model underestimates that likelihood of "near-simultaneous" sharing for very small time differences (between 0 and 15 seconds), meaning that our estimated threshold is conservative.

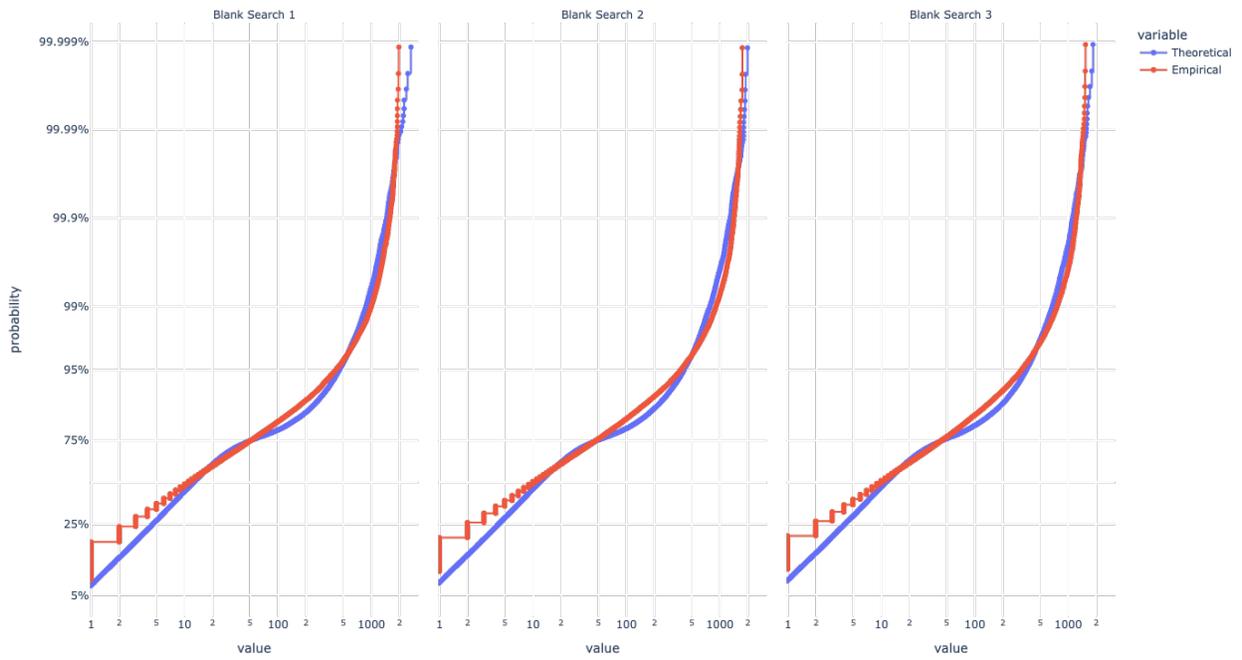

**Fig S4-1. Comparison of empirical and theoretical interarrival time cumulative distributions for each of three CrowdTangle blank searches.**

Probabilities are displayed on a logit axis in order to display differences in the distributions' tails.

## 5. Interactions Over Time

We examined how the relative proportions of different types of interactions (shares, comments, reactions) changed over time.

*"Non-Toxic" Reactions*

**Likes.** We found that the proportion of likes increased over time for anti-vaccine, but not pro-vaccine pages, and groups.

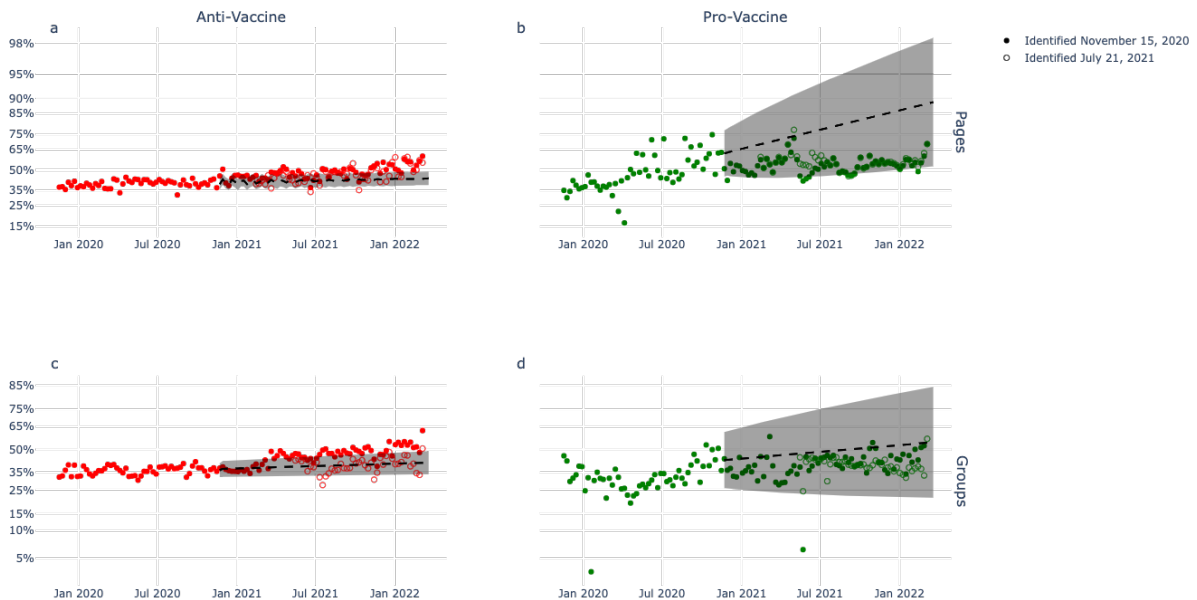

**Fig S5-1. Changes in the proportion of "Likes" over time compared to pre-policy projections**

ARIMA model results show that, compared to projections (black dashed line), Facebook's policies led to an increase in "Likes" **a.** anti-vaccine pages (14% average increase, $\chi^2(66)= 367.71, p<0.001$) **b.** anti-vaccine groups (16% average increase, $\chi^2(66)= 241.03, p<0.001$). In contrast, the proportion of "Likes" decreased significantly in **c.** pro-vaccine pages (30% average decrease, $\chi^2(66)= 118.86, p<0.001$). We did not detect a significant difference in **d.** pro-vaccine groups ($\chi^2(66)= 40.15, p=1.00$).

**Love.** We found that the proportion of "Love" reactions increased over time for anti-vaccine content and for pro-vaccine pages, but not pro-vaccine groups.

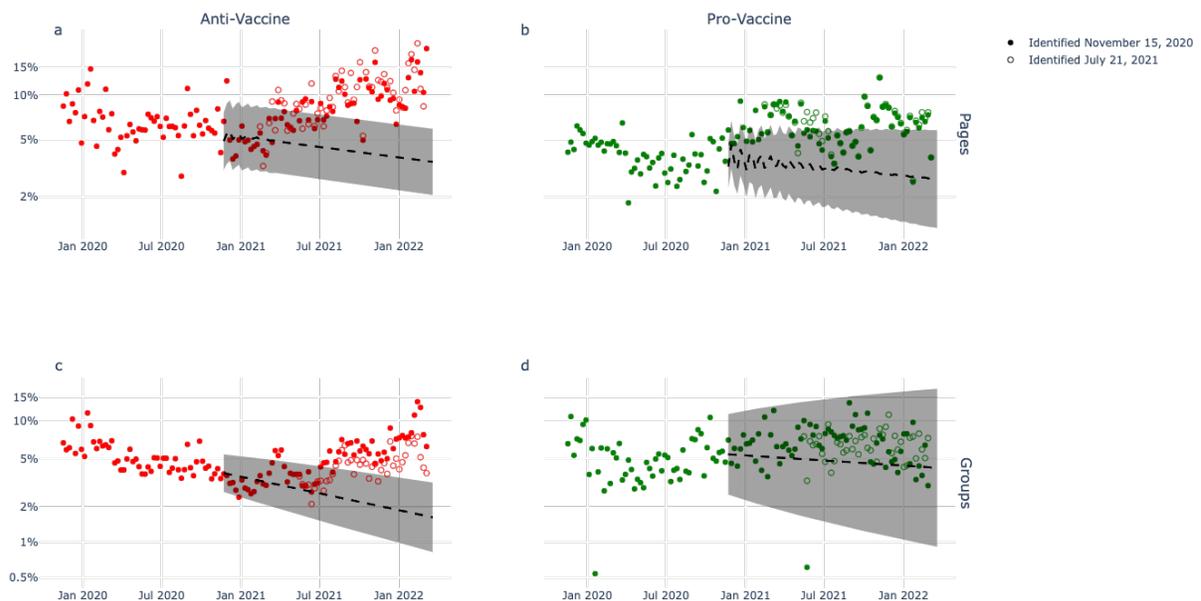

**Fig S5-2. Changes in the proportion of "Love" reactions over time compared to pre-policy projections**

ARIMA model results show that, compared to projections (black dashed line), Facebook's policies led to an increase in "Love" reactions **a.** anti-vaccine pages (96% average increase, $\chi^2(66)= 382.98, p<0.001$) **b.** anti-vaccine groups (126% average increase, $\chi^2(66)= 397.37, p<0.001$). In addition, the proportion of "Love" reactions increased significantly in **c.** pro-vaccine pages (100% average increase, $\chi^2(66)= 258.57, p<0.001$). We did not detect a significant difference in **d.** pro-vaccine groups ($\chi^2(66)= 37.74, p=1.00$).

**Sad.** We found that the proportion of "Sad" reactions increased over time for anti-vaccine content and for pro-vaccine groups, but not pro-vaccine pages.

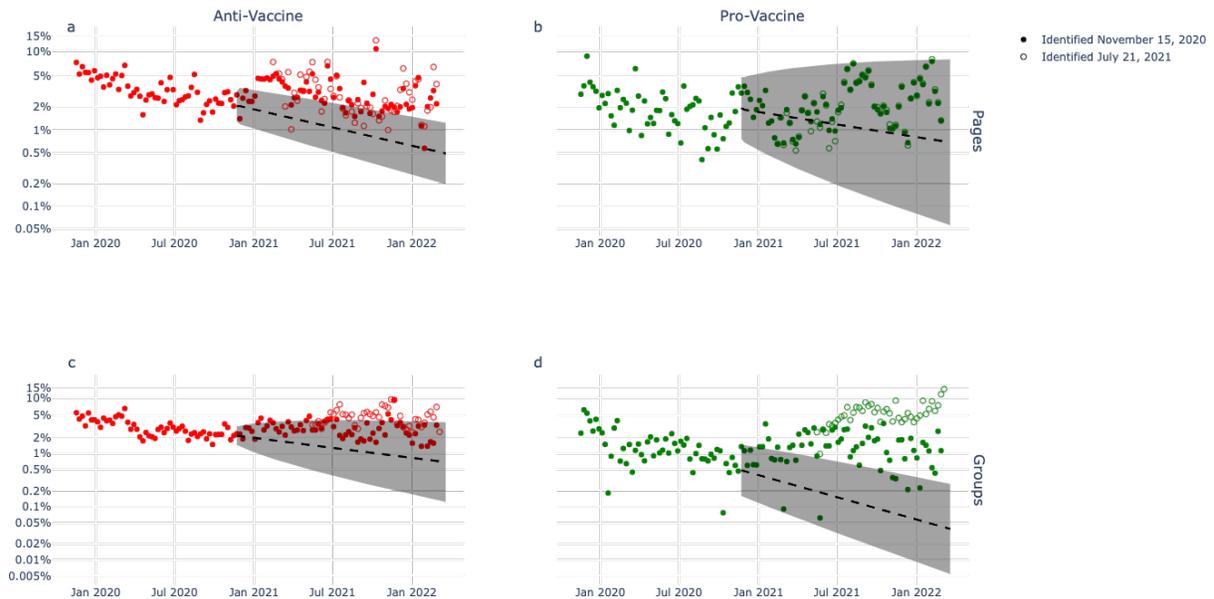

**Fig S5-3. Changes in the proportion of "Sad" reactions over time compared to pre-policy projections**

ARIMA model results show that, compared to projections (black dashed line), Facebook's policies led to an increase in "Sad" reactions **a.** anti-vaccine pages (195% average increase, $\chi^2(66)= 371.09, p<0.001$) **b.** anti-vaccine groups (146% average increase, $\chi^2(66)= 101.82, p<0.001$). In contrast, the proportion of "Sad" reactions did not change significantly in **c.** pro-vaccine pages ($\chi^2(66)= 42.37, p=0.99$), but did increase significantly in **d.** pro-vaccine groups (1102% average increase, $\chi^2(66)= 367.23, p<0.001$).

**Care.** We found that the proportion of "Care" reactions increased over time for anti-vaccine pages.

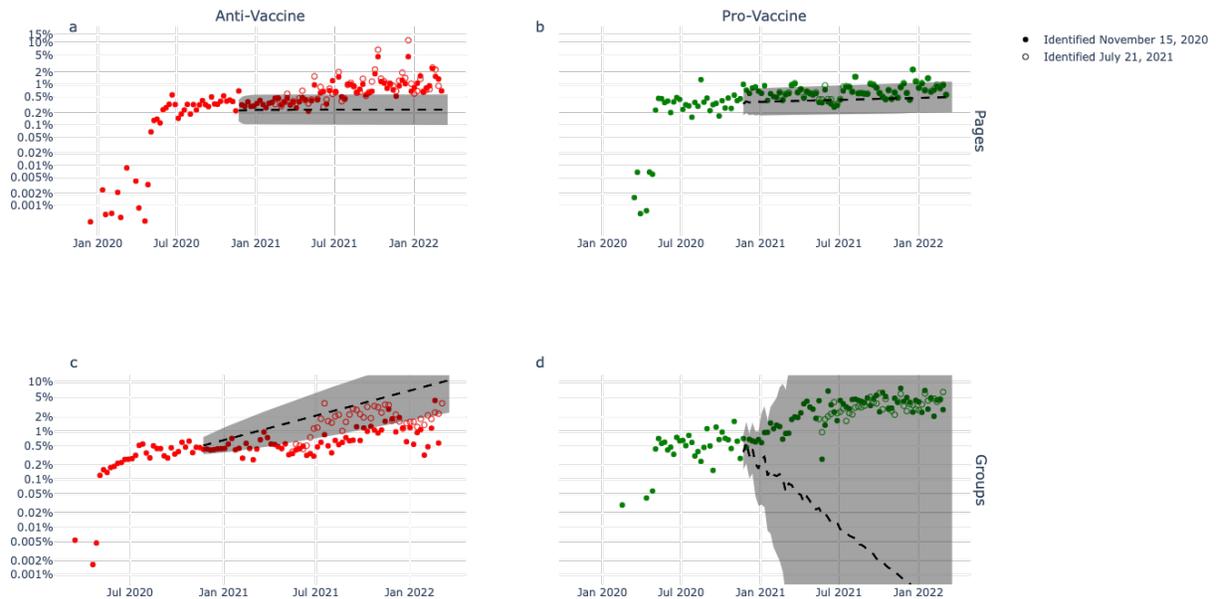

**Fig S5-4. Changes in the proportion of "Care" reactions over time compared to pre-policy projections**

ARIMA model results show that, compared to projections (black dashed line), Facebook's policies led to an increase in "Care" reactions **a.** anti-vaccine pages (253% average increase, $\chi^2(66)= 343.74, p<0.001$) **b.** but a significant decrease in anti-vaccine groups (74% average decrease, $\chi^2(66)= 261.19, p<0.001$). In addition, the proportion of "Care" reactions increased significantly in **c.** pro-vaccine pages (68% average increase, $\chi^2(66)= 90.61, p=0.02$), but did change significantly in **d.** pro-vaccine groups ($\chi^2(66)= 35.26, p=1.00$). All ARIMA models for the "Care" reaction were fit to data from May 3, 2020, forward since this reaction was not widely used before this date.

Taken together, these results suggest an overall increase in the proportion of "non-toxic" interactions in anti-vaccine, but not pro-vaccine venues (see Extended Data Fig. 4).

*"Toxic" Reactions*

**Angry.** We found that the proportion of "Angry" reactions decreased over time for anti-vaccine, but not pro-vaccine pages. "Angry" reactions decreased for anti-vaccine and pro-vaccine groups.

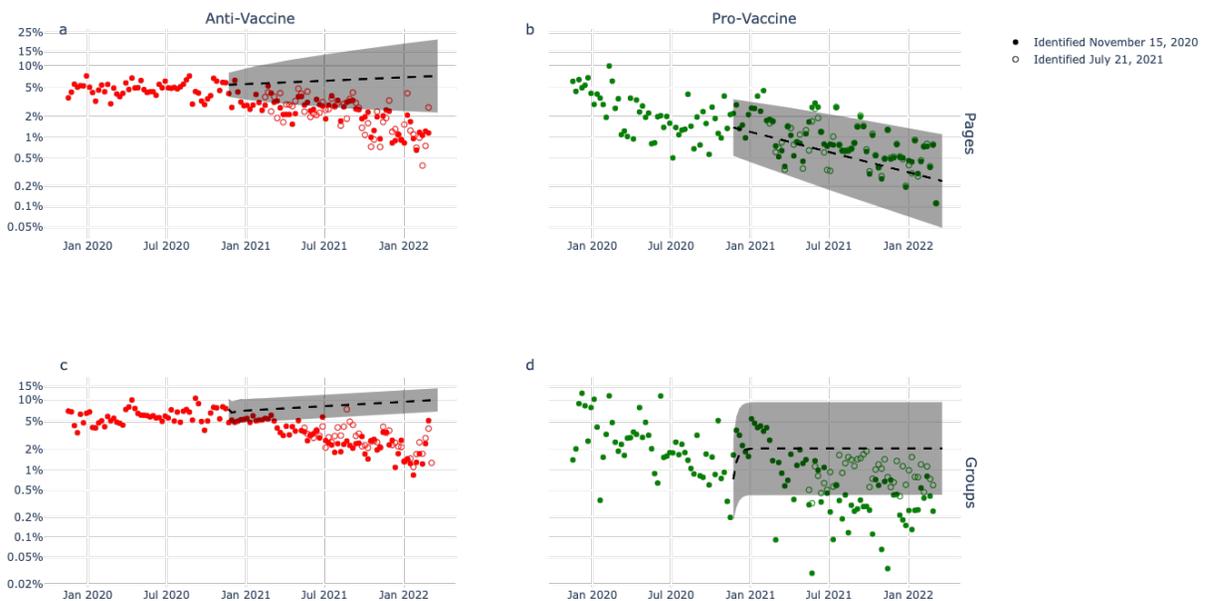

**Fig S5-5. Changes in the proportion of "Angry" reactions over time compared to pre-policy projections**

ARIMA model results show that, compared to projections (black dashed line), Facebook's policies led to an decrease in "Angry" reactions **a.** anti-vaccine pages (60% average decrease, $\chi^2(66) = 296.90, p<0.001$) **b.** anti-vaccine groups (58% average decrease, $\chi^2(66) = 1701.99, p<0.001$). In contrast, the proportion of "Angry" reactions did not change significantly in **c.** pro-vaccine pages ($\chi^2(66)= 60.92, p=0.65$), although we did detect a significant difference in **d.** pro-vaccine groups (41% average decrease, $\chi^2(66)= 232.79, p<0.001$).

**Haha.** We found that the proportion of "Haha" reactions decreased over time for anti-vaccine pages, and pro-vaccine groups.

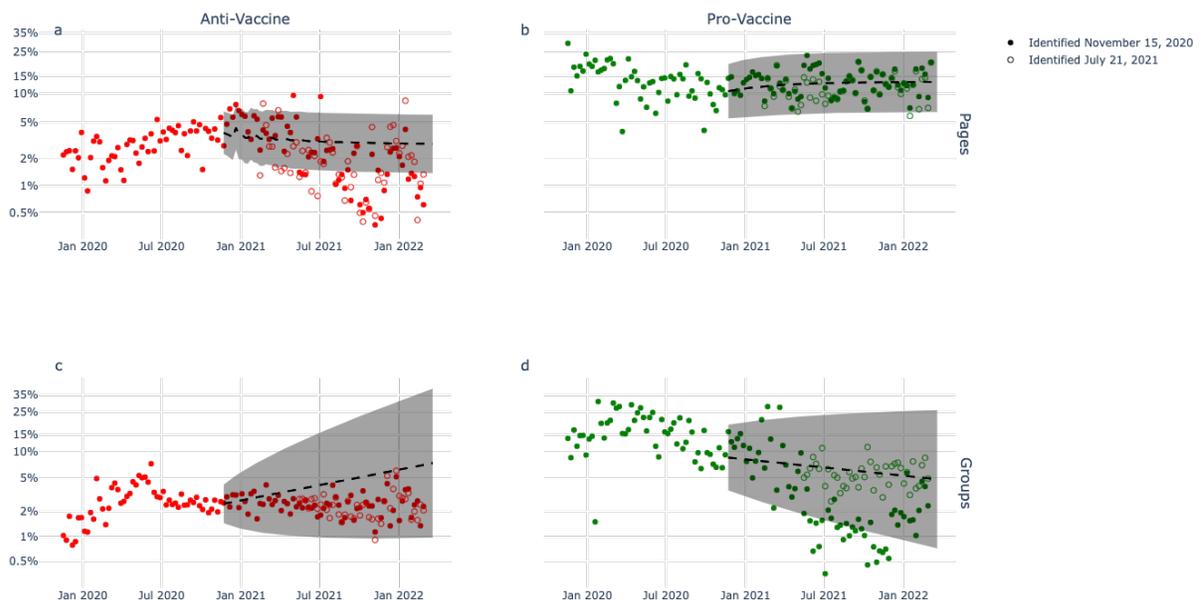

**Fig S5-6. Changes in the proportion of "Haha" reactions over time compared to pre-policy projections**

ARIMA model results show that, compared to projections (black dashed line), Facebook's policies led to a decrease in "Haha" reactions in **a.** anti-vaccine pages (9% average decrease, $\chi^2(66)= 233.88, p<0.001$), **b.** but no change in anti-vaccine groups ($\chi^2(66)= 34.58, p=1.00$). Similarly, we did not detect a change the proportion of "Haha" reactions in **c.** pro-vaccine pages ($\chi^2(66)= 31.78, p=1.00$). We did detect a significant decrease in **d.** pro-vaccine groups (33% average decrease, $\chi^2(66)= 142.98, p<0.001$).

**Wow.** We found that the proportion of "Wow" reactions decreased over time for anti-vaccine pages and increased for pro-vaccine pages.

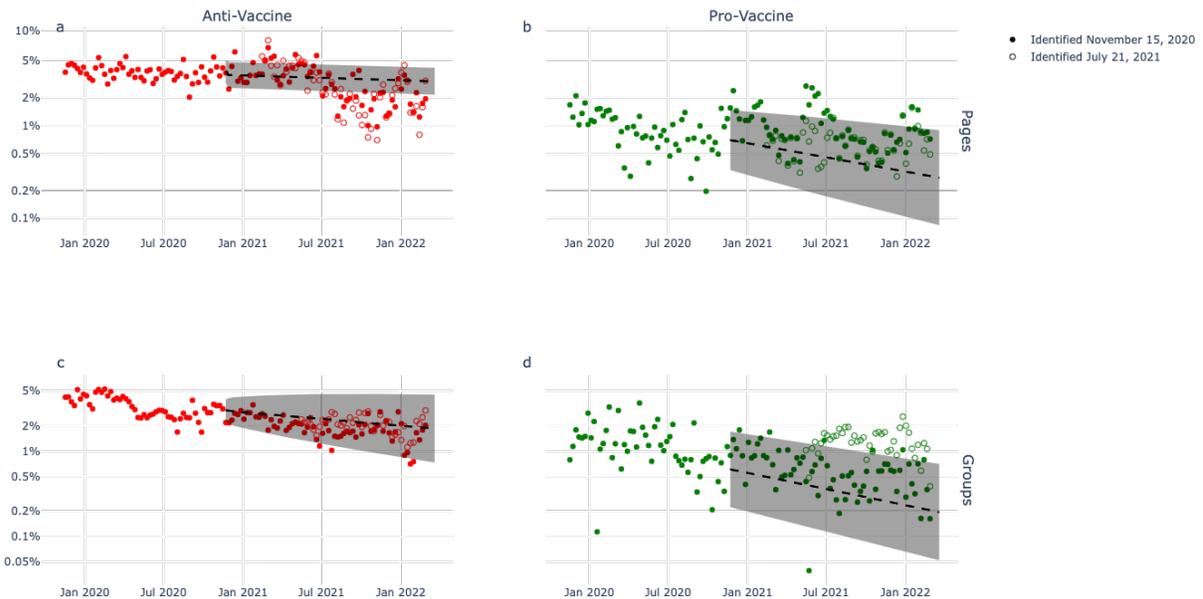

**Fig S5-7. Changes in the proportion of "Sad" reactions over time compared to pre-policy projections**

ARIMA model results show that, compared to projections (black dashed line), Facebook's policies led to an decrease in "Wow" reactions **a.** anti-vaccine pages (8% average decrease, $\chi^2(66)= 362.75, p<0.001$) **b.** but no significant change in anti-vaccine groups ($\chi^2(66)= 44.81, p=0.98$). In contrast, the proportion of "Wow" reactions increased significantly in **c.** pro-vaccine pages (119% average increase, $\chi^2(66)= 130.83, p<0.001$), but did change significantly in **d.** pro-vaccine groups ($\chi^2(66)= 72.39, p=0.28$).

Taken together, these results suggest an overall decrease in the proportion of "toxic" interactions in anti-vaccine, but not pro-vaccine venues (see Extended Data Fig. 4).

*Other Interactions*

**Comments.** We found that the proportion of comments increased slightly in anti-vaccine pages, and decreased slightly in pro-vaccine venues.

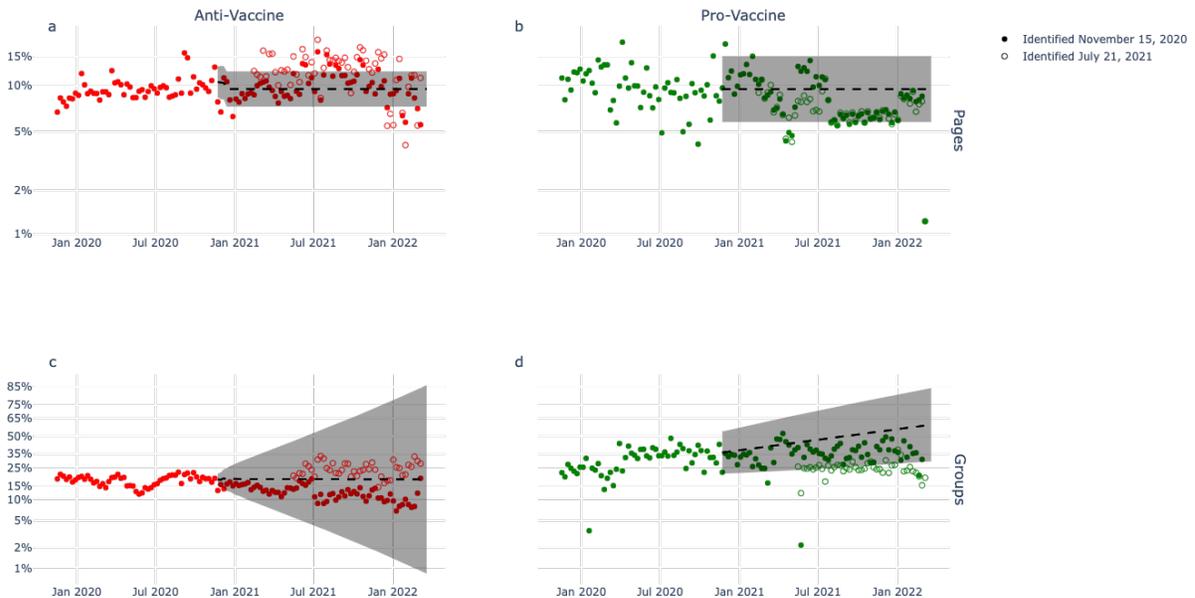

**Fig S5-8. Changes in the proportion of comments over time compared to pre-policy projections**

ARIMA model results show that, compared to projections (black dashed line), Facebook's policies led to an in comments in **a.** anti-vaccine pages (5% average increase, $\chi^2(66)= 131.29$, $p<0.001$), **b.** but no change in anti-vaccine groups ($\chi^2(66)= 21.37, p=1.00$). Additionally, we observed a slight decrease in comments in **c.** pro-vaccine pages (9% average decrease, $\chi^2(66)= 90.60, p=0.02$) and in **d.** pro-vaccine groups (24% average decrease, $\chi^2(66)= 95.76, p=0.01$).

**Shares.** We found that the proportion of shares decreased over time for anti-vaccine pages and increased for pro-vaccine pages.

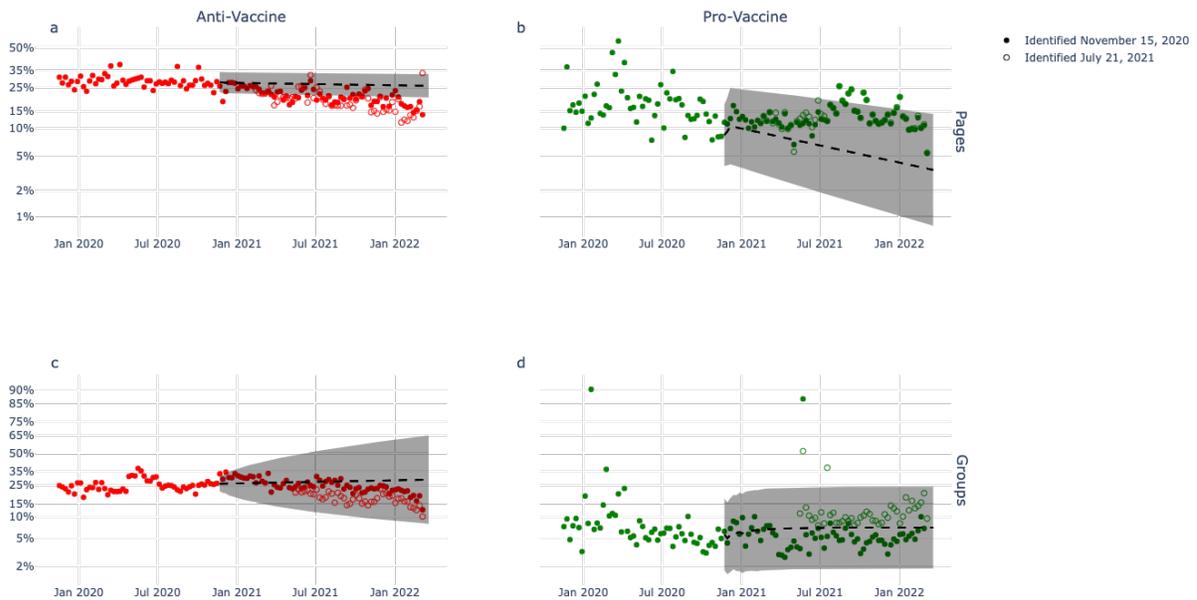

**Fig S5-9. Changes in the proportion of Shares over time compared to pre-policy projections**

ARIMA model results show that, compared to projections (black dashed line), Facebook's policies led to a decrease in Shares **a.** anti-vaccine pages (21% average decrease, $\chi^2(66)= 260.82, p<0.001$) although the proportion of shares repeatedly returned to baseline **b.** but no significant change in anti-vaccine groups ($\chi^2(66)= 16.48, p=1.00$). In contrast, the proportion of Shares did not change significantly in **c.** pro-vaccine pages ($\chi^2(66)= 82.92, p=0.08$) or in **d.** pro-vaccine groups ($\chi^2(66)= 45.83, p=0.97$).

These results are consistent with an overall decrease in supply of anti-vaccine content, but a repeated recovery in demand.

## 6. Reaction-Weighted Topic Analysis

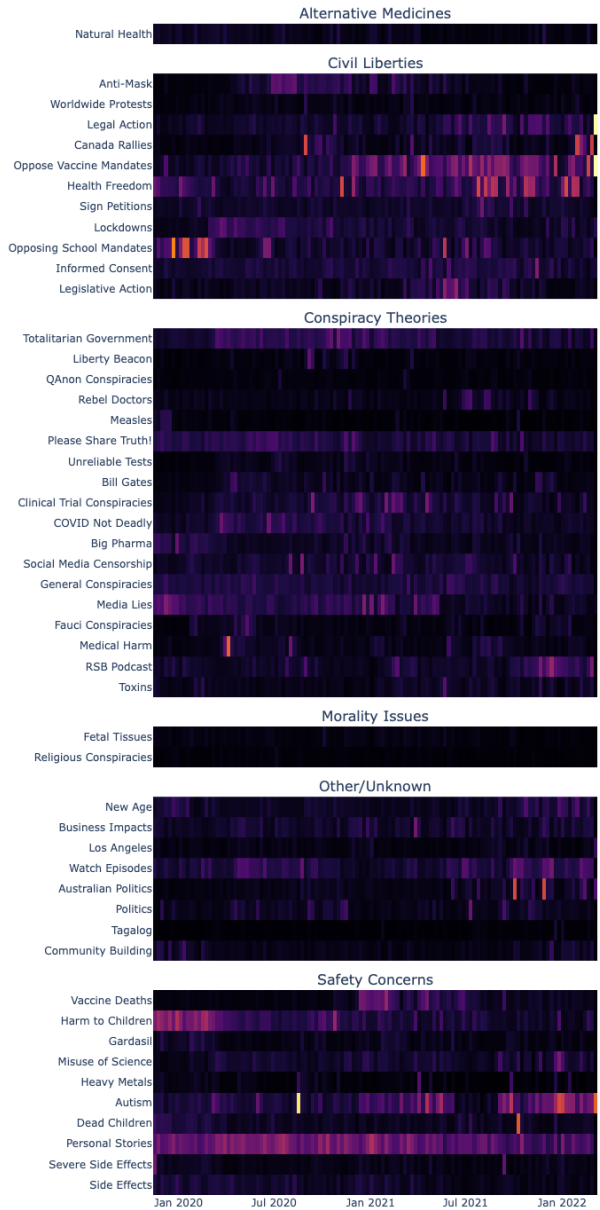

**Fig. S6-1.** All 50 anti-vaccine topics weighted by their proportion of "likes" in the sample of posts in pages identified on November 15, 2020.

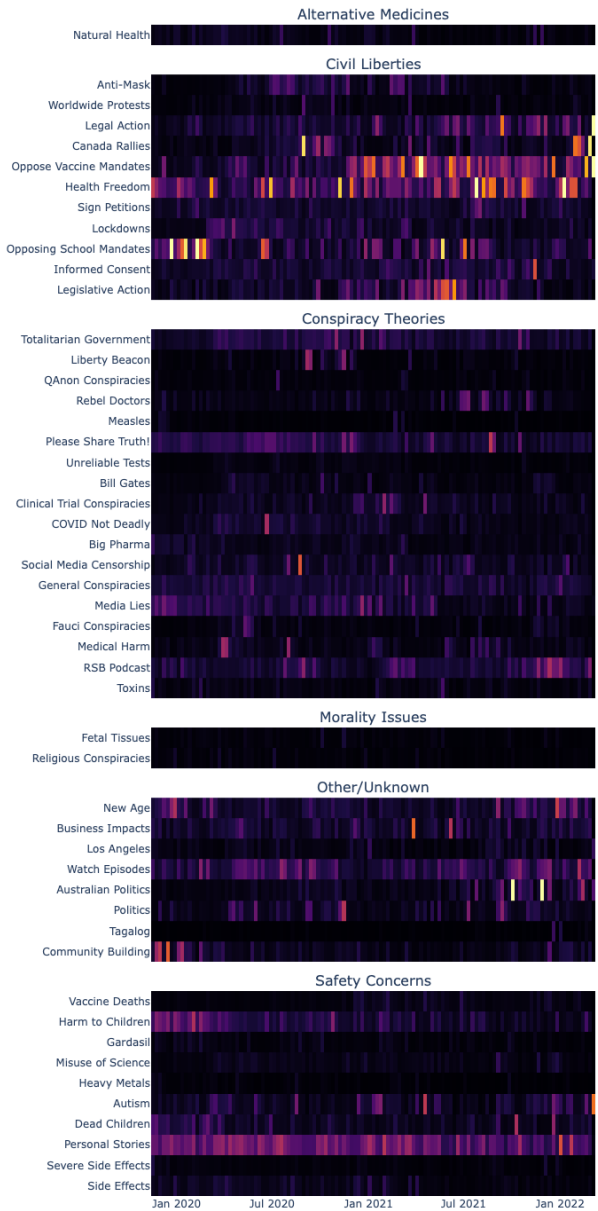

**Fig. S6-2.** All 50 anti-vaccine topics weighted by their proportion of "Love" reactions in the sample of posts in pages identified on November 15, 2020.

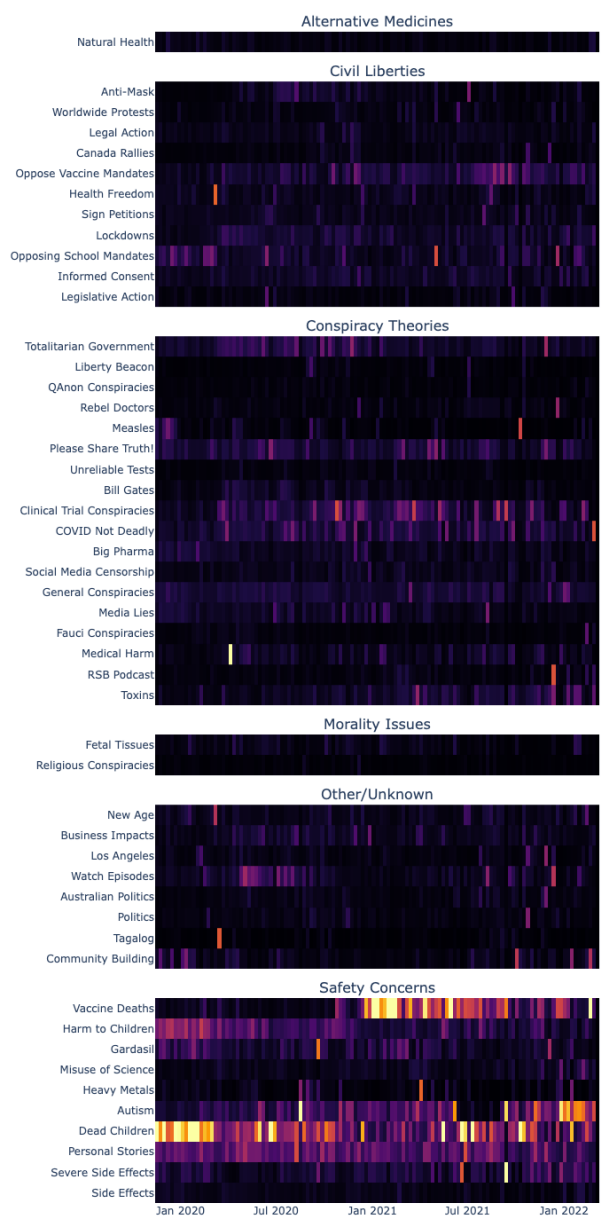

**Fig. S6-3.** All 50 anti-vaccine topics weighted by their proportion of "Sad" reactions in the sample of posts in pages identified on November 15, 2020.

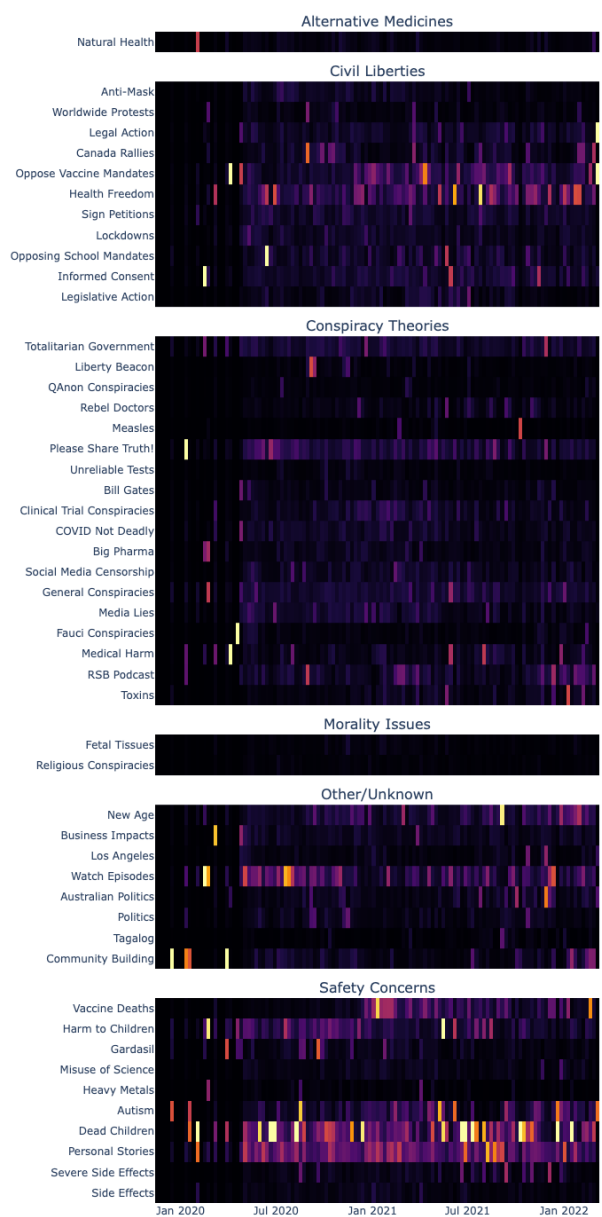

**Fig. S6-4.** All 50 anti-vaccine topics weighted by their proportion of "Care" reactions in the sample of posts in pages identified on November 15, 2020.

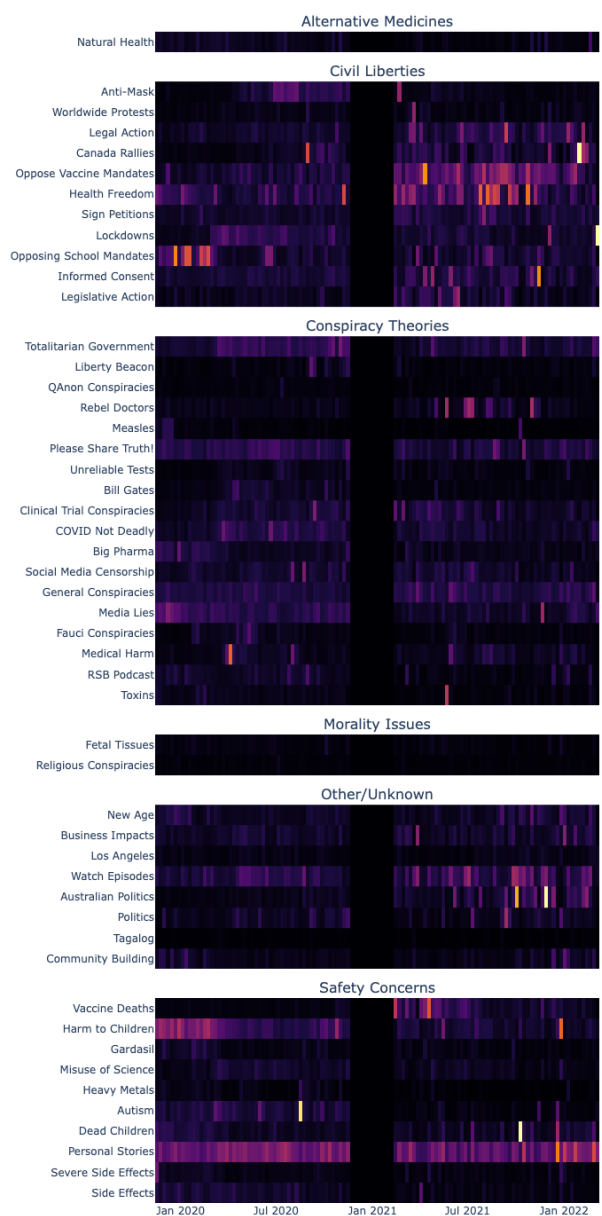

**Fig. S6-5.** All 50 anti-vaccine topics weighted by their proportion of "likes" in the sample of posts in pages identified on July 21, 2021.

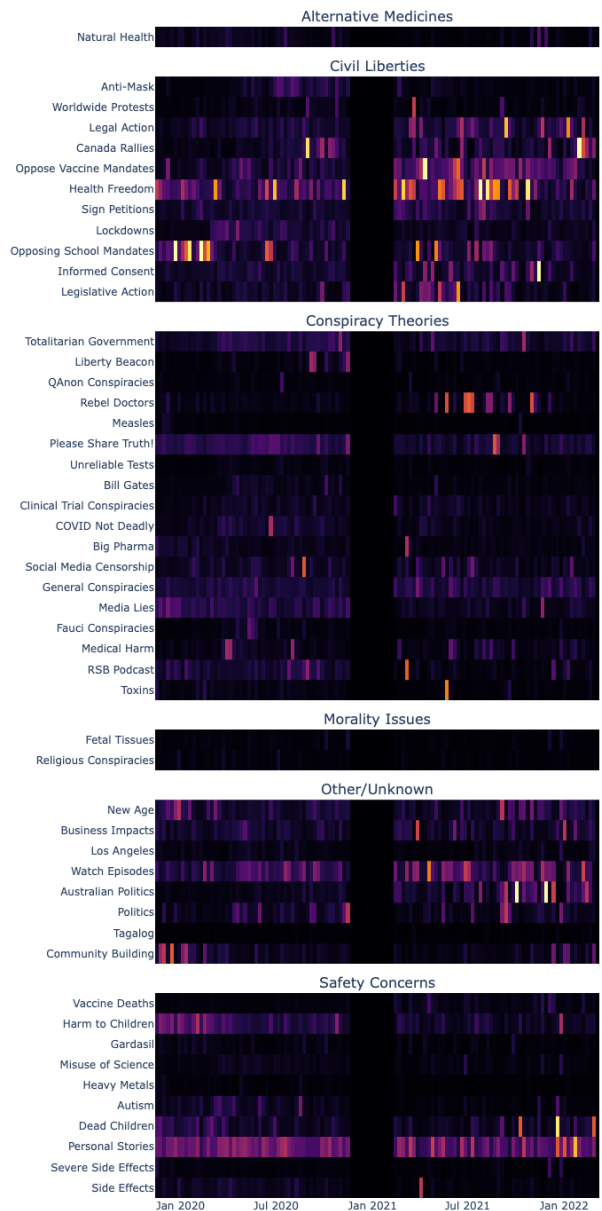

**Fig. S6-6.** All 50 anti-vaccine topics weighted by their proportion of "Love" reactions in the sample of posts in pages identified on July 21, 2021.

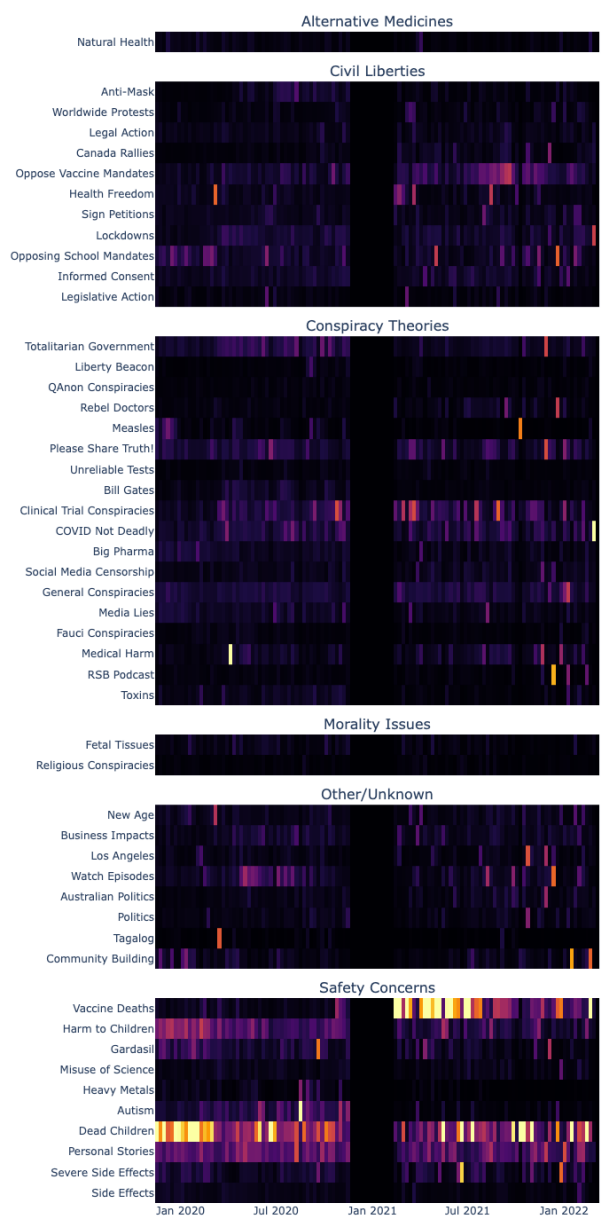

**Fig. S6-7.** All 50 anti-vaccine topics weighted by their proportion of "Sad" reactions in the sample of posts in pages identified on July 21, 2021.

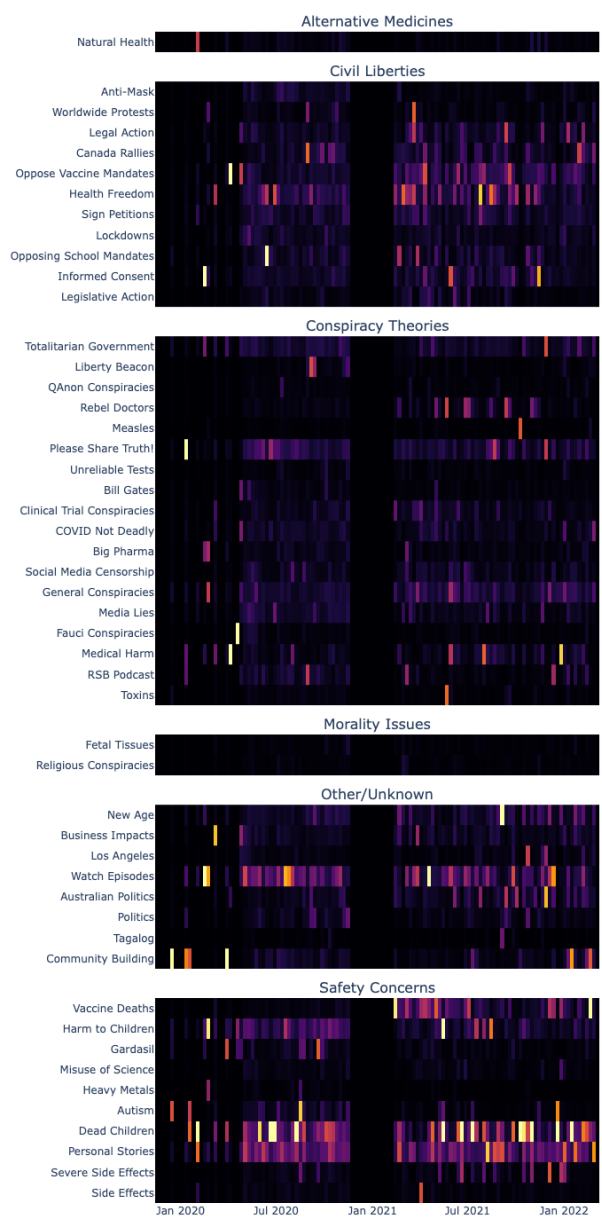

**Fig. S6-8.** All 50 anti-vaccine topics weighted by their proportion of "Care" reactions in the sample of posts in pages identified on July 21, 2021.

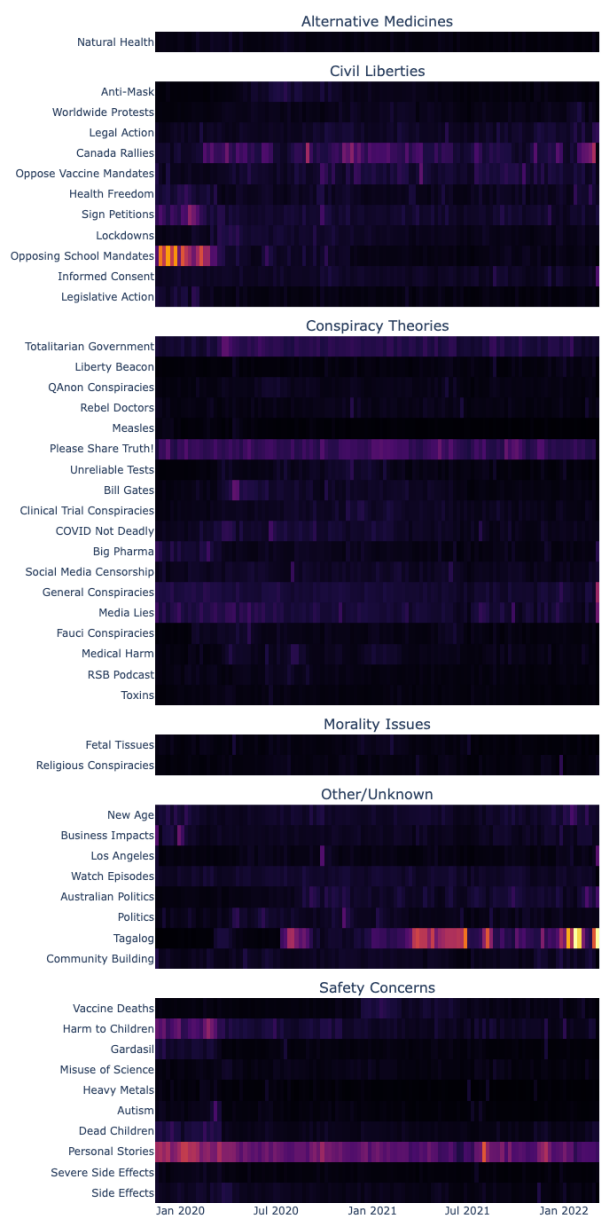

**Fig. S6-9.** All 50 anti-vaccine topics weighted by their proportion of "likes" in the sample of posts in groups identified on November 15, 2020.

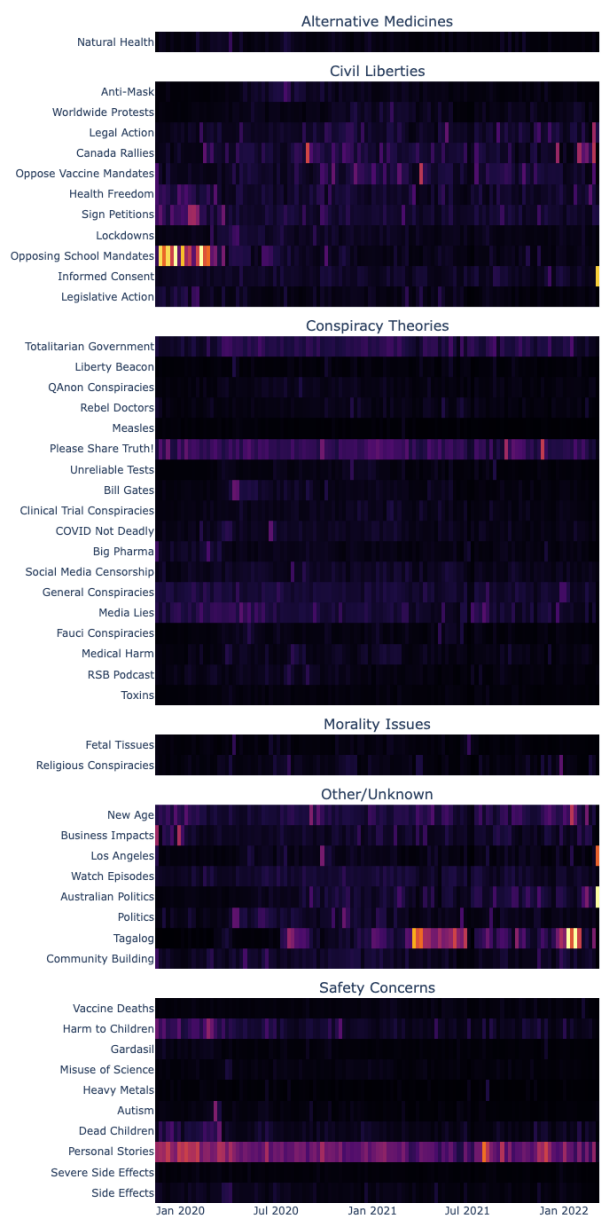

**Fig. S6-10.** All 50 anti-vaccine topics weighted by their proportion of "Love" reactions in the sample of posts in groups identified on November 15, 2020.

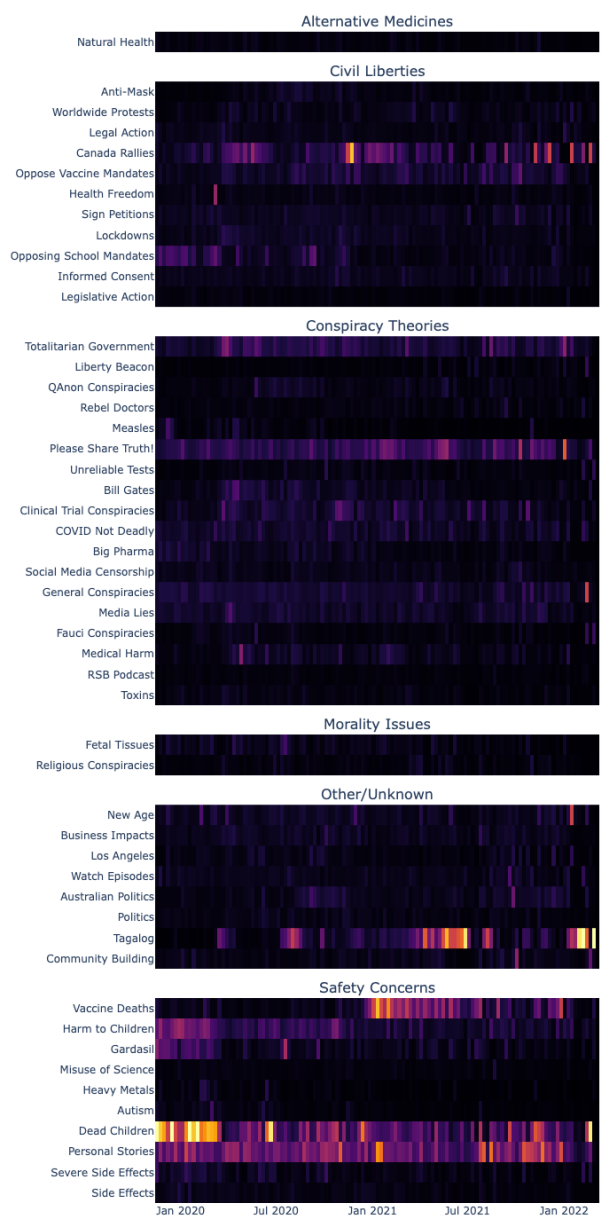

**Fig. S6-11.** All 50 anti-vaccine topics weighted by their proportion of "Sad" reactions in the sample of posts in groups identified on November 15, 2020.

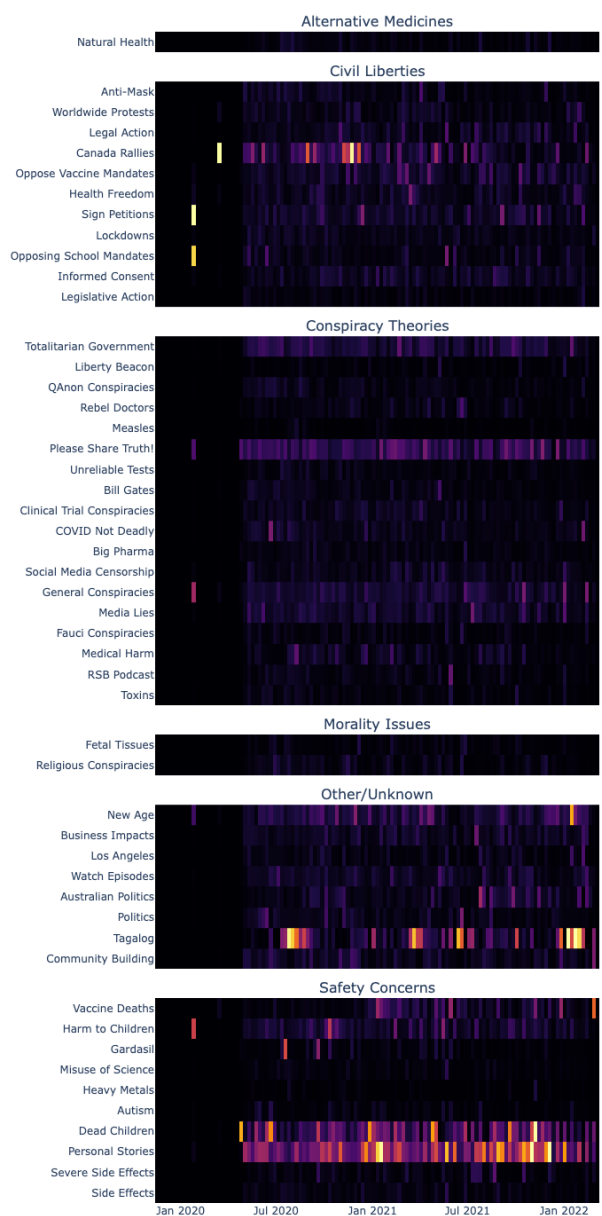

**Fig. S6-12.** All 50 anti-vaccine topics weighted by their proportion of "Care" reactions in the sample of posts in groups identified on November 15, 2020.

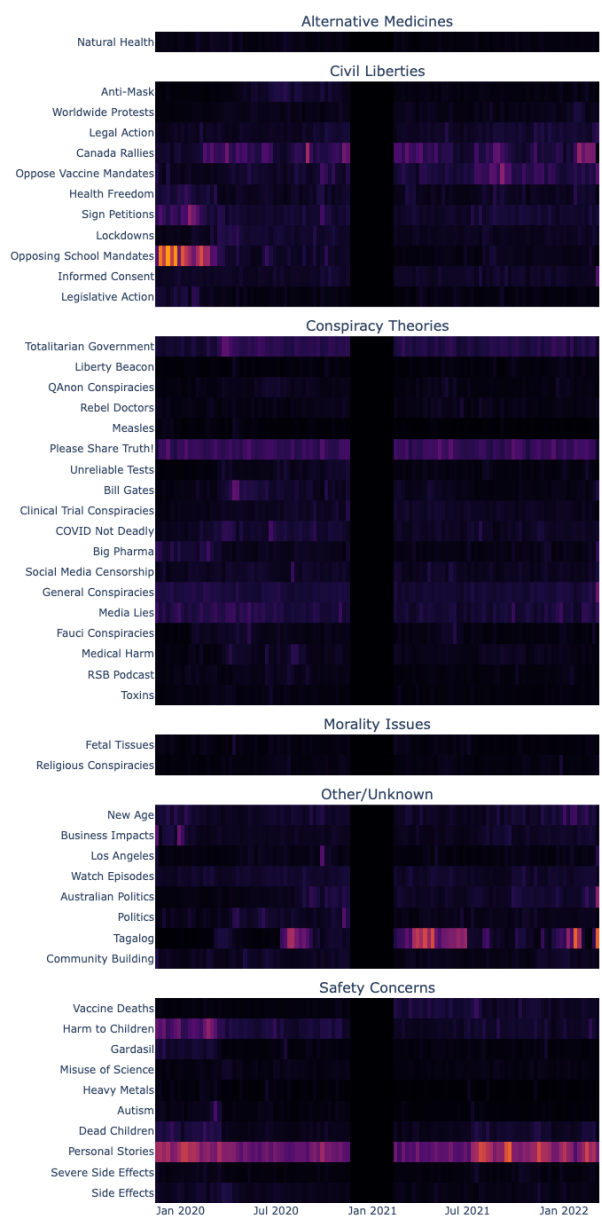

**Fig. S6-13.** All 50 anti-vaccine topics weighted by their proportion of "likes" in the sample of posts in groups identified on July 21, 2021.

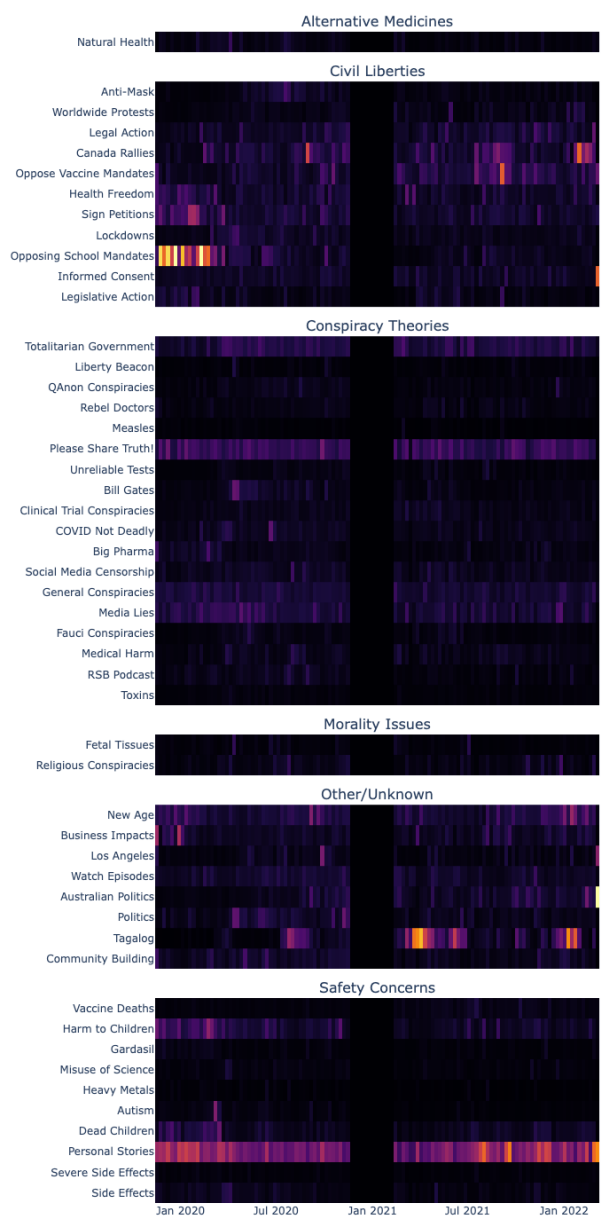

**Fig. S6-14.** All 50 anti-vaccine topics weighted by their proportion of "Love" reactions in the sample of posts in groups identified on July 21, 2021.

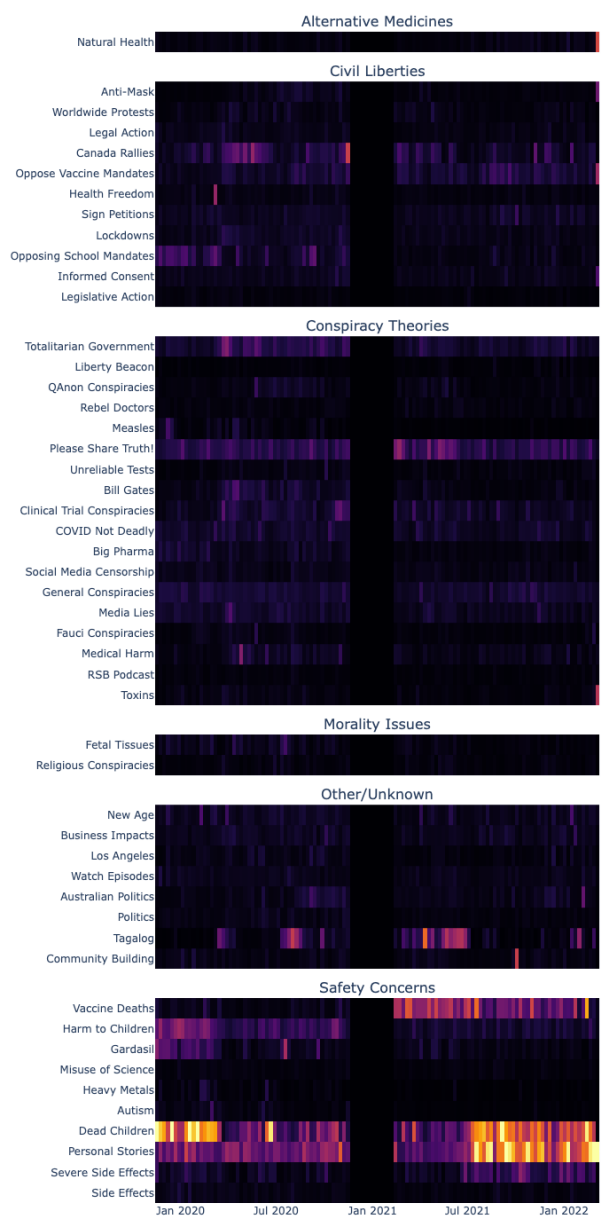

**Fig. S6-15.** All 50 anti-vaccine topics weighted by their proportion of "Sad" reactions in the sample of posts in groups identified on July 21, 2021.

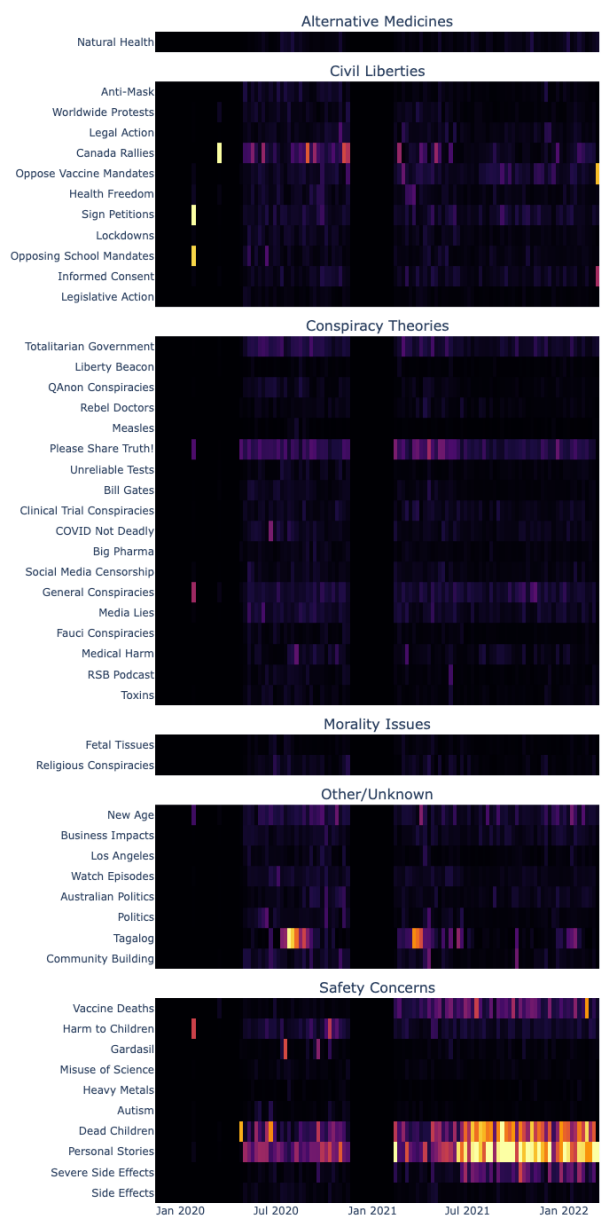

**Fig. S6-16.** All 50 anti-vaccine topics weighted by their proportion of "Care" reactions in the sample of posts in groups identified on July 21, 2021.

## 7. Venue and Post Removal Analyses

Our collection of retrospective data at several points in time allows us to estimate the number and topics of posts that were deleted (Fig. S7-1).

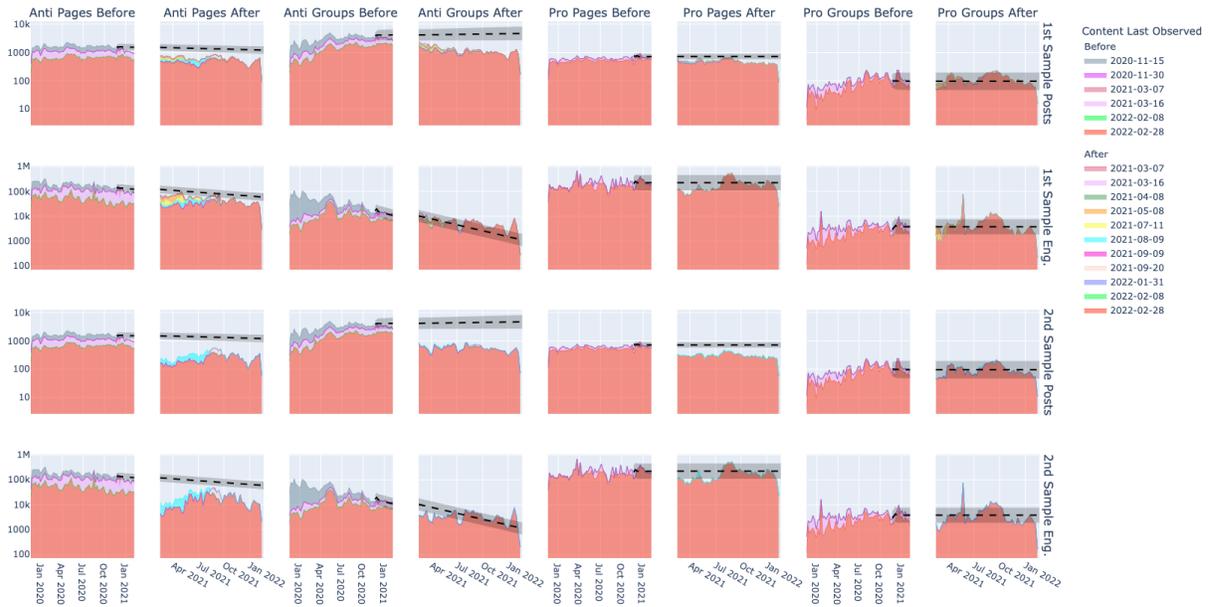

**Fig. S7-1. Area plot showing the number of posts that were last seen in each dataset.**

"1st Sample" refers to the venues identified on November 15, 2020. "2nd Sample" refers to the venues identified on July 21, 2021. All plots use a logarithmic axis.

Specifically, in order to determine which types of content were targeted for removal by Facebook, we examined which venues were present in earlier datasets, but not present in later datasets (Fig. S7-2).

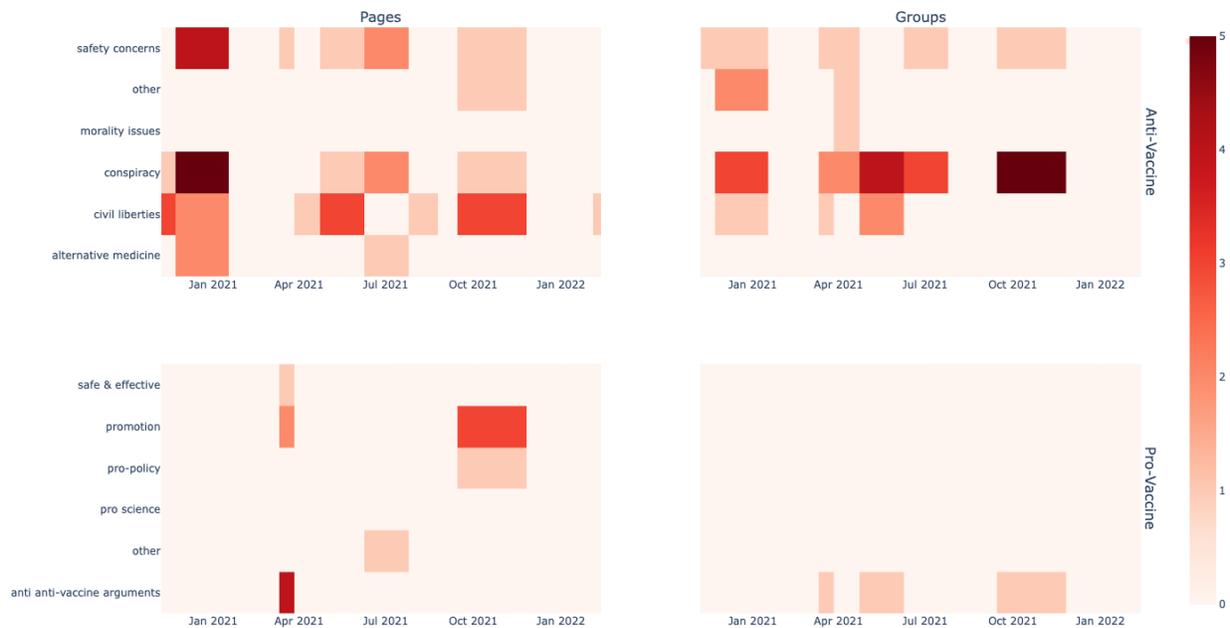

**Fig. S7-2. Heatmap showing which venues were removed from each type of venue.**

Pages presenting as discussing safety concerns and conspiracy theories were removed prior to March, 2021. After March, 2021, venues discussing civil liberties were removed more frequently. Pro-vaccine venues also appear to have been removed following Facebook's February 8, 2021 policy.

In addition, to examine which posts were removed, we assigned each post a unique ID corresponding to its timestamp, and the venue in which it was published. Posts were considered to have been deleted by Facebook if their venue had not been deleted, and if they were present in earlier, but not later datasets. Using our LDA model, we assigned each post a distribution over topics, and then calculated the proportion of all removed posts for each topic. We then aggregated these proportions for each topic category (Fig S7-3).

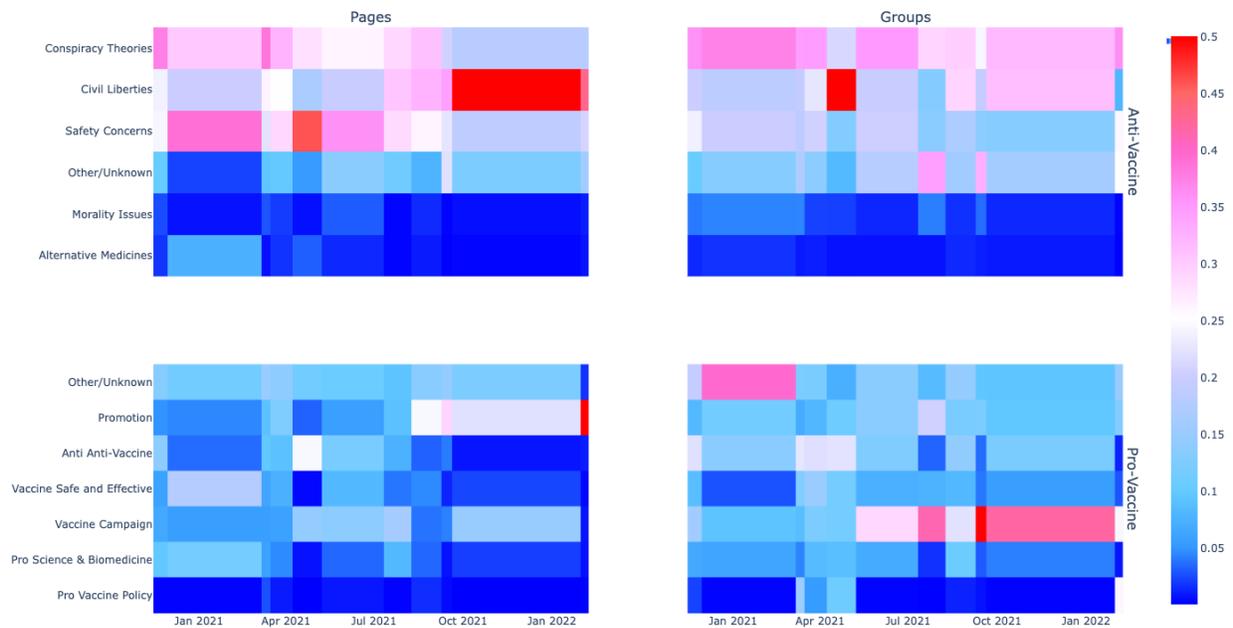

**Fig. S7-3. Heatmap showing which topics were removed from each type of venue.**

Posts discussing conspiracy theories and safety concerns were removed prior to April, 2021, in anti-vaccine pages. Posts discussing civil liberties appear to have been removed more frequently after July, 2021. In anti-vaccine groups, removals largely focused on conspiracy theories and civil liberties issues. Finally, in pro-vaccine groups, topics discussing vaccination campaigns appear to have been removed at the same time as posts discussing civil liberties in anti-vaccine pages.

## 8. Sentiment Analysis

We examined whether Facebook's new policies changed the overall sentiment of the content posted in the venues in our first sample. To do so, we extracted the text from each Facebook post by combining the "Message", "Image Text", "Link Text" and "Description" fields. After identifying and removing all non-English-language posts using the langdetect Python package,[13] we evaluated the sentiment of each post using the DistilBERT model[14] fine tuned on the Stanford Sentiment Treebank 2 corpus – the default sentiment analysis model implemented in the huggingface transformers library ([100]) on June 9, 2022. This model classified each post as expressing either POSITIVE or NEGATIVE sentiment, with a probability score expressing confidence in this classification ranging from 0.5 (no confidence) to 1.0 (complete confidence). We rescaled the scores for NEGATIVE sentiment posts by subtracting them from 1, such that they ranged from 0.0 (complete confidence) to 0.5 (no confidence). We next calculated weekly sentiment scores for each dataset by averaging across positive and rescaled negative scores. Interpreting these as probabilities, we next conducted interrupted time series analyses on these weekly scores after applying a logit transform to control for floor and ceiling effects (Fig. S8-1).

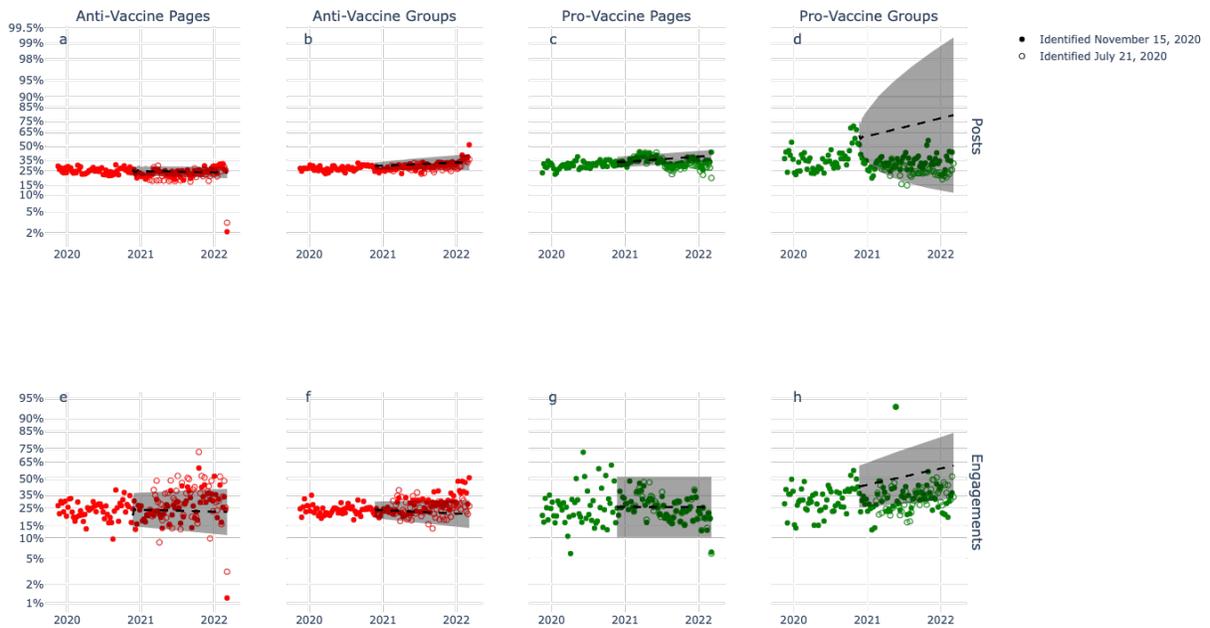

**Fig. S8-1. Weekly average sentiment scores for pro- and anti-vaccine pages and groups.**

We examined whether Facebook's policies led to a significant change in the average weekly sentiment of content posted and engagements with that content. We found **a.** a slight, yet significant, decrease in anti-vaccine pages: 1% average decrease, $\chi^2(66)= 101.34, p=0.003$, but **b.** no change in anti-vaccine groups: $\chi^2(66)= 78.91, p=0.13$. **c.** We also detected a significant decrease of 6% in pro-vaccine pages, $\chi^2(66)= 163.54, p<0.001$, **d.** and in pro-vaccine groups – 53% on average compared to pre-policy projections, $\chi^2(66)= 103.77, p<0.001$. **e.** We did not detect a significant change in engagement with these posts $\chi^2(66)= 79.63, p=0.12$. **f.** However, we did detect a significant increase – 26% – in the sentiment of posts garnering engagements in anti-vaccine groups $\chi^2(66)= 162.05, p=0.87$. **g.** We did not detect a significant change in the sentiment of engagements with pro-vaccine pages, $\chi^2(66)= 23.35, p=1.00$. **h.** In contrast, engagement rates in pro-vaccine groups decreased 37%, $\chi^2(66)= 183.49, p<0.001$.

## 9. Toxicity

We also examined whether Facebook's new policies changed the overall toxicity of the content posted in the venues in our first sample. To do so, we extracted the text from each Facebook post by combining the "Message", "Image Text", "Link Text" and "Description" fields. We evaluated the toxicity of each post using Google's Perspective API[15] between June 9 and June 14, 2022, which assigned each post a toxicity score ranging from 0 (non-toxic) to 1 (toxic). We calculated weekly average scores for each dataset. We next conducted interrupted time series analyses on these weekly scores after applying a logit transform to control for floor and ceiling effects.

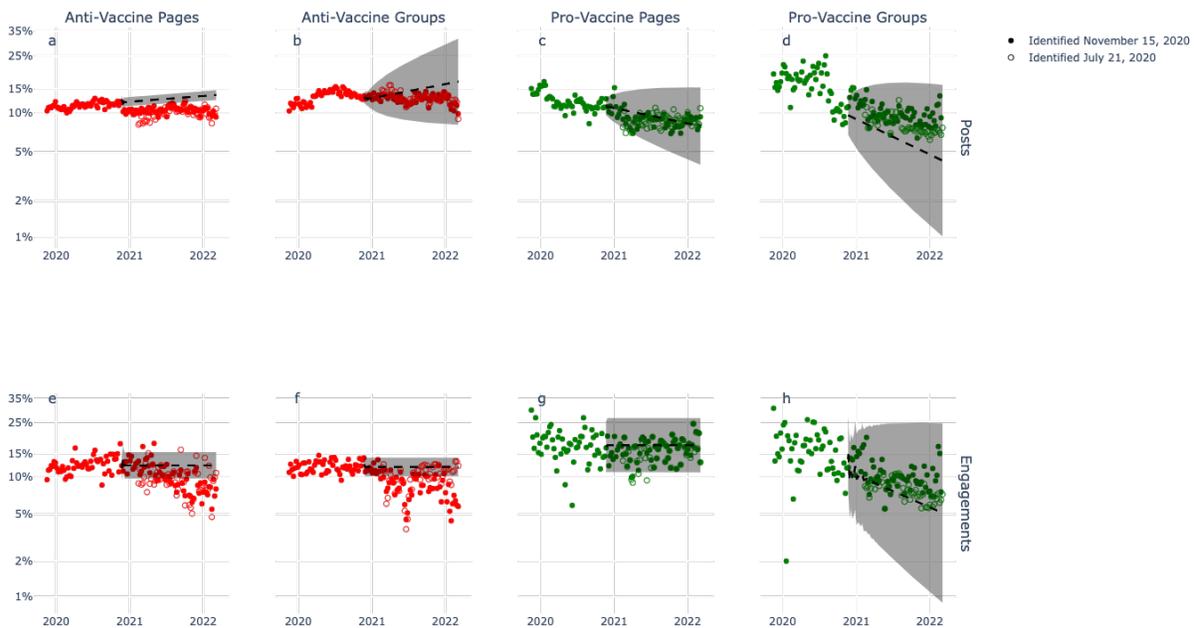

**Fig. S9-1. Weekly average toxicity scores for pro- and anti-vaccine pages and groups.**

We examined whether Facebook's policies led to a significant change in the average weekly toxicity of content posted and engagements with that content. We found **a.** a significant decrease in anti-vaccine pages: 18% average decrease, $\chi^2(66)= 1127.91, p<0.001$, but **b.** no change in anti-vaccine groups: $\chi^2(66)= 18.25, p=1.00$. **c.** We did not detect a significant change in the toxicity of pro-vaccine pages, $\chi^2(66)= 24.92, p=1.00$, **d.** or in pro-vaccine groups, $\chi^2(66)= 50.52, p=0.91$. **e.** The toxicity of the average engagement decreased by 16% in anti-vaccine pages $\chi^2(66)= 323.91, p<0.001$. **f.** and decreased by 25% in anti-vaccine groups $\chi^2(66)= 979.64, p<0.001$. **g.** We did not detect a significant change in the toxicity of engagements with pro-vaccine pages, $\chi^2(66)= 34.77, p=1.00$. **h.** or in pro-vaccine groups, $\chi^2(66) = 20.36, p=1.00$.

In sum, Facebook's policies appear to have led to a sustained decline in the toxicity of anti-vaccine, but not pro-vaccine content, and engagement with that content.

# 10. ARIMA Model Tables

**Table S10-1. Summary statistics for ARIMA models fit throughout this study.**

| Position | Venue | | ADF | p | d | q | Q | H | JB | ADF2 | $\chi^2$ (dof) | p |
|---|---|---|---|---|---|---|---|---|---|---|---|---|
| | | | Counts (controlling for number of venues) | | | | | | | | | |
| Anti-Vaccine | Pages | Posts | 0.41 | 0 | 1 | 3 | 1.91 | 0.56 | 0.29 | 0.01 | 1425.15 (66) | <0.001 |
| | | Engagements | 0.19 | 2 | 1 | 1 | 0.03 | 0.99 | 0.19 | 0.01 | 468.37 (66) | <0.001 |
| | Groups | Posts | 0.43 | 0 | 1 | 4 | 0.47 | 0.40 | 1.67 | 0.01 | 1030.71 (66) | <0.001 |
| | | Engagements | 0.09 | 1 | 1 | 4 | 0.09 | 1.28 | 0.02 | 0.03 | 164.27 (66) | <0.001 |
| Pro-Vaccine | Pages | Posts | 0.02 | 0 | 0 | 3 | 0.08 | 0.78 | 0.22 | 0.01 | 612.17 (66) | <0.001 |
| | | Engagements | 0.04 | 0 | 0 | 2 | 0.20 | 3.43* | 1.85 | 0.02 | 111.43 (66) | <0.001 |
| | Groups | Posts | 0.04 | 1 | 0 | 0 | 0.16 | 0.70 | 1.37 | 0.01 | 61.15 (66) | 0.65 |
| | | Engagements | 0.04 | 0 | 0 | 3 | 0.01 | 0.37 | 7.33* | 0.03 | 164.64 (66) | <0.001 |
| | | | Links to In-Sample Pages (Combined Sample) | | | | | | | | | |
| Anti-Vaccine | Pages | Posts | 0.47 | 1 | 1 | 1 | 0.98 | 2.02 | 1.38 | 0.01 | 192.43 (66) | <0.001 |
| | | Engagements | 0.18 | 0 | 1 | 3 | 0.89 | 3.13 | 9.38* | 0.01 | 234.15 (66) | <0.001 |
| Pro-Vaccine | Pages | Posts | 0.56 | 2 | 1 | 0 | 0.00 | 1.15 | 2.19 | 0.14 | 200.37 (66) | <0.001 |
| | | Engagements | 0.06 | 1 | 1 | 2 | 0.07 | 1.80 | 3.88 | 0.08 | 175.70 (66) | <0.001 |
| | | | Coordinated Venues (Combined Sample) | | | | | | | | | |
| Anti-Vaccine | | | 0.21 | 0 | 1 | 2 | 0.04 | 0.38 | 32.63*** | 0.05 | 267.00 (66) | <0.001 |
| Pro-Vaccine | | | N/A | 2 | 1 | 2 | 0.00 | 0.53 | 1.08 | 0.04 | 27.93 (66) | 1.00 |
| | | | Reactions | | | | | | | | | |
| Anti-Vaccine | Pages | Toxic | 0.21 | 0 | 0 | 1 | 0.00 | 1.10 | 9.84* | 0.02 | 816.74 (66) | <0.001 |
| | | Non-toxic | 0.21 | 3 | 1 | 1 | 0.01 | 0.97 | 3.04 | 0.11 | 463.39 (66) | <0.001 |
| Anti-Vaccine | Groups | Toxic | 0.27 | 0 | 1 | 2 | 0.01 | 1.11 | 0.16 | 0.01 | 423.24 (66) | <0.001 |
| | | Non-toxic | 0.27 | 0 | 1 | 1 | 0.62 | 0.52 | 1.81 | 0.04 | 280.21 (66) | <0.001 |
| Pro-Vaccine | Pages | Toxic | 0.3 | 0 | 1 | 3 | 0.02 | 1.46 | 13.39*** | 0.03 | 398.87 (66) | <0.001 |
| | | Non-toxic | 0.29 | 0 | 1 | 2 | 0.39 | 4.89*** | 0.27 | 0.07 | 73.53 (66) | 0.25 |
| Pro-Vaccine | Groups | Toxic | 0.09 | 0 | 1 | 1 | 0.16 | 0.25 | 137.53*** | 0.01 | 86.34 (66) | 0.05 |
| | | Non-toxic | 0.09 | 0 | 1 | 1 | 0.01 | 0.21 | 625.68*** | 0.01 | 21.07 (66) | 1.00 |
| | | | Iffy URLs | | | | | | | | | |
| Anti-Vaccine | Pages | Posts | 0.01 | 1 | 0 | 0 | 0.01 | 1.10 | 1.26 | 0.01 | 934.55 (66) | <0.001 |
| | | Engagements | 0.36 | 0 | 1 | 2 | 0.03 | 0.92 | 2.22 | 0.03 | 247.64 (66) | <0.001 |
| | Groups | Posts | 0.68 | 1 | 1 | 0 | 0.03 | 1.04 | 1.09 | 0.01 | 24.43 (66) | 1.00 |
| | | Engagements | 0.55 | 0 | 1 | 1 | 0.23 | 2.43 | 0.41 | 0.01 | 70.96 (66) | 0.32 |
| | | | Government and Academic URLs | | | | | | | | | |
| Anti-Vaccine | Pages | Posts | 0.07 | 0 | 1 | 1 | 1.54 | 0.41 | 0.68 | 0.01 | 94.44 (66) | 0.01 |
| | | Engagements | 0.49 | 0 | 1 | 1 | 0.16 | 1.50 | 0.56 | 0.01 | 72.88 (66) | 0.26 |
| | Groups | Posts | 0.04 | 1 | 0 | 0 | 0.35 | 2.45 | 2.32 | 0.02 | 161.72 (64) | <0.001 |
| | | Engagements | 0.46 | 0 | 1 | 1 | 0.04 | 1.09 | 3.58 | 0.08 | 50.32 (63) | 0.88 |
| Pro-Vaccine | Pages | Posts | 0.17 | 1 | 1 | 3 | 0.01 | 0.37 | 1.55 | 0.01 | 246.88 (66) | <0.001 |
| | | Engagements | 0.02 | 1 | 0 | 0 | 0.26 | 1.53 | 0.47 | 0.02 | 187.26 (66) | <0.001 |
| | Groups | Posts | N/A | 0 | 1 | 2 | 0.27 | 0.94 | 1.51 | 0.07 | 100.09 (55) | <0.001 |
| | | Engagements | N/A | 0 | 1 | 5 | 0.31 | 0.44 | 5.66 | 0.03 | 54.75 (55) | 0.48 |
| | | | Average Fact Rating | | | | | | | | | |
| Anti-Vaccine | Pages | Posts | 0.19 | 0 | 1 | 4 | 0.09 | 0.50 | 0.82 | 0.01 | 48.52 (66) | 0.94 |
| | | Engagements | 0.02 | 3 | 0 | 0 | 0.02 | 1.61 | 1.54 | 0.08 | 183.36 (66) | <0.001 |
| | Groups | Posts | 0.5 | 0 | 1 | 1 | 0.33 | 0.55 | 5.14 | 0.04 | 43.76 (66) | 0.98 |
| | | Engagements | 0.49 | 0 | 1 | 1 | 0.04 | 0.71 | 0.27 | 0.05 | 130.82 (66) | <0.001 |
| Pro-Vaccine | Pages | Posts | 0.01 | 2 | 0 | 0 | 0.04 | 0.64 | 3.90 | 0.01 | 110.06 (66) | <0.001 |
| | | Engagements | 0.2 | 0 | 1 | 3 | 0.09 | 0.86 | 7.88* | 0.21 | 56.73 (66) | 0.78 |
| | Groups | Posts | 0.65 | 1 | 1 | 1 | 0.43 | 0.89 | 0.92 | 0.17 | 15.16 (66) | 1.00 |
| | | Engagements | 0.29 | 3 | 1 | 0 | 0.04 | 0.93 | 0.35 | 0.05 | 20.43 (66) | 1.00 |
| | | | Average Bias Rating | | | | | | | | | |
| Anti-Vaccine | Pages | Posts | 0.01 | 1 | 0 | 0 | 0.15 | 0.69 | 1.31 | 0.01 | 222.24 (66) | <0.001 |

| | | | | | | | | | | | |
|---|---|---|---|---|---|---|---|---|---|---|---|
| | | Engagements | 0.01 | 2 | 0 | 2 | 0.15 | 0.73 | 4.31 | 0.04 | 114.21 (66) | <0.001 |
| | Groups | Posts | 0.08 | 0 | 1 | 1 | 0.48 | 0.68 | 1.22 | 0.01 | 106.48 (66) | <0.001 |
| | | Engagements | 0.23 | 0 | 1 | 2 | 0.19 | 0.81 | 1.14 | 0.10 | 402.15 (66) | <0.001 |
| Pro-Vaccine | Pages | Posts | 0.07 | 1 | 1 | 3 | 0.04 | 0.79 | 3.92 | 0.01 | 95.86 (66) | 0.01 |
| | | Engagements | 0.02 | 0 | 0 | 0 | 0.00 | 0.60 | 8.47* | 0.02 | 42.21 (66) | 0.99 |
| | Groups | Posts | 0.2 | 1 | 1 | 2 | 0.03 | 0.17 | 15.13** | 0.28 | 75.52 (66) | 0.17 |
| | | Engagements | 0.01 | 2 | 0 | 0 | 0.01 | 0.17 | 86.71*** | 0.02 | 24.02 (66) | 1.00 |

*Note. ADF1 = p-value of Augmented Dickey-Fuller test fit to raw data; ADF2 = p-value of Augmented Dickey-Fuller test fit to ARIMA model residuals; p=autoregressive order; N/A = not available because of missing data; d = differencing order; q = moving average order; Q = Ljung-Box statistic; H = Heteroskedasticity; JB = Jarque-Bera statistic; \* = p<0.05, \*\*=p<0.01, \*\*\*=p<0.001.*

**Table S10-2: ARIMA models fit throughout this study**

Anti-Vaccine Page Posts: ARIMA(0,1,3), Q=1.91, p=0.17; H=0.56, p=0.24, JB=0.29, p=0.87

| | coef | std err | z | P>|z| | [0.025 | 0.975] |
|---|---|---|---|---|---|---|
| intercept | 0.00 | 0.00 | -1.20 | 0.23 | -0.01 | 0.00 |
| ma.L1 | 0.09 | 0.18 | 0.51 | 0.61 | -0.26 | 0.44 |
| ma.L2 | -0.42 | 0.22 | -1.94 | 0.05 | -0.85 | 0.00 |
| ma.L3 | -0.53 | 0.17 | -3.05 | 0.00 | -0.88 | -0.19 |
| $s^2$ | 0.01 | 0.00 | 3.53 | 0.00 | 0.00 | 0.01 |

Anti-Vaccine Page Engagements: ARIMA(2,1,1), Q=0.03, p=0.85; H=0.88, p=0.79, JB=0.19, p=0.91

| | coef | std err | z | P>|z| | [0.025 | 0.975] |
|---|---|---|---|---|---|---|
| intercept | -0.01 | 0.00 | -1.94 | 0.05 | -0.01 | 0.00 |
| ar.L1 | 0.45 | 0.15 | 2.92 | 0.00 | 0.15 | 0.75 |
| ar.L2 | 0.03 | 0.14 | 0.22 | 0.83 | -0.24 | 0.30 |
| ma.L1 | -0.99 | 0.65 | -1.51 | 0.13 | -2.27 | 0.29 |
| $s^2$ | 0.03 | 0.02 | 1.69 | 0.09 | -0.01 | 0.07 |

Anti-Vaccine Group Posts: ARIMA(0,1,4), Q=0.47, p=0.49; H=0.40, p=0.07, JB=1.67, p=0.43

| | coef | std err | z | P>|z| | [0.025 | 0.975] |
|---|---|---|---|---|---|---|
| intercept | 0.00 | 0.01 | 0.25 | 0.80 | -0.02 | 0.02 |
| ma.L1 | -0.02 | 0.18 | -0.14 | 0.89 | -0.37 | 0.32 |
| ma.L2 | -0.05 | 0.20 | -0.27 | 0.79 | -0.44 | 0.33 |
| ma.L3 | -0.45 | 0.19 | -2.35 | 0.02 | -0.82 | -0.07 |
| ma.L4 | -0.30 | 0.14 | -2.20 | 0.03 | -0.58 | -0.03 |
| $s^2$ | 0.02 | 0.01 | 3.74 | 0.00 | 0.01 | 0.04 |

Anti-Vaccine Group Engagements: ARIMA(1,1,4), Q=0.09, p=0.77; H=1.28, p=0.62, JB=0.02, p=0.99

| | coef | std err | z | P>|z| | [0.025 | 0.975] |
|---|---|---|---|---|---|---|
| intercept | -0.06 | 0.02 | -3.83 | 0.00 | -0.10 | -0.03 |
| ar.L1 | -0.53 | 0.29 | -1.85 | 0.06 | -1.09 | 0.03 |
| ma.L1 | 0.43 | 0.55 | 0.77 | 0.44 | -0.66 | 1.51 |
| ma.L2 | -0.05 | 0.73 | -0.07 | 0.94 | -1.49 | 1.38 |
| ma.L3 | -0.58 | 0.69 | -0.85 | 0.40 | -1.92 | 0.76 |
| ma.L4 | -0.74 | 0.42 | -1.77 | 0.08 | -1.55 | 0.08 |
| $s^2$ | 0.04 | 0.02 | 2.04 | 0.04 | 0.00 | 0.08 |

Pro-Vaccine Page Posts: ARIMA(0,0,3), Q=0.08, p=0.77; H=0.78, p=0.62, JB=0.22, p=0.90

| | coef | std err | z | P>|z| | [0.025 | 0.975] |
|---|---|---|---|---|---|---|
| intercept | 1.95 | 0.03 | 57.16 | 0.00 | 1.89 | 2.02 |
| ma.L1 | 0.76 | 0.16 | 4.77 | 0.00 | 0.45 | 1.07 |
| ma.L2 | 0.35 | 0.15 | 2.29 | 0.02 | 0.05 | 0.64 |
| ma.L3 | 0.40 | 0.17 | 2.38 | 0.02 | 0.07 | 0.73 |
| $s^2$ | 0.01 | 0.00 | 5.16 | 0.00 | 0.01 | 0.01 |
| Pro-Vaccine Page Engagements: ARIMA(0,0,2), Q=0.20, p=0.66; H=3.43, p=0.02, JB=1.85, p=0.40 | | | | | | |
| intercept | 7.69 | 0.09 | 87.78 | 0.00 | 7.52 | 7.86 |
| ma.L1 | 0.04 | 0.15 | 0.29 | 0.77 | -0.25 | 0.33 |
| ma.L2 | 0.52 | 0.12 | 4.38 | 0.00 | 0.29 | 0.75 |
| $s^2$ | 0.15 | 0.03 | 4.70 | 0.00 | 0.09 | 0.21 |
| Pro-Vaccine Group Posts: ARIMA(1,0,0), Q=0.16, p=0.69; H=0.70, p=0.47, JB=1.37, p=0.50 | | | | | | |
| intercept | 0.77 | 0.23 | 3.30 | 0.00 | 0.32 | 1.23 |
| ar.L1 | 0.69 | 0.10 | 7.15 | 0.00 | 0.50 | 0.88 |
| $s^2$ | 0.10 | 0.02 | 4.71 | 0.00 | 0.06 | 0.14 |
| Pro-Vaccine Group Engagements: ARIMA(0,0,3), Q=0.01, p=0.90; H=0.37, p=0.05, JB=7.33, p=0.03 | | | | | | |
| intercept | 6.16 | 0.08 | 82.15 | 0.00 | 6.01 | 6.31 |
| ma.L1 | 0.25 | 0.19 | 1.32 | 0.19 | -0.12 | 0.63 |
| ma.L2 | 0.26 | 0.21 | 1.29 | 0.20 | -0.14 | 0.66 |
| ma.L3 | -0.25 | 0.15 | -1.67 | 0.10 | -0.55 | 0.04 |
| $s^2$ | 0.15 | 0.03 | 5.82 | 0.00 | 0.10 | 0.20 |
| Anti-Vaccine Page Links: ARIMA(1,1,1), Q=0.00, p=0.98; H=2.02, p=0.16, JB=1.38, p=0.50 | | | | | | |
| intercept | -0.01 | 0.01 | -0.55 | 0.58 | -0.03 | 0.02 |
| ar.L1 | 0.32 | 0.23 | 1.44 | 0.15 | -0.12 | 0.77 |
| ma.L1 | -0.82 | 0.15 | -5.35 | 0.00 | -1.12 | -0.52 |
| $s^2$ | 0.14 | 0.03 | 4.60 | 0.00 | 0.08 | 0.21 |
| Pro-Vaccine Page Links: ARIMA(2,1,0), Q=0.00, p=1.00; H=1.15, p=0.78, JB=2.19, p=0.34 | | | | | | |
| intercept | -0.02 | 0.06 | -0.36 | 0.72 | -0.14 | 0.10 |
| ar.L1 | -0.27 | 0.16 | -1.68 | 0.09 | -0.59 | 0.04 |
| ar.L2 | -0.39 | 0.11 | -3.53 | 0.00 | -0.61 | -0.18 |
| $s^2$ | 0.15 | 0.04 | 3.95 | 0.00 | 0.08 | 0.23 |
| Anti-Vaccine Coordinated Links: ARIMA(0,1,2), Q=0.04, p=0.83; H=0.38, p=0.05, JB=32.63, p<0.001 | | | | | | |
| intercept | 0.01 | 0.02 | 0.34 | 0.73 | -0.03 | 0.04 |
| ma.L1 | -0.26 | 0.12 | -2.18 | 0.03 | -0.49 | -0.03 |
| ma.L2 | -0.49 | 0.11 | -4.65 | 0.00 | -0.70 | -0.29 |
| $s^2$ | 0.09 | 0.01 | 7.56 | 0.00 | 0.07 | 0.12 |
| Pro-Vaccine Coordinated Links: ARIMA(2,1,2), Q=0.00, p=1.00; H=0.53, p=0.33, JB=1.08, p=0.58 | | | | | | |
| ntercept | 0.03 | 0.11 | 0.24 | 0.81 | -0.20 | 0.25 |
| ar.L1 | 0.11 | 0.20 | 0.54 | 0.59 | -0.29 | 0.51 |
| ar.L2 | -0.74 | 0.20 | -3.65 | 0.00 | -1.14 | -0.34 |
| ma.L1 | -0.80 | 0.27 | -2.97 | 0.00 | -1.33 | -0.27 |
| ma.L2 | 0.68 | 0.27 | 2.49 | 0.01 | 0.14 | 1.21 |
| $s^2$ | 0.35 | 0.12 | 3.01 | 0.00 | 0.12 | 0.58 |

### Anti-Vaccine Page "Toxic" Reactions: ARIMA(0,0,1), Q=0.00, p=0.99; H=1.10, p=0.84, JB=9.84, p=0.01

| | | | | | | |
|---|---|---|---|---|---|---|
| intercept | -2.03 | 0.04 | -52.36 | 0.00 | -2.11 | -1.96 |
| ma.L1 | 0.26 | 0.16 | 1.65 | 0.10 | -0.05 | 0.57 |
| $s^2$ | 0.04 | 0.01 | 5.46 | 0.00 | 0.02 | 0.05 |

### Anti-Vaccine Page "Nontoxic" Reactions: ARIMA(3,1,1), Q=0.01, p=0.93; H=0.97, p=0.96, JB=3.04, p=0.22

| | | | | | | |
|---|---|---|---|---|---|---|
| intercept | -0.01 | 0.01 | -0.58 | 0.56 | -0.02 | 0.01 |
| ar.L1 | -0.36 | 0.36 | -0.99 | 0.33 | -1.06 | 0.35 |
| ar.L2 | -0.46 | 0.27 | -1.75 | 0.08 | -0.98 | 0.06 |
| ar.L3 | -0.19 | 0.26 | -0.72 | 0.48 | -0.70 | 0.33 |
| ma.L1 | -0.59 | 0.31 | -1.91 | 0.06 | -1.20 | 0.02 |
| $s^2$ | 0.02 | 0.00 | 4.90 | 0.00 | 0.01 | 0.03 |

### Anti-Vaccine Group "Toxic" Reactions: ARIMA(0,1,2), Q=0.01, p=0.94; H=1.11, p=0.83, JB=0.16, p=0.93

| | | | | | | |
|---|---|---|---|---|---|---|
| intercept | 0.00 | 0.01 | 0.38 | 0.70 | -0.01 | 0.01 |
| ma.L1 | -0.58 | 0.17 | -3.30 | 0.00 | -0.92 | -0.23 |
| ma.L2 | -0.26 | 0.17 | -1.49 | 0.14 | -0.59 | 0.08 |
| $s^2$ | 0.03 | 0.01 | 5.25 | 0.00 | 0.02 | 0.05 |

### Anti-Vaccine Group "Nontoxic" Reactions: ARIMA(0,1,1), Q=0.62, p=0.43; H=0.52, p=0.19, JB=1.81, p=0.41

| | | | | | | |
|---|---|---|---|---|---|---|
| intercept | 0.00 | 0.01 | -0.38 | 0.70 | -0.02 | 0.01 |
| ma.L1 | -0.73 | 0.12 | -6.30 | 0.00 | -0.95 | -0.50 |
| $s^2$ | 0.02 | 0.01 | 4.09 | 0.00 | 0.01 | 0.03 |

### Pro-Vaccine Page "Toxic" Reactions: ARIMA(0,1,1), Q=0.16, p=0.69; H=0.25, p=0.01, JB=137.53, p<0.001

| | | | | | | |
|---|---|---|---|---|---|---|
| intercept | -0.02 | 0.02 | -0.72 | 0.48 | -0.06 | 0.03 |
| ma.L1 | -0.81 | 0.10 | -7.84 | 0.00 | -1.01 | -0.61 |
| $s^2$ | 0.30 | 0.04 | 7.62 | 0.00 | 0.22 | 0.37 |

### Pro-Vaccine Page "Nontoxic" Reactions: ARIMA(0,1,1), Q=0.09, p=0.77; H=0.21, p=0.00, JB=625.67, p<0.001

| | | | | | | |
|---|---|---|---|---|---|---|
| intercept | 0.00 | 0.03 | 0.07 | 0.95 | -0.06 | 0.06 |
| ma.L1 | -0.75 | 0.12 | -6.18 | 0.00 | -0.99 | -0.51 |
| $s^2$ | 0.26 | 0.03 | 9.13 | 0.00 | 0.20 | 0.31 |

### Anti-Vaccine Page "Iffy": ARIMA(1,0,0), Q=0.01, p=0.92; H=1.10, p=0.84, JB=1.26, p=0.53

| | | | | | | |
|---|---|---|---|---|---|---|
| intercept | -1.14 | 0.30 | -3.78 | 0.00 | -1.73 | -0.55 |
| ar.L1 | 0.49 | 0.13 | 3.70 | 0.00 | 0.23 | 0.76 |
| $s^2$ | 0.01 | 0.00 | 3.91 | 0.00 | 0.01 | 0.02 |

| Anti-Vaccine "Iffy Engagements: ARIMA(0,1,2), Q=0.03, p=0.95; H=0.92, p=0.87, JB=2.22, p=0.33 | | | | | | |
|---|---|---|---|---|---|---|
| intercept | -0.01 | 0.02 | -0.84 | 0.40 | -0.04 | 0.02 |
| ma.L1 | -0.70 | 0.17 | -4.22 | 0.00 | -1.02 | -0.37 |
| ma.L2 | -0.01 | 0.14 | -0.08 | 0.94 | -0.28 | 0.26 |
| $s^2$ | 0.11 | 0.03 | 3.98 | 0.00 | 0.05 | 0.16 |
| Anti-Vaccine Group "Iffy" Engagements: ARIMA(0,1,1), Q=0.23, p=0.63; H=2.43, p=0.08, JB=0.41, p=0.81 | | | | | | |
| intercept | -0.01 | 0.01 | -0.58 | 0.56 | -0.03 | 0.02 |
| ma.L1 | -0.76 | 0.09 | -8.51 | 0.00 | -0.93 | -0.58 |
| $s^2$ | 0.10 | 0.02 | 4.49 | 0.00 | 0.06 | 0.15 |
| Anti-Vaccine Page HQHS: ARIMA(0,1,1), Q=1.54, p=0.22; H=0.41, p=0.08, JB=0.68, p=0.71 | | | | | | |
| intercept | 0.00 | 0.01 | -0.41 | 0.68 | -0.02 | 0.02 |
| ma.L1 | -0.80 | 0.08 | -9.83 | 0.00 | -0.97 | -0.64 |
| $s^2$ | 0.08 | 0.02 | 4.29 | 0.00 | 0.05 | 0.12 |
| Anti-Vaccine HQHS Engagements: ARIMA(0,1,1), Q=0.16, p=0.69; H=1.50, p=0.41, JB=0.56, p=0.76 | | | | | | |
| intercept | -0.02 | 0.03 | -0.53 | 0.59 | -0.07 | 0.04 |
| ma.L1 | -0.64 | 0.11 | -5.70 | 0.00 | -0.85 | -0.42 |
| $s^2$ | 0.33 | 0.08 | 4.43 | 0.00 | 0.19 | 0.48 |
| Anti-Vaccine Group HQHS: ARIMA(1,0,0), Q=0.35, p=0.56; H=2.45, p=0.06, JB=2.32, p=0.31 | | | | | | |
| intercept | -1.64 | 0.49 | -3.37 | 0.00 | -2.59 | -0.68 |
| ar.L1 | 0.61 | 0.11 | 5.49 | 0.00 | 0.39 | 0.83 |
| $s^2$ | 0.14 | 0.03 | 5.61 | 0.00 | 0.09 | 0.19 |
| Anti-Vaccine Group HQHS Engagements: ARIMA(0,1,1), Q=0.04, p=0.85; H=1.09, p=0.86, JB=3.58, p=0.17 | | | | | | |
| intercept | -0.02 | 0.02 | -1.04 | 0.30 | -0.07 | 0.02 |
| ma.L1 | -0.73 | 0.12 | -6.16 | 0.00 | -0.97 | -0.50 |
| $s^2$ | 0.25 | 0.05 | 4.75 | 0.00 | 0.15 | 0.36 |
| Pro-Vaccine Page HQHS: ARIMA(1,1,3), Q=0.01, p=0.94; H=0.37, p=0.05, JB=1.55, p=0.46 | | | | | | |
| intercept | 0.01 | 0.01 | 0.57 | 0.57 | -0.02 | 0.03 |
| ar.L1 | -0.76 | 0.25 | -3.06 | 0.00 | -1.25 | -0.27 |
| ma.L1 | 0.18 | 0.25 | 0.72 | 0.47 | -0.31 | 0.67 |
| ma.L2 | -0.58 | 0.29 | -1.99 | 0.05 | -1.14 | -0.01 |
| ma.L3 | -0.43 | 0.15 | -2.79 | 0.01 | -0.73 | -0.13 |
| $s^2$ | 0.07 | 0.01 | 5.23 | 0.00 | 0.04 | 0.09 |
| colspan=7 | Pro-Vaccine Page HQHS Engagements: ARIMA(1,0,0), Q=0.26, p=0.61; H=1.53, p=0.38, JB=0.47, p=0.79 |
| intercept | -3.86 | 0.99 | -3.90 | 0.00 | -5.80 | -1.92 |
| ar.L1 | 0.36 | 0.16 | 2.31 | 0.02 | 0.06 | 0.67 |
| $s^2$ | 1.22 | 0.23 | 5.24 | 0.00 | 0.76 | 1.68 |
| colspan=7 | Pro-Vaccine Group HQHS: ARIMA(0,1,2), Q=0.27, p=0.61; H=0.94, p=0.92, JB=1.51, p=0.47 |
| intercept | -0.02 | 0.03 | -0.88 | 0.38 | -0.07 | 0.03 |
| ma.L1 | -0.76 | 0.16 | -4.63 | 0.00 | -1.08 | -0.44 |
| ma.L2 | -0.02 | 0.18 | -0.14 | 0.89 | -0.38 | 0.33 |
| $s^2$ | 0.38 | 0.11 | 3.33 | 0.00 | 0.16 | 0.60 |
| colspan=7 | Pro-Vaccine Group HQHS Engagements: ARIMA(0,1,5), Q=0.31, p=0.58; H=0.44, p=0.17, JB=5,66, p=0.06 |
| intercept | -0.01 | 0.05 | -0.14 | 0.89 | -0.10 | 0.09 |
| ma.L1 | -0.47 | 0.30 | -1.58 | 0.11 | -1.06 | 0.11 |
| ma.L2 | -0.17 | 0.24 | -0.72 | 0.47 | -0.64 | 0.30 |
| ma.L3 | -0.43 | 0.23 | -1.88 | 0.06 | -0.87 | 0.02 |
| ma.L4 | 0.59 | 0.27 | 2.16 | 0.03 | 0.05 | 1.12 |
| ma.L5 | -0.37 | 0.31 | -1.18 | 0.24 | -0.97 | 0.24 |
| $s^2$ | 0.86 | 0.24 | 3.58 | 0.00 | 0.39 | 1.33 |
| colspan=7 | Anti-Vaccine Page Fact Ratings: ARIMA(0,1,4), Q=0.09, p=0.77; H=0.50, p=0.17, JB=0.82, p=0.67 |
| intercept | 0.00 | 0.01 | -0.19 | 0.85 | -0.01 | 0.01 |
| ma.L1 | -0.45 | 0.17 | -2.72 | 0.01 | -0.77 | -0.12 |
| ma.L2 | -0.55 | 0.14 | -4.06 | 0.00 | -0.81 | -0.28 |
| ma.L3 | -0.19 | 0.17 | -1.18 | 0.24 | -0.52 | 0.13 |
| ma.L4 | 0.53 | 0.15 | 3.48 | 0.00 | 0.23 | 0.83 |
| $s^2$ | 0.02 | 0.01 | 3.52 | 0.00 | 0.01 | 0.03 |
| colspan=7 | Anti-Vaccine Fact Rating Engagements: ARIMA(0,0,3), Q=0.00, p=0.96; H=0.83, p=0.70, JB=12.53, p=0.00 |
| intercept | 2.52 | 0.03 | 84.99 | 0.00 | 2.46 | 2.58 |
| ma.L1 | 0.17 | 0.15 | 1.12 | 0.26 | -0.13 | 0.46 |
| ma.L2 | -0.12 | 0.15 | -0.84 | 0.40 | -0.41 | 0.16 |
| ma.L3 | -0.41 | 0.17 | -2.47 | 0.01 | -0.73 | -0.08 |
| $s^2$ | 0.07 | 0.01 | 5.16 | 0.00 | 0.04 | 0.10 |
| colspan=7 | Anti-Vaccine Group Fact Ratings: ARIMA(0,1,1), Q=0.33, p=0.56; H=0.55, p=0.23, JB=5.14, p=0.08 |
| intercept | -0.01 | 0.01 | -0.77 | 0.44 | -0.03 | 0.01 |
| ma.L1 | -0.50 | 0.13 | -3.75 | 0.00 | -0.76 | -0.24 |
| $s^2$ | 0.02 | 0.00 | 5.92 | 0.00 | 0.01 | 0.02 |
| colspan=7 | Anti-Vaccine Group Fact Rating Engagements: ARIMA(0,0,3), Q=0.00, p=0.96; H=0.83, p=0.70, JB=12.53, p=0.00 |

| | | | | | | |
|---|---|---|---|---|---|---|
| intercept | 2.52 | 0.03 | 84.99 | 0.00 | 2.46 | 2.58 |
| ma.L1 | 0.17 | 0.15 | 1.12 | 0.26 | -0.13 | 0.46 |
| ma.L2 | -0.12 | 0.15 | -0.84 | 0.40 | -0.41 | 0.16 |
| ma.L3 | -0.41 | 0.17 | -2.47 | 0.01 | -0.73 | -0.08 |
| $s^2$ | 0.07 | 0.01 | 5.16 | 0.00 | 0.04 | 0.10 |
| Pro-Vaccine Page Fact Ratings: ARIMA(2,0,0), Q=0.04, p=0.85; H=0.64, p=0.36, JB=3.90, p=0.14 | | | | | | |
| intercept | 2.16 | 0.54 | 4.00 | 0.00 | 1.10 | 3.22 |
| ar.L1 | 0.17 | 0.17 | 1.01 | 0.31 | -0.16 | 0.50 |
| ar.L2 | 0.24 | 0.20 | 1.19 | 0.24 | -0.15 | 0.63 |
| $s^2$ | 0.01 | 0.00 | 4.42 | 0.00 | 0.01 | 0.01 |
| Pro-Vaccine Page Fact Rating Engagements: ARIMA(0,1,3), Q=0.09, p=0.77; H=0.86, p=0.75, JB=7.88, p=0.02 | | | | | | |
| intercept | 0.00 | 0.01 | -0.13 | 0.90 | -0.01 | 0.01 |
| ma.L1 | -0.91 | 0.20 | -4.52 | 0.00 | -1.31 | -0.52 |
| ma.L2 | -0.06 | 0.27 | -0.23 | 0.82 | -0.58 | 0.46 |
| ma.L3 | 0.02 | 0.15 | 0.13 | 0.90 | -0.27 | 0.30 |
| $s^2$ | 0.08 | 0.02 | 4.47 | 0.00 | 0.04 | 0.11 |
| Pro-Vaccine Group Fact Ratings: ARIMA(1,1,1), Q=0.43, p=0;51; H=0.89, p=0.81, JB=0.92, p=0.63 | | | | | | |
| intercept | 0.00 | 0.03 | 0.02 | 0.99 | -0.06 | 0.06 |
| ar.L1 | -0.30 | 0.22 | -1.41 | 0.16 | -0.73 | 0.12 |
| ma.L1 | -0.58 | 0.19 | -3.06 | 0.00 | -0.95 | -0.21 |
| $s^2$ | 0.20 | 0.05 | 4.27 | 0.00 | 0.11 | 0.29 |
| Pro-Vaccine Group Fact Rating Engagements: ARIMA(3,1,0), Q=0.04, p=0.85; H=0.93, p=0.87, JB=0.35, p=0.84 | | | | | | |
| intercept | 0.01 | 0.08 | 0.18 | 0.86 | -0.14 | 0.16 |
| ar.L1 | -0.78 | 0.12 | -6.30 | 0.00 | -1.02 | -0.54 |
| ar.L2 | -0.60 | 0.16 | -3.76 | 0.00 | -0.91 | -0.29 |
| ar.L3 | -0.56 | 0.13 | -4.42 | 0.00 | -0.81 | -0.31 |
| $s^2$ | 0.25 | 0.06 | 4.49 | 0.00 | 0.14 | 0.36 |
| Anti-Vaccine Page Bias Ratings: ARIMA(1,0,0), Q=0.15, p=0.70; H=0.69, p=0.44, JB=1.31, p=0.52 | | | | | | |
| intercept | 0.31 | 0.08 | 3.83 | 0.00 | 0.15 | 0.46 |
| ar.L1 | 0.50 | 0.13 | 3.86 | 0.00 | 0.25 | 0.75 |
| $s^2$ | 0.03 | 0.01 | 4.04 | 0.00 | 0.01 | 0.04 |
| Anti-Vaccine Bias Rating Engagements: ARIMA(2,0,2), Q=0.15, p=0.70; H=0.73, p=0.52, JB=4.31, p=0.12 | | | | | | |
| intercept | 0.40 | 0.10 | 4.25 | 0.00 | 0.22 | 0.59 |
| ar.L1 | 0.57 | 0.20 | 2.78 | 0.01 | 0.17 | 0.97 |
| ar.L2 | -0.64 | 0.18 | -3.49 | 0.00 | -1.00 | -0.28 |
| ma.L1 | -0.25 | 0.16 | -1.53 | 0.13 | -0.56 | 0.07 |
| ma.L2 | 0.87 | 0.14 | 6.34 | 0.00 | 0.60 | 1.14 |
| $s^2$ | 0.09 | 0.02 | 5.28 | 0.00 | 0.06 | 0.13 |
| Anti-Vaccine Group Bias Ratings: ARIMA(0,1,1), Q=0.48, p=0.49; H=0.68, p=0.44, JB=1.22, p=0.54 | | | | | | |
| intercept | 0.02 | 0.01 | 2.75 | 0.01 | 0.01 | 0.03 |
| ma.L1 | -0.82 | 0.10 | -8.24 | 0.00 | -1.01 | -0.62 |
| $s^2$ | 0.04 | 0.01 | 4.00 | 0.00 | 0.02 | 0.07 |
| Anti-Vaccine Group Bias Rating Engagements: ARIMA(0,1,2), Q=0.19, p=0.67; H=0.81, p=0.67, JB=1.14, p=0.57 | | | | | | |
| intercept | 0.01 | 0.01 | 1.16 | 0.25 | 0.00 | 0.02 |
| ma.L1 | -0.93 | 0.15 | -6.43 | 0.00 | -1.21 | -0.65 |
| ma.L2 | -0.01 | 0.14 | -0.07 | 0.95 | -0.29 | 0.27 |
| $s^2$ | 0.11 | 0.03 | 4.43 | 0.00 | 0.06 | 0.16 |
| Pro-Vaccine Page Fact Ratings: ARIMA(1,1,3), Q=0.04, p=0.84; H=0.79, p=0.64, JB=3.92, p=0.14 | | | | | | |
| intercept | 0.00 | 0.00 | -0.13 | 0.90 | -0.01 | 0.01 |
| ar.L1 | -0.59 | 0.28 | -2.17 | 0.03 | -1.13 | -0.06 |
| ma.L1 | -0.15 | 0.37 | -0.40 | 0.69 | -0.88 | 0.58 |
| ma.L2 | -0.37 | 0.33 | -1.15 | 0.25 | -1.01 | 0.26 |
| ma.L3 | -0.44 | 0.24 | -1.82 | 0.07 | -0.92 | 0.04 |
| $s^2$ | 0.02 | 0.00 | 4.26 | 0.00 | 0.01 | 0.03 |
| Pro-Vaccine Page Bias Rating Engagements: ARIMA(0,0,0), Q=0.00, p=0.97; H=0.60, p=0.29, JB=8.47, p=0.01 | | | | | | |
| intercept | -0.73 | 0.05 | -15.07 | 0.00 | -0.83 | -0.64 |
| $s^2$ | 0.07 | 0.02 | 4.13 | 0.00 | 0.04 | 0.11 |
| Pro-Vaccine Group Bias Ratings: ARIMA(1,1,2), Q=0.03, p=0.85; H=0.17, p=0.00, JB=15.13, p<0.001 | | | | | | |
| intercept | -0.02 | 0.01 | -1.26 | 0.21 | -0.04 | 0.01 |
| ar.L1 | -0.73 | 0.19 | -3.75 | 0.00 | -1.11 | -0.35 |
| ma.L1 | -0.07 | 0.27 | -0.26 | 0.80 | -0.61 | 0.47 |
| ma.L2 | -0.87 | 0.26 | -3.42 | 0.00 | -1.37 | -0.37 |
| $s^2$ | 0.16 | 0.04 | 3.87 | 0.00 | 0.08 | 0.24 |
| Pro-Vaccine Group Bias Rating Engagements: ARIMA(2,0,0), Q=0.01, p=0.94; H=0.17, p=0.00, JB=86.71, p<0.001 | | | | | | |
| intercept | -0.75 | 0.17 | -4.44 | 0.00 | -1.08 | -0.42 |
| ar.L1 | 0.18 | 0.21 | 0.90 | 0.37 | -0.22 | 0.59 |
| ar.L2 | -0.39 | 0.19 | -2.07 | 0.04 | -0.76 | -0.02 |
| $s^2$ | 0.27 | 0.05 | 5.04 | 0.00 | 0.17 | 0.38 |

Note. ar = autoregressive term; ma = moving average term; $s^2$ = variance; "Iffy" = Proportion of URLs found on the Iffy Index of Unreliable Sources; HQHS = Proportion of URLs that are "High Quality Health Sources" (i.e., Government and Academic Sources); Q = Ljung-Box statistic; H = Heteroskedasticity; JB = Jarque-Bera statistic

# References

1. Gu, J. *et al.* The impact of Facebook's vaccine misinformation policy on user endorsements of vaccine content: An interrupted time series analysis. *Vaccine* **40**, 2209–2214 (2022).

2. Dias, N., Pennycook, G. & Rand, D. G. Emphasizing publishers does not effectively reduce susceptibility to misinformation on social media. *Harvard Kennedy School Misinformation Review* **1**, (2020).

3. Curry Jansen, S. & Martin, B. The Streisand Effect and Censorship Backfire. *International Journal of Communication* **9**, 656–671 (2015).

4. Pennycook, G., Bear, A., Collins, E. T. & Rand, D. G. The Implied Truth Effect: Attaching Warnings to a Subset of Fake News Headlines Increases Perceived Accuracy of Headlines Without Warnings. *Management Science* **66**, 4944–4957 (2020).

5. Bail, C. A. *et al.* Exposure to opposing views on social media can increase political polarization. *Proceedings of the National Academy of Sciences* **115**, 9216–9221 (2018).

6. Bak-Coleman, J. B. *et al.* Combining interventions to reduce the spread of viral misinformation. *Nat Hum Behav* 1–9 (2022) doi:10.1038/s41562-022-01388-6.

7. Pennycook, G. *et al.* Shifting attention to accuracy can reduce misinformation online. *Nature* **592**, 590–595 (2021).

8. Roozenbeek, J., Freeman, A. L. J. & van der Linden, S. How Accurate Are Accuracy-Nudge Interventions? A Preregistered Direct Replication of Pennycook et al. (2020). *Psychol Sci* 095679762110245 (2021) doi:10.1177/09567976211024535.

9. Fighting Coronavirus Misinformation and Disinformation. *Center for American Progress* https://www.americanprogress.org/article/fighting-coronavirus-misinformation-disinformation/.

10. Okada, M., Yamanishi, K. & Masuda, N. Long-tailed distributions of inter-event times as mixtures of exponential distributions. *Royal Society Open Science* **7**, 191643.

11. Akaike, H. A new look at the statistical model identification. *IEEE transactions on automatic control* **19**, 716–723 (1974).

12. Schwarz, G. Estimating the dimension of a model. *The annals of statistics* **6**, 461–464 (1978).

13. Danilak, M. M. langdetect: Language detection library ported from Google's language-detection.

14. Sanh, V., Debut, L., Chaumond, J. & Wolf, T. DistilBERT, a distilled version of BERT: smaller, faster, cheaper and lighter. Preprint at http://arxiv.org/abs/1910.01108 (2020).

15. Perspective API. https://www.perspectiveapi.com/.